\documentclass[aps,prl,twocolumn,showpacs,superscriptaddress,letterpaper]{revtex4-2}
\usepackage[utf8]{inputenc}
\usepackage{graphicx}
\usepackage{textcomp}
\usepackage{amsmath,amssymb}
\usepackage{color}
\usepackage{soul}
\usepackage[normalem]{ulem}
\usepackage{mathtools}
\usepackage{multirow}
\usepackage{array}
\usepackage{float}
\usepackage{multirow}
\usepackage[colorlinks=true, urlcolor=blue, citecolor=blue, linkcolor=blue]{hyperref}
\usepackage{orcidlink}
\setstcolor{red}
\usepackage[absolute,overlay]{textpos}

\begin{document}

\title{Many-Body \emph{Structural Effects} in Periodically Driven Quantum Batteries}

\author{Rohit Kumar Shukla\,\orcidlink{0000-0003-2693-8745}}
\email[]{rohitkrshukla.rs.phy17@itbhu.ac.in}
\affiliation{Department of Chemistry, Institute of Nanotechnology and Advanced Materials, Center for Quantum Entanglement Science and Technology, Bar-Ilan University, Ramat-Gan 5290002, Israel}

\author{Cheng Shang\,\orcidlink{0000-0001-8393-2329}}
\email[]{cheng.shang@riken.jp}
\affiliation{Analytical Quantum Complexity RIKEN Hakubi Research Team, RIKEN Center for Quantum Computing (RQC), Wako, Saitama 351-0198, Japan}

\date{\today}

\begin{abstract}
While quantum batteries have been widely studied under static driving, their performance under periodic driving in many-body systems has received only limited attention. In this Letter, we uncover structural principles establishing that many-body structure fundamentally determines the charging performance of a collective spin-1/2 quantum battery driven by a periodic Ising charger. In particular, interaction range, boundary conditions, system size, and integrability—capturing the graph connectivity, geometry, even-odd effect, and dynamics of the many-body system—emerge as critical factors in enhancing stored energy and charging power. First, we analyze how connectivity scaling and boundary geometry shape quantum battery performance. We show that long-range interacting chargers exhibit super-extensive energy storage, approaching the well-known fundamental upper bound across broad ranges of driving periods and system sizes. In contrast, nearest-neighbor chargers achieve optimal charging only under finely tuned commensurability conditions. Moreover, we indicate that open boundary conditions (OBC) enhance robustness relative to periodic boundary conditions (PBC). Second, we further examine the impact of integrability and periodic driving on charging dynamics. We demonstrate that nonintegrability consistently enhances energy storage by suppressing conserved quantities and promoting ergodic Floquet dynamics, thereby enabling efficient population of the many-body spectrum. Through systematic structural optimization across multiple parameters, we identify long-range nonintegrability as a central resource for fast, scalable, and robust charging of collective quantum batteries. Our results elucidate how the complex structural features of many-body systems, together with periodic driving, can be harnessed to achieve efficient collective charging dynamics.
\end{abstract}

\maketitle
\textit{Introduction.}—Motivated by the fundamental limitations of classical chemical batteries and the ongoing miniaturization of quantum devices, quantum batteries (QBs) have emerged as a promising paradigm for energy storage and work extraction at microscopic scales. Since the seminal proposal by Alicki and Fannes~\cite{alicki2013entanglement}, a central question has been whether quantum effects provide genuine operational advantages over classical batteries. Extensive studies have shown that uniquely quantum features, including entanglement, give rise to energy storage and release processes that outperform their classical counterparts~\cite{campaioli2017enhancing,seah2021quantum,Zhu2023Charging,downing2024hyperbolic,friis2018precision,Lu2025topological,rossini2020quantum,gyhm2022quantum,gyhm2024beneficial,rosa2020ultra,rodriguez2021collective,konar2024quantum,mazzoncini2023optimal,zhang2019powerful,yang2023battery,shukla2026collective,ferraro2018high,andolina2019quantum}.  While entanglement naturally arises during the charging process and plays an important role in quantum battery (QB)~\cite{alicki2013entanglement,binder2015quantacell,campaioli2017enhancing}, it is now understood that, especially in quantum many-body systems where ``more is different,'' entanglement is not the exclusive origin of quantum advantage. In fact, recent studies have demonstrated that it is neither a necessary nor a sufficient resource for enhanced charging power in many-body batteries, as maximal performance can be achieved even in the absence of strong multipartite correlations~\cite{andolina2019quantum,andolina2019extractable,julia2020bounds,kamin2020entanglement,shi2022entanglement,perarnau2015extractable}. This indicates that collective effects and coherence in many-body systems provide additional operational resources beyond entanglement. In particular, the emergence of collective quantum effects in interacting many-body systems can further enhance QB performance, leading to faster charging rates, increased stored energy, and greater extractable work~\cite{binder2015quantacell,campaioli2017enhancing,ferraro2018high,andolina2019quantum,andolina2019extractable,le2018spin}. These collective advantages have motivated substantial theoretical efforts to seek optimal charging protocols and to elucidate the fundamental physical resources governing many-body QB performance~\cite{ferraro2018high,andolina2019quantum,PhysRevE.100.032107,rossini2020quantum,julia2020bounds,le2018spin,ghosh2020enhancement,PhysRevA.104.032207,PhysRevA.105.022628,PhysRevA.104.L030402,PhysRevB.102.245407,PhysRevResearch.4.033216,PhysRevA.106.022618,PhysRevA.110.022226,konar2024quantum,chaki2024positive,PhysRevA.106.062609,PhysRevA.107.032203,cyrc-ms34,chaki2024universal,PhysRevA.108.042618,PhysRevA.110.012227,PhysRevApplied.19.064069,PhysRevLett.132.090401,PhysRevLett.133.180401,mitra2025bound,perciavalle2025extractable,PhysRevLett.131.240401,PhysRevB.100.115142,PhysRevE.108.064106,gyhm2024beneficial}. Many of these proposals have subsequently been implemented or explored experimentally in platforms including quantum dots~\cite{PhysRevLett.131.260401}, transmons~\cite{PhysRevA.107.023725,hu2022optimal,gemme2022ibm}, organic semiconductors~\cite{quach2022superabsorption}, superconducting circuits~\cite{hu2026quantum}, and nuclear magnetic resonance systems~\cite{PhysRevA.106.042601}. In light of recent theoretical and experimental advances, QBs have become a field rich in both opportunities and challenges~\cite{ferraro2026opportunities}.

Collective and coherence advantages render many-body QBs not only powerful platforms for energy storage but also versatile windows into complex many-body dynamics~\cite{puri2024floquet,romero2025kicked}. However, despite these appealing features, optimizing QB performance remains highly nontrivial due to the intrinsic structural complexity of interacting many-body systems. The charging dynamics and energy landscape depend sensitively on the interaction range (including short-range, power-law, and all-to-all forms), boundary conditions (e.g., periodic or open), system size, and integrability, each of which reshapes the underlying graph connectivity and spectral structure. These intertwined structural features introduce competing length and energy scales, thereby rendering optimization a genuine multi-parameter problem. The challenge is further amplified in the presence of external driving, where the driving protocol must be optimized in conjunction with system size, interaction structure, and boundary geometry~\cite{andolina2019quantum,mondal2022periodically}. While early works primarily considered quench-based unitary charging~\cite{binder2015quantacell,campaioli2017enhancing}, it has become clear that periodic and time-dependent driving provide a powerful way to engineer effective interactions, stabilize nonequilibrium steady states, and access dynamical regimes unattainable in static settings~\cite{bukov2015universal,eckardt2017colloquium}. In this context, Floquet or time-periodic driving has also become a powerful tool for probing nonequilibrium many-body quantum dynamics, enabling phenomena such as topological phases, prethermalization, dynamical localization, and artificial gauge fields~\cite{eckardt2015high,mori2023floquet,PhysRevB.106.224306,PhysRevResearch.5.043008,bloch2008many,PhysRevLett.111.185301}. Such protocols are experimentally accessible in platforms including cold atoms in optical lattices~\cite{wintersperger2020realization,PhysRevLett.130.043201}, trapped ions~\cite{katz2025floquet,PhysRevLett.123.213605}, and superconducting qubits~\cite{zhang2022digital}, positioning Floquet engineering as a well-established framework for introducing nonequilibrium dynamics into many-body quantum systems~\cite{PhysRevResearch.6.013116,fauseweh2023quantum,PhysRevApplied.21.044050,PhysRevE.106.024110,PhysRevE.92.042123,niedenzu2018cooperative,mondal2022periodically,PhysRevB.111.024315}. Moreover, from the perspective of open quantum systems, Floquet protocols suppress decoherence and mitigate aging effects through the formation of bound states in the quasienergy spectrum of many-body systems coupled to environments~\cite{bai2020floquet}, thereby strengthening robustness against dissipation.

Combining the advantages of many-body effects with advances in periodic driving, periodically driven many-body QBs have attracted increasing attention~\cite{mondal2022periodically,romero2025kicked,bai2020floquet,puri2024floquet,downing2023quantum,crescente2020charging,schmid2026superextensive,dvs.202520281}. Periodically driven QBs were first introduced in~\cite{mondal2022periodically}. More recently, the kicked-Ising QB model was proposed, establishing a Floquet framework to analyze the energy-injection dynamics at the self-dual point~\cite{romero2025kicked}. However, these prior studies are largely restricted to short-range interactions and integrable regime. Notably, recent works on driven spin systems have further shown that long-range interactions can induce super-extensive scaling of charging power with system size~\cite{schmid2026superextensive}, while resonant tunneling in integrable transverse-field Ising models enhances energy transfer and stabilizes stored energy~\cite{puri2024floquet,mondal2022periodically}. These results highlight the important roles of interaction range and driving protocol in determining QB performance, yet their implications in the nonintegrable regime remain poorly understood. Therefore, a systematic understanding of how structural effects interplay with time-dependent Floquet driving remains crucial for advancing high-performance QBs.

In this Letter, building upon the prior work \cite{romero2025kicked,mondal2022periodically}, we investigate the \emph{structural effects} in periodically driven Ising QBs within a general Floquet framework with independently tunable interaction and field durations, covering both integrable and integrability-breaking regimes. Our analysis transcends the kicked and self-dual limits, enabling a unified and comprehensive exploration of long-range Ising interactions with power-law decay and their nearest-neighbor counterparts under both OBC and PBC, in integrable and nonintegrable regimes. By systematically analyzing how many-body \emph{structural effects} and periodic driving protocols jointly shape the charging dynamics, we characterize the stored energy and charging power across a broad parameter space. We find that periodic driving robustly and scalably enhances QB performance, enabling the QB to attain optimal storage energy through appropriate choices of system and driving parameters. In particular, integrability breaking and long-range interactions emerge as key points for enhancing QB charging performance, underscoring the central role of \emph{structural effects} in periodically driven Ising QBs. Our results provide valuable insights into the design and optimization of complex periodically driven many-body QBs.

\textit{Model and charging protocol.—}We consider a quantum battery (QB) composed of $N$ spin-$1/2$ units with bare Hamiltonian $H_B = \omega \sum_{j=1}^N \sigma_j^z$, where $\omega$ denotes the energy splitting and $\sigma_j^z$ is the Pauli-$z$ operator acting on the $j$th spin. The QB is initially prepared in its ground state, corresponding to a fully discharged configuration. Charging is realized by applying a time-periodic driving Hamiltonian $H_{\rm C}(t)$, which is switched on at $t=0$ and repeats with period $T$, i.e., $H_{\rm C}(t+T)=H_{\rm C}(t)$ $\forall t$. The Floquet nature of the charger enables stroboscopic evolution of the system at discrete times $t=nT$, where $n\in\mathbb{Z}$. During each period $T=\tau_0+\tau_1$, the Hamiltonian remains piecewise constant and takes the form
\begin{equation}
\!H_{\rm C}(t) =
\begin{cases}
J H_{xx} + h_x H_x, & t \in [nT, nT+\tau_0),\\[1ex]
h_z H_z, & t \in [nT+\tau_0, (n+1)T),
\end{cases}
\label{Charger_H}
\end{equation}
where $H_{x(z)}=\sum_{j=1}^N \sigma_j^{x(z)}$ and $J$ is the interaction strength. Here, $\tau_0$ and $\tau_1$ are independently tunable durations of the two consecutive intervals within each period, over which the Hamiltonian remains constant. The internal interactions of the charger are governed by a $\mathcal{K}$-local long-range Ising coupling along the $x$ direction. We consider two common geometric boundary conditions: PBC and OBC. Specifically, for chains with both ends connected, PBC are defined by $\sigma_{N+k}^x \coloneqq \sigma_k^x$, which preserve translational invariance and eliminate edge effects, under which the interaction Hamiltonian takes the form $H_{xx} = \sum_{j=1}^{N} \sum_{k=1}^{\mathcal{K}} \frac{1}{2^{k-1}} \, \sigma_j^x \sigma_{j+k}^x,$ where $\mathcal{K}$ denotes the maximum interaction range measured from site $j$, with $\mathcal{K} = (N-1)/2$ for odd $N$ and $\mathcal{K} = N/2$ for even $N$ to avoid double counting. For chains with open ends, OBC restrict interactions within the chain, yielding $H_{xx} = \sum_{j=1}^{N-1} \sum_{k=1}^{\mathcal{K}} \frac{1}{2^{k-1}} \, \sigma_j^x \sigma_{j+k}^x,$ where $\mathcal{K}=N-j$ ensures that only spin pairs within the chain are coupled. As a limiting case, restricting to $k=1$ reduces the long-range interaction to the standard nearest-neighbor Ising coupling, thereby recovering conventional Ising QB models, while finite $\mathcal{K}$ enables exploration of long-range effects on periodically driven charging dynamics. The total Hamiltonian is 
\begin{equation}
H(t) = H_B + \lambda(t)[H_{\rm C}(t) - H_B],\label{total Hamiltonian}
\end{equation}
where $\lambda(t)=1$ during the charging interval $[0,nT]$ and vanishes otherwise, ensuring that $[H_B,H_{\rm C}(t)]\neq 0$ so as to enable energy transfer. The QB evolves unitarily under driven quantum dynamics as $\rho(t) = U(t) \rho(0) U^\dagger(t)$, where $U(t) = \mathcal{T}\exp[-i \int_0^t H_{\rm C}(t') dt'/\hbar]$ and $\rho(0)$ denotes the ground state of the bare QB Hamiltonian $H_B$. In the following, we set $\hbar=1$, such that the stroboscopic evolution at integer multiples of the period $T$ is governed by the Floquet operator $U_F = \exp(-i \tau_1 h_z H_z) \exp[-i \tau_0 (J H_{xx} + h_x H_x)]$, and after $n$ periods the state becomes $\rho(nT)=U_F^n \rho(0) U_F^{\dagger n}$. This approach allows the charging protocol to be freely tuned via the parameters $\tau_0$, $\tau_1$, $J$, $h_x$, and $h_z$, thereby extending beyond the constraints of the self-dual point. Whereas earlier work~\cite{romero2025kicked} explored the instantaneous-kick limit ($\tau_1 \to 0$) at the self-dual point, our protocol retains finite durations for both steps, thereby capturing a broader and more realistic range of QB dynamics under piecewise-constant driving.

\textit{Charging performance metrics of QB.}—To quantify the charging performance of QBs, we introduce two crucial thermodynamic indicators. One is the stored energy, which for the QB at time $t$ is defined as the change in the expectation value of the QB Hamiltonian, $\Delta E(t)=\mathrm{Tr}[H_B\rho(t)]-\mathrm{Tr}[H_B\rho(0)]$. Based on the stored energy, the other is the charging power. The corresponding average charging power of the QB is defined as $P(t)=\Delta E(t)/t$, where $t=nT$ with $n\in\mathbb{Z}$ and $T=\tau_0+\tau_1$.

\begin{figure}
    \centering
\includegraphics[width=0.49\linewidth,height=0.40\linewidth]{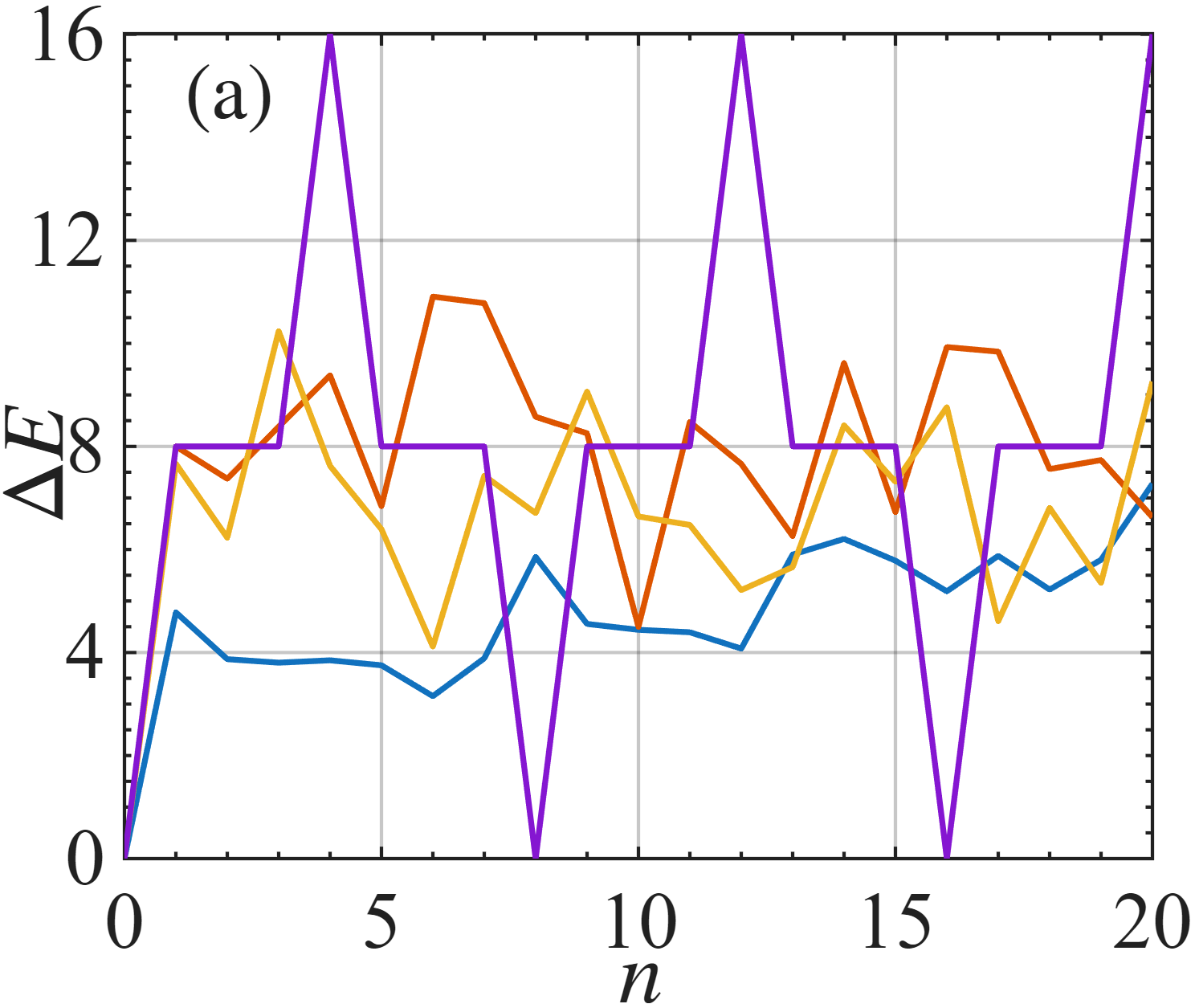}
\includegraphics[width=0.49\linewidth,height=0.40\linewidth]{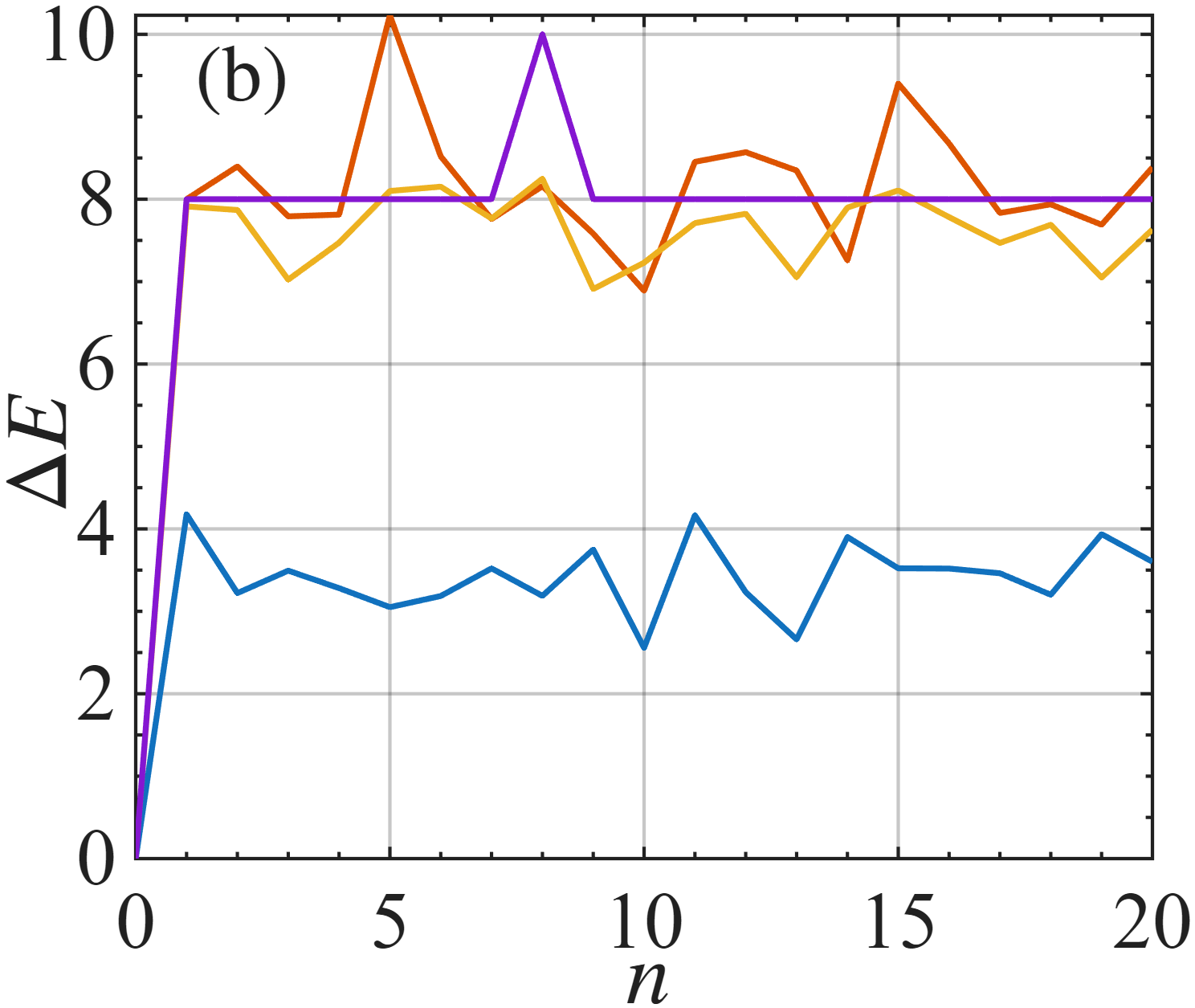}
\includegraphics[width=0.49\linewidth,height=0.40\linewidth]{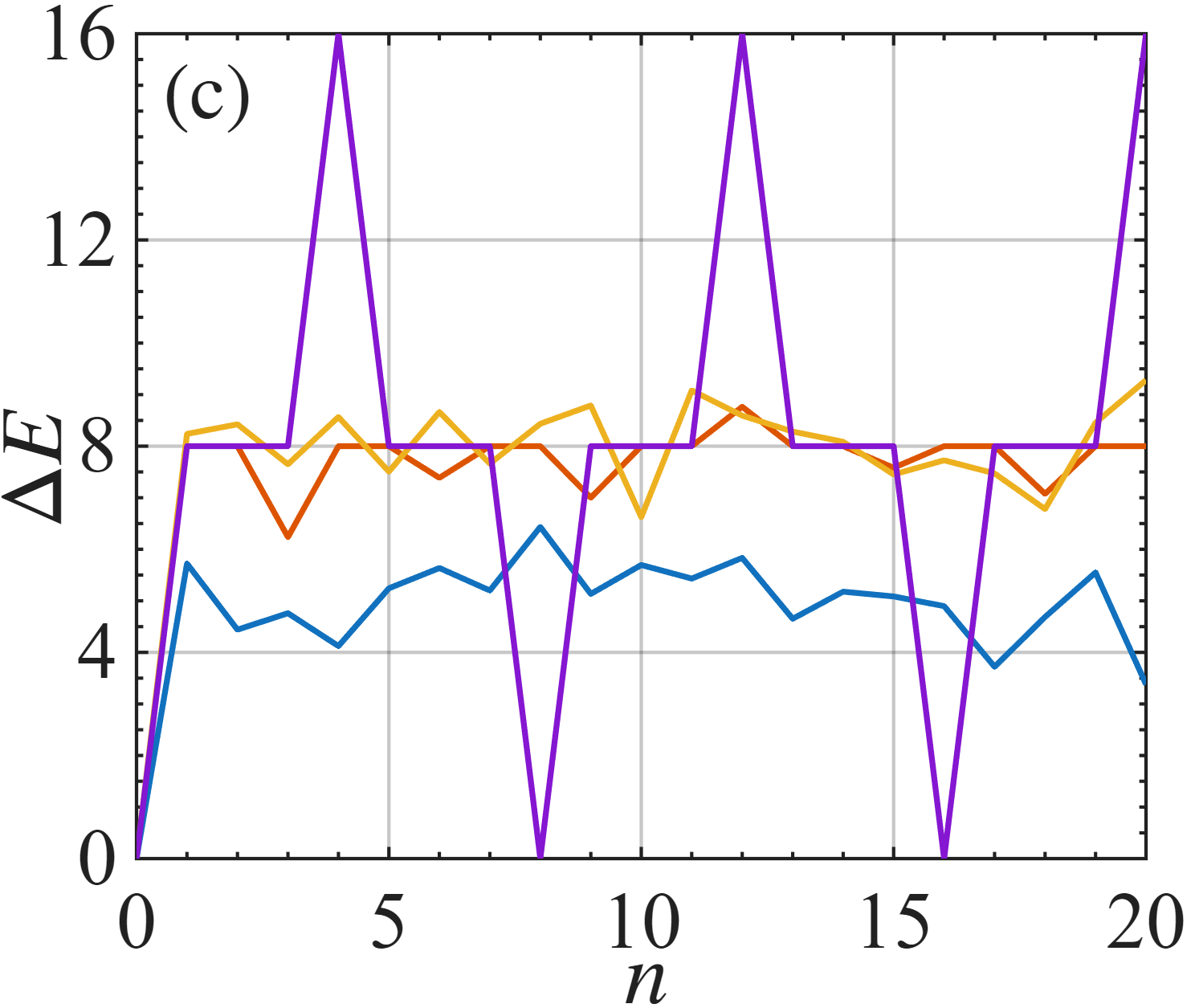}
\includegraphics[width=0.49\linewidth,height=0.40\linewidth]{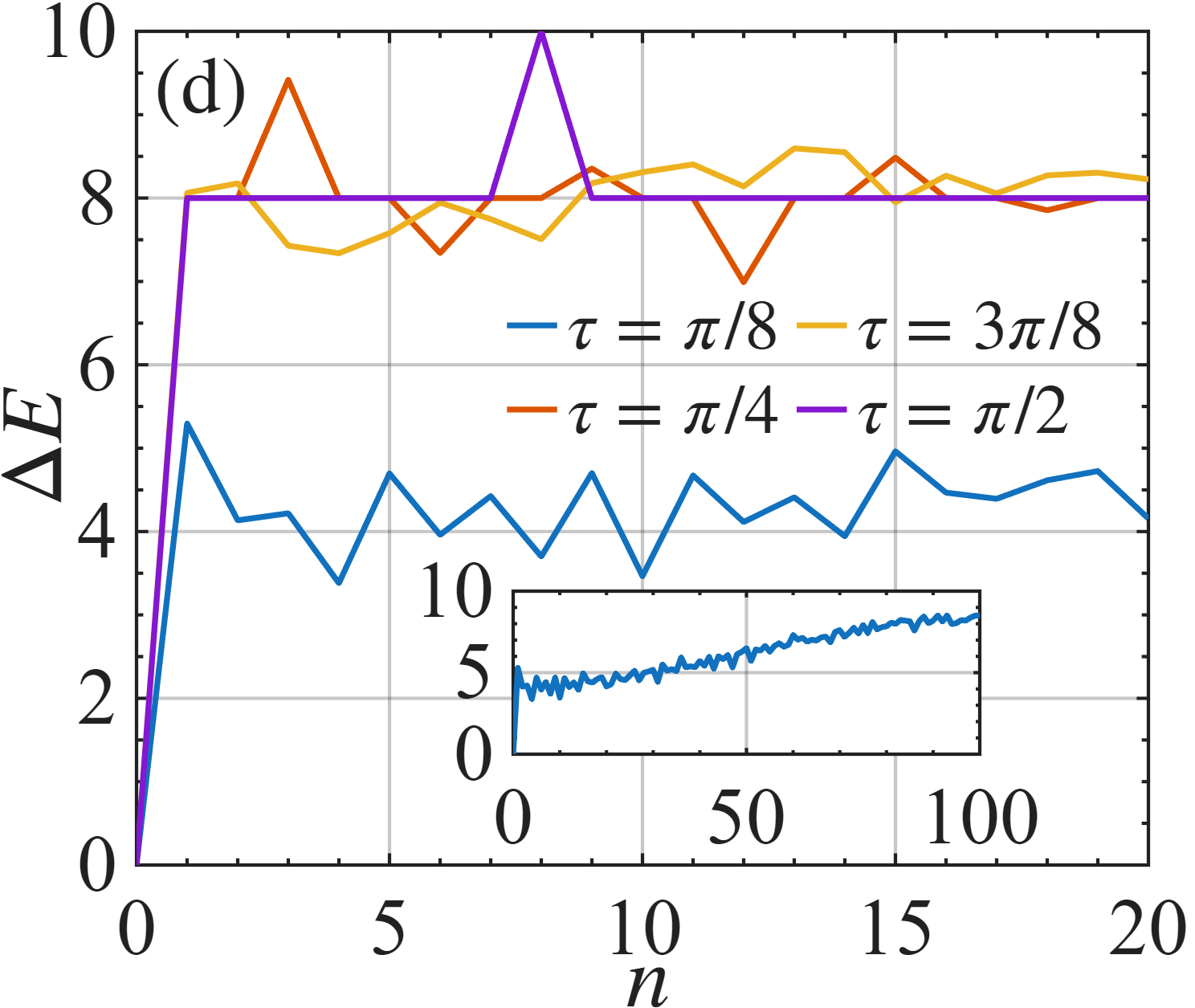}
\caption{Stored energy $\Delta E$ as a function of the number of kicks $n$ for a noninteracting QB driven by a time-periodic long-range Ising charger, shown for different driving periods with $\tau_0=\tau_1=\tau$ (see legend). Panels (a,b) correspond to the integrable charger, whereas panels (c,d) represent the nonintegrable case. PBC are used in (a,c), and OBC in (b,d). The inset of panel (d) highlights the long-time behavior of the stored energy for $\tau=\pi/8$. The system parameters are chosen as $N=8$, $J=1$, $h_x=0$ (integrable) or $h_x=1$ (nonintegrable), $h_z=1$, and $\omega=1$.}
 \label{long_E_tau_int}
\end{figure}

\begin{figure}
    \centering
\includegraphics[width=0.49\linewidth,height=0.40\linewidth]{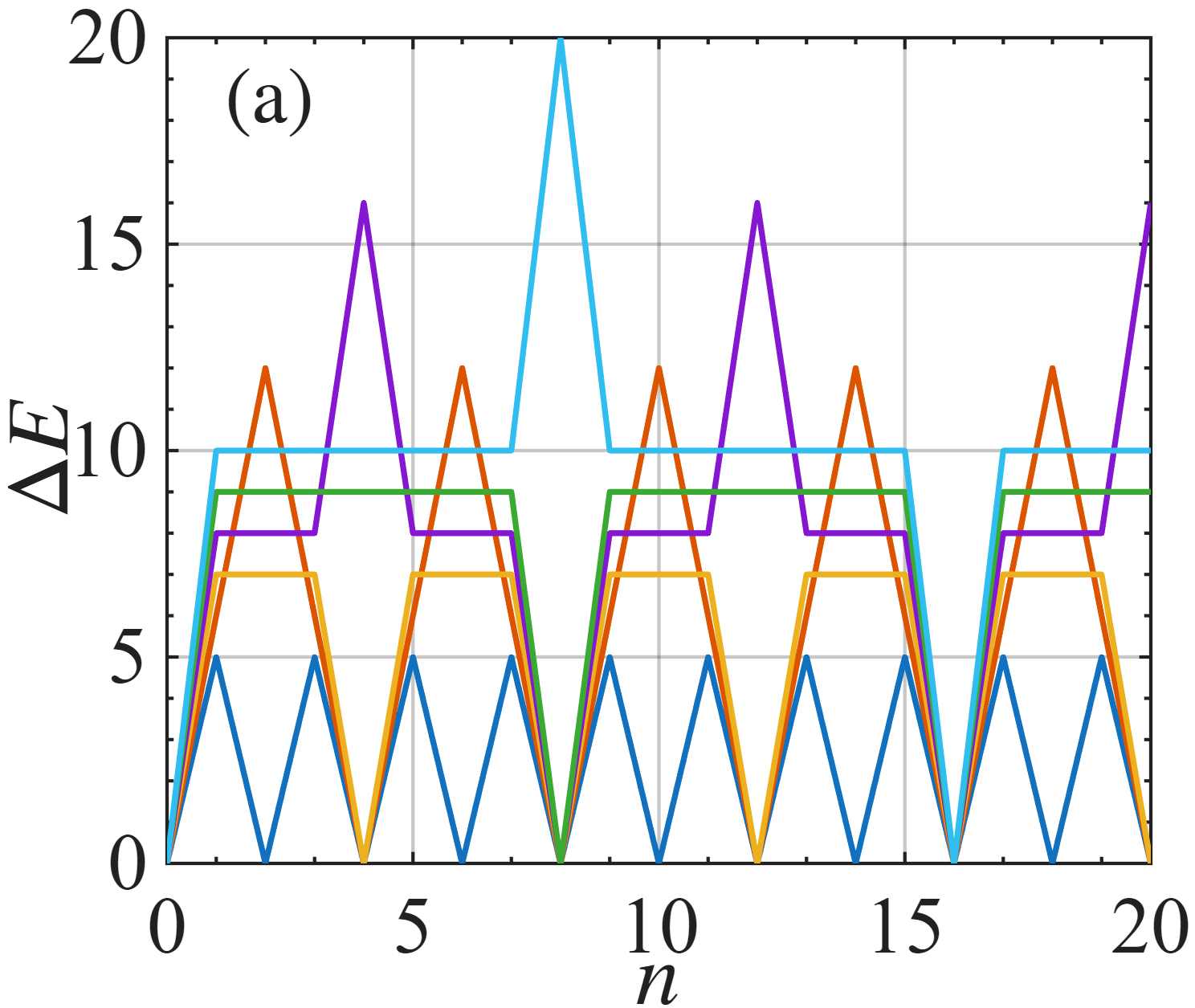}
\includegraphics[width=0.49\linewidth,height=0.40\linewidth]{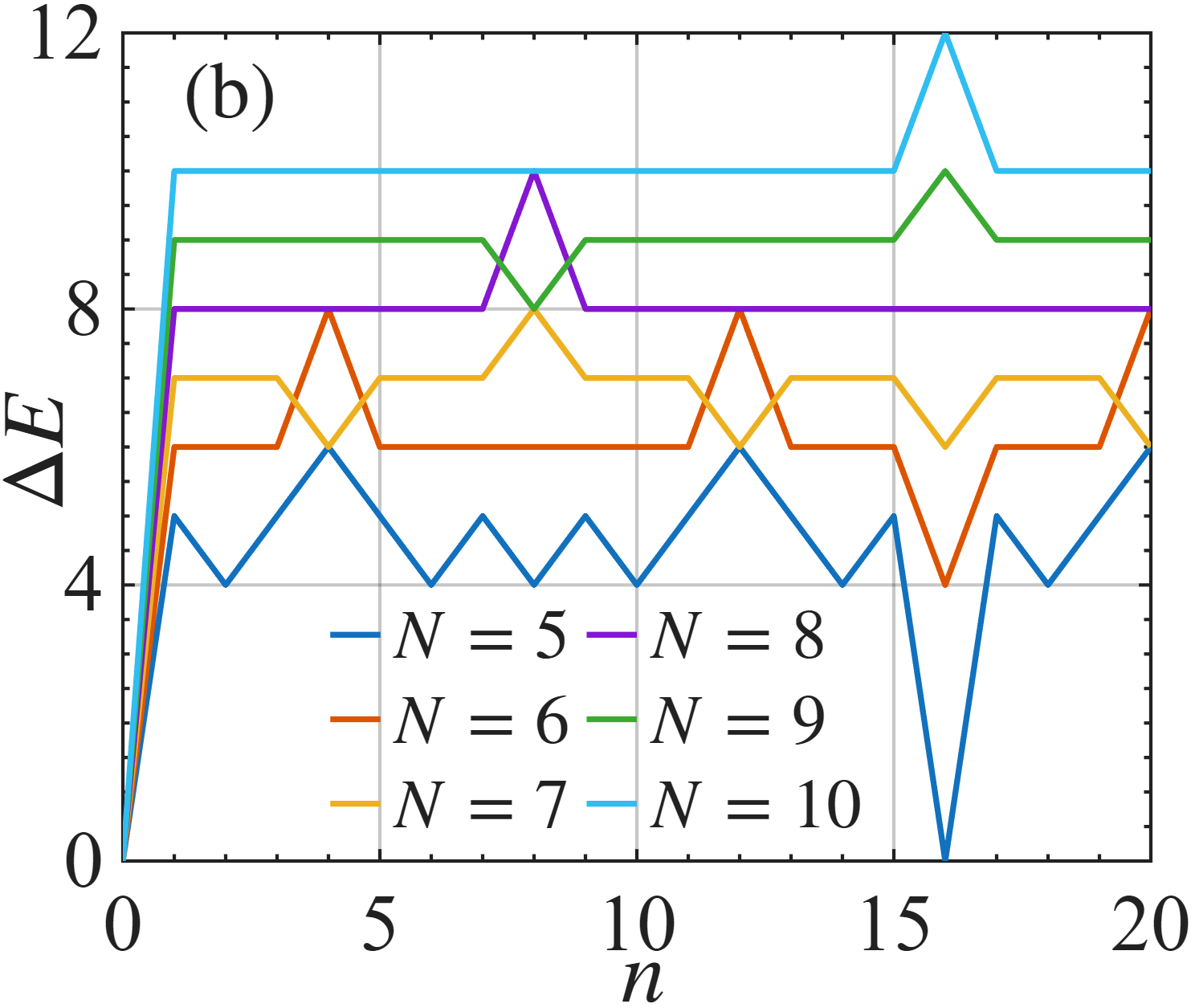}
\caption{Stored energy $\Delta E$ as a function of the number of kicks $n$ for a noninteracting QB driven by a time-periodic long-range interacting charger, shown for different system sizes $N$ (see legend). Panels (a,b) correspond to the integrable regime, whereas panels (c,d) represent the nonintegrable regime. PBC are used in panels (a,c), while OBC are used in panels (b,d). The system parameters are chosen as $\tau_0=\tau_1=\pi/2$, $J=1$, $h_x=0$ (integrable) or $h_x=1$ (nonintegrable), $h_z=1$, and $\omega=1$.}
    \label{E_ATA_N_taupi2_int}

\end{figure}

\textit{QB optimization strategy.—}We investigate how structural features of a periodically driven Ising charger control the performance of QBs. Specifically, we demonstrate that the interplay among interaction range, boundary conditions, and integrability qualitatively determines both the achievable stored energy and charging efficiency. The QB consists of $N$ spin-$1/2$ subsystems with level spacing $\omega$, whose maximum storage capacity is $E_{\max}=2\omega N$~\cite{shukla2025optimizing}.~Charging is implemented via a Floquet Ising protocol in which interactions and transverse fields act sequentially within each driving period. By tuning the interaction profile from nearest-neighbor to long-range and selectively breaking integrability via a longitudinal field, we identify structural regimes in which optimal charging is either realized or fundamentally suppressed.

\textit{Long-range interacting charger.—}We first focus on the stored energy in a general setting, namely a noninteracting QB driven by a long-range Ising charger.~To clarify the role of the driving protocol, we consider equal-duration steps $\tau_0=\tau_1 =\tau$ and analyze representative driving periods $\tau=\pi/8,\ \pi/4,\ 3\pi/8$, and $\pi/2$ at fixed system size $N=8$. Both periodic and open boundary conditions (PBC and OBC) are considered. This setting enables a direct assessment of how structural symmetries, integrability, and boundary effects jointly shape energy accumulation under repeated Floquet driving. In the following, we discuss how \emph{structural effects} influence QB stored energy across different scenarios.

We begin with the integrable charger.~Under PBC, the charging dynamics are highly sensitive to the driving period. For $\tau=\pi/2$, the QB reaches its maximum storage energy, $\Delta E_{\max}=2\omega N$, corresponding to perfectly efficient charging. The energy exhibits a clear periodic dependence on the number of kicks, with a period equal to the system size $N$. As a result, the QB becomes fully charged after $n=N/2$ kicks, indicating a resonance between the Floquet driving and the intrinsic many-body structure of the charger. For other driving periods, this coherent buildup does not occur. In particular, for intermediate periods $\tau=\pi/4$ and $\tau=3\pi/8$, the stored energy increases but saturates below the optimal value, remaining in the range $\omega N<\Delta E_{\max}<2\omega N$. For shorter periods $\tau=\pi/8$, the QB is only weakly charged, with the stored energy remaining below half of its maximum value, $\Delta E_{\max}<\omega N$, as shown in Fig.~\ref{long_E_tau_int}(a). In contrast, the impact of boundary conditions becomes evident under OBC. Unlike the PBC case, the QB never reaches full charge for any of the driving periods studied. Instead, the stored energy is systematically reduced and remains bounded within $\omega N<\Delta E_{\max}<2\omega N$ for all $0<\tau\leq\pi/2$, as shown in Fig.~\ref{long_E_tau_int}(b). This suppression underscores the essential role of translational invariance and closed-loop correlations in facilitating optimal energy storage in the integrable regime.

We next turn to the nonintegrable regime by introducing a transverse field $h_x=1$, which breaks integrability and induces chaotic dynamics in the charger. In this case, the dependence on the driving period becomes less structured but more robust. Under PBC, the stored energy exceeds $\omega N$ for all driving periods except the shortest one, $\tau=\pi/8$, and attains the optimal value $\Delta E_{\max}=2\omega N$ at $\tau=\pi/2$, as shown in Fig.~\ref{long_E_tau_int}(c). This behavior indicates that chaotic dynamics enhance energy redistribution and reduce the fine-tuning required for efficient charging. Under OBC, the stored energy displays a nonmonotonic dependence on $\tau$, reaching its maximum at $\tau=\pi/2$. Remarkably, for all driving periods considered, the stored energy remains above half of the maximum value, $\Delta E_{\max}>\omega N$, as shown in Fig.~\ref{long_E_tau_int}(d). Compared with the integrable case, the nonintegrable charger is less sensitive to boundary effects and provides more uniform charging across a broad range of driving parameters. More broadly, relative to the integrable charger, the nonintegrable system exhibits a systematic enhancement of energy storage.~Chaotic many-body dynamics suppress extensive conserved quantities and promote rapid operator spreading together with strong multipartite entanglement, allowing the driving protocol to access a significantly larger portion of the Hilbert space.~Consequently, energy absorption becomes more ergodic, preventing early saturation and enabling storage beyond the integrable bound. Notably, extending this analysis to unequal driving periods $\tau_0\neq\tau_1$ further enriches the charging dynamics and reveals a pronounced sensitivity to many-body \emph{structural effects}. Nevertheless, efficient charging remains robust over a broad region of the $(\tau_0,\tau_1)$ parameter space, with the stored energy typically exceeding half of the optimal value. This behavior highlights a strong resilience against temporal asymmetry in the driving protocol, as detailed in the Supplemental Material (SM)~\cite{Structural_SM}.

To summarize, many-body \emph{structural effects} play a central role in shaping the optimization landscape of QB charging, as reflected in the pronounced dependence of the maximum stored energy of a long-range interacting charger on the driving period $\tau$. For small $\tau$, energy absorption is strongly suppressed due to the near-identity Floquet evolution. A sharp crossover emerges around $\tau\simeq 0.4$, beyond which the stored energy increases rapidly and exceeds $\omega N$. Strikingly, for all boundary conditions and for both integrable and nonintegrable dynamics, the energy absorption is maximized at the resonant point $\tau=\pi/2$. At this point, the charger reaches its optimal performance, yielding $\Delta E_{\max}=2\omega N$ under PBC and $\Delta E_{\max}=\omega N+2$ under OBC for even $N$. Details of the full $\tau$ dependence are presented in the SM~\cite{Structural_SM}, and summarized in Table~\ref{tab:optimization_landscape_structured} for the long-range case over the full Floquet period range $\tau_0=\tau_1=\tau\in[0,\pi/2]$.

\textit{Long-range finite-size effects—}Before concluding the analysis of the long-range interacting regime and turning to the nearest-neighbor setting, we highlight an interesting finding for QBs, namely the emergence of finite-size effects. Specifically, we examine the system-size dependence of the stored energy at the fixed driving period $\tau = \pi/2$. For the integrable charger under PBC, Fig.~\ref{E_ATA_N_taupi2_int}(a) demonstrates that even system sizes attain the optimal stored energy, $\Delta E_{\max} = 2 \omega N$, whereas odd system sizes are restricted to $\Delta E_{\max} = \omega N$. Small systems with $N=2$ and $N=3$ constitute exceptions, exhibiting vanishing stored energy. More generally, the charging dynamics display a periodic structure, with the period scaling as $n = 2^{N/2-1}$ for even $N$ and $n = 2^{(N-1)/2}$ for odd $N$. Under OBC, a similar even–odd effect persists, but the maximal stored energy is reduced: even $N$ reach $\Delta E_{\max} = \omega N + 2$, whereas odd $N$ attain $\Delta E_{\max} = \omega N + 1$, as shown in Fig.~\ref{E_ATA_N_taupi2_int}(b). In this case, the charging dynamics remain periodic for all system sizes, with a period scaling as $n = 2^{N-1}$.

Strikingly, the same universal charging behavior emerges in both integrable and nonintegrable regimes. Under both PBC and OBC, the system exhibits coherent stroboscopic evolution that maps the initial ground state onto highly excited states. This robustness originates from the structure of the Floquet operator,
\begin{equation}
U_F = \exp[-i \tau h_z H_z] \, \exp[-i \tau (J H_{xx} + h_x H_x)],\label{Floquet-operator}
\end{equation}
which generates a collective rotation of the spin ensemble by $\tau = \pi/2$. In the integrable case with $h_x = 0$, this leads to perfectly periodic energy buildup. In the nonintegrable case with $h_x \neq 0$, chaotic dynamics redistribute energy while preserving the net rotation over a full period. Hence, $\tau = \pi/2$ establishes a universal resonance condition that maximizes energy absorption, irrespective of interaction range, boundary conditions, or integrability. Departing from this resonant point, we next consider shorter and asymmetric driving protocols. For equal but shorter periods with $\tau_0=\tau_1=\pi/4$, the maximum stored energy satisfies $\omega N < \Delta E_{\max} < 2\omega N$ for all system sizes and boundary geometries, thereby interpolating smoothly between half of the optimal value and full capacity, as detailed in the SM~\cite{Structural_SM}.

The system-size dependence of the maximum stored energy reveals a clear distinction between off-resonant and resonant driving. For $\tau=\pi/4$, the stored energy increases approximately linearly with system size under all boundary conditions and for both integrable and nonintegrable dynamics, with only minor deviations at specific sizes. By contrast, at the resonant point $\tau=\pi/2$, a pronounced odd-even effect emerges. Although both PBC and OBC enable optimal charging for even system sizes, the stored energy is systematically reduced for odd sizes. Under OBC, the maximum stored energy remains above $\omega N$ for all $N$, taking the values $\Delta E_{\max}=\omega N+1$ for odd $N$ and $\Delta E_{\max}=\omega N+2$ for even $N$, thereby demonstrating a robust finite-size enhancement at resonance, as detailed in the SM~\cite{Structural_SM}.

\begin{figure}
    \centering
\includegraphics[width=0.49\linewidth,height=0.35\linewidth]{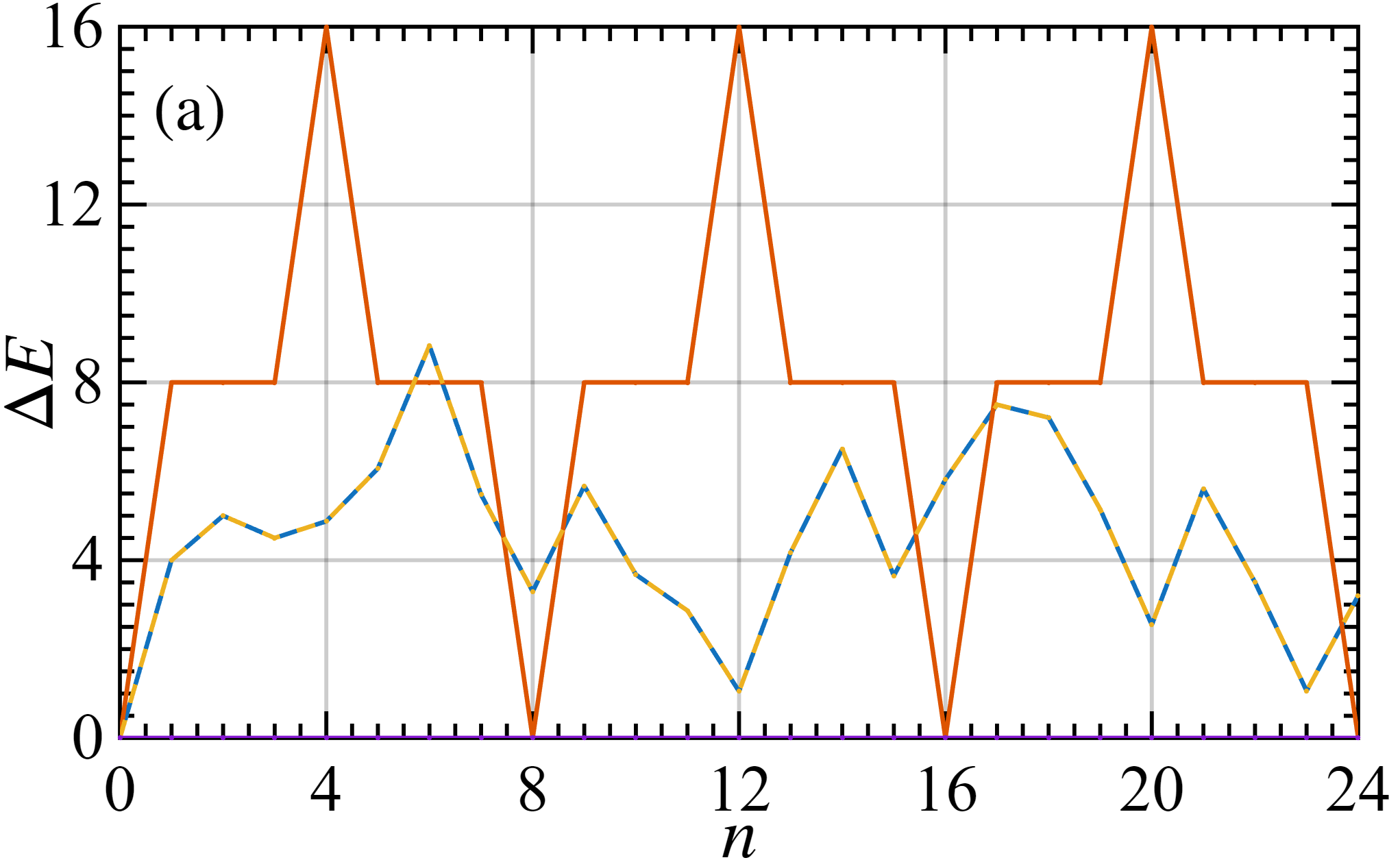}
\includegraphics[width=0.49\linewidth,height=0.35\linewidth]{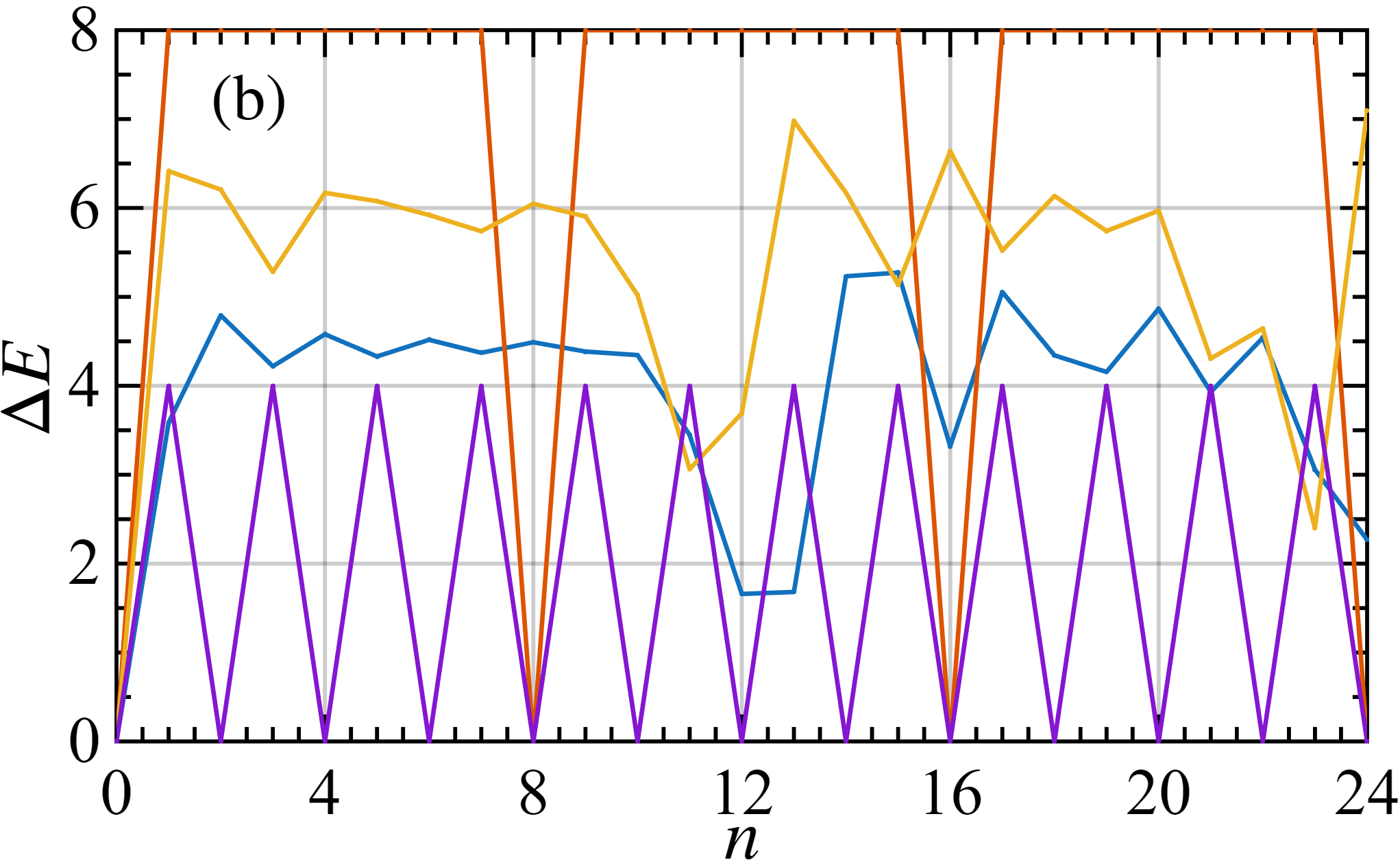}
\includegraphics[width=0.49\linewidth,height=0.35\linewidth]{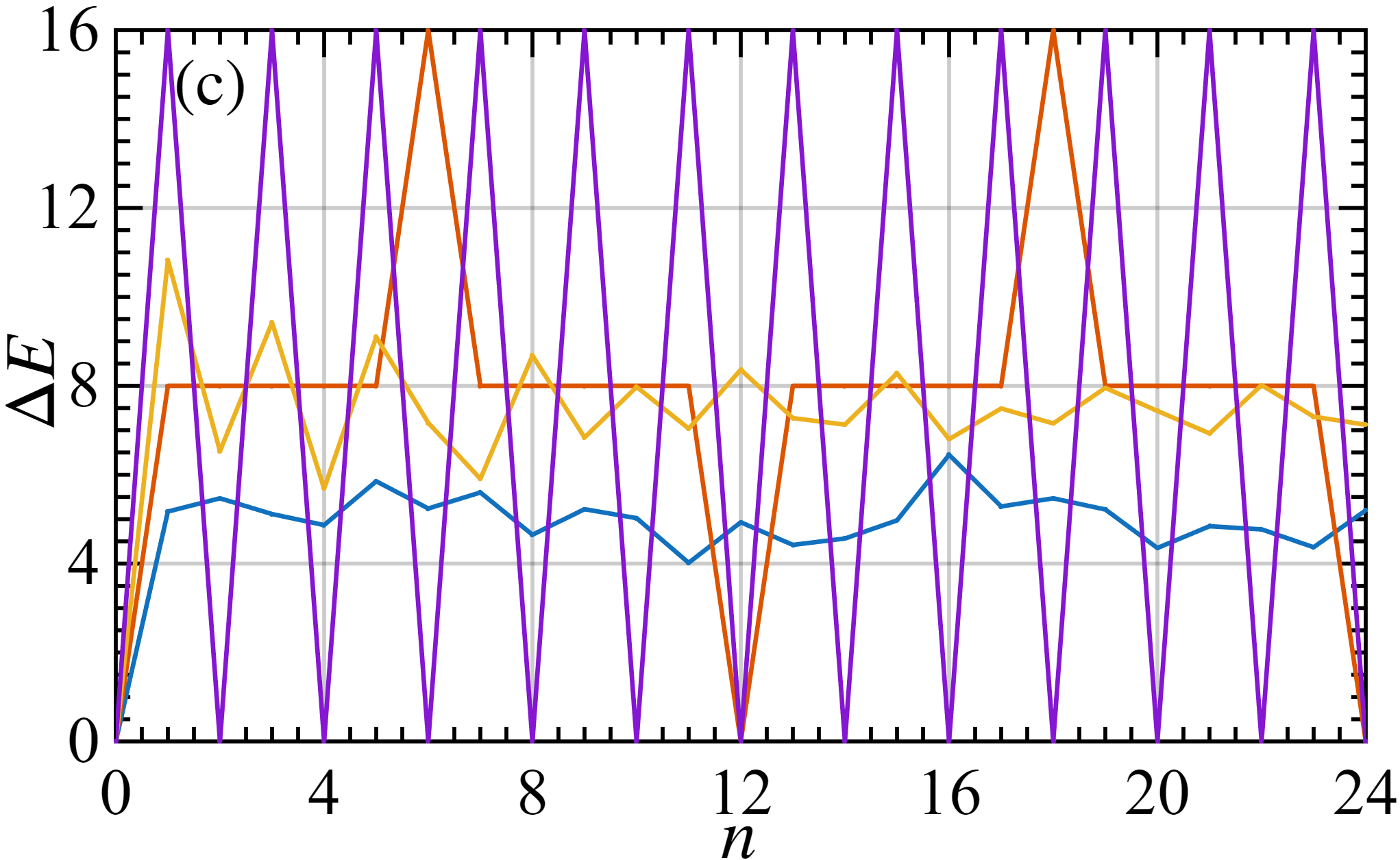}
\includegraphics[width=0.49\linewidth,height=0.35\linewidth]{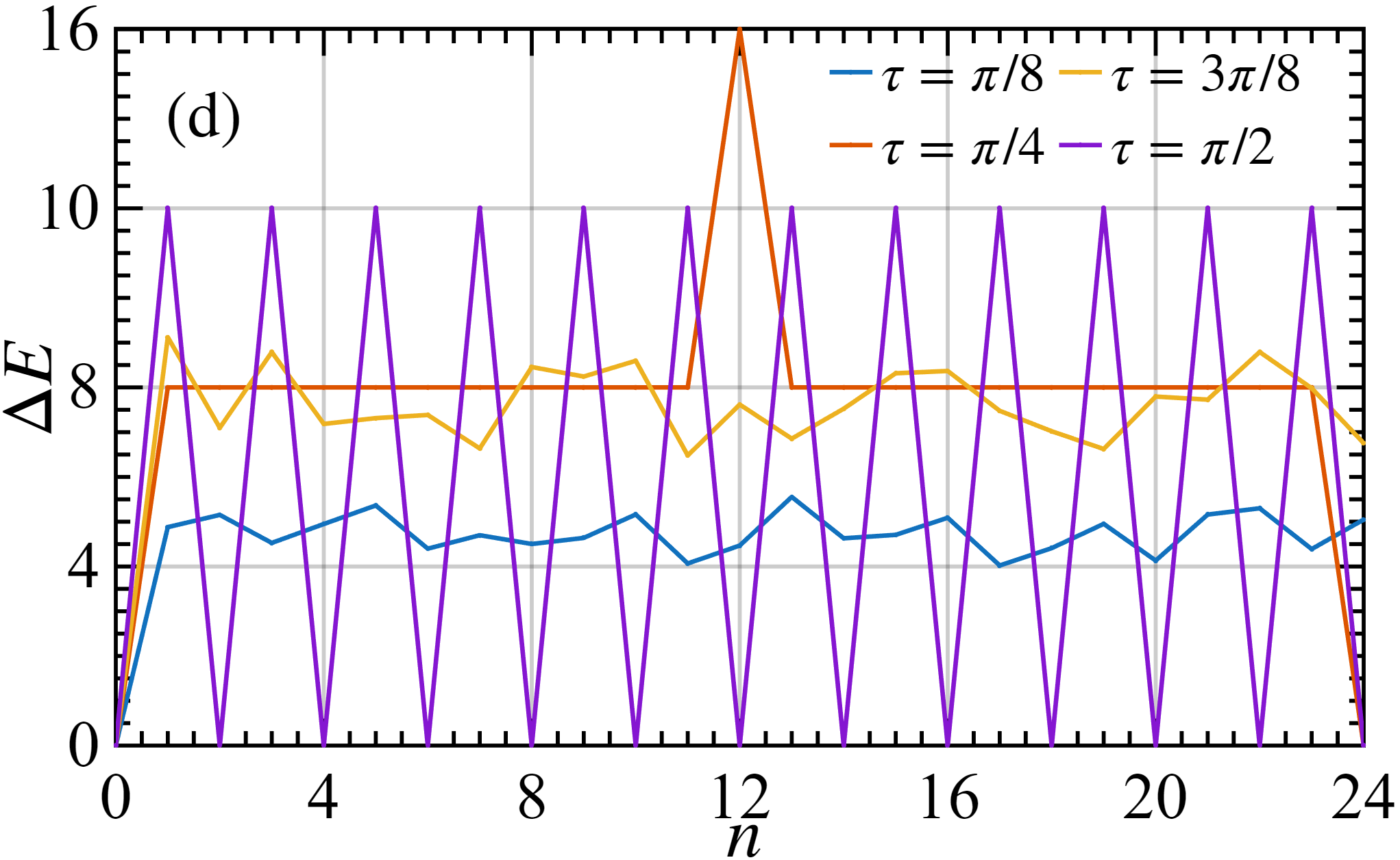}
   \caption{Stored energy $\Delta E$ as a function of the number of kicks $n$ for a noninteracting QB driven by a time-periodic nearest-neighbor Ising charger at different driving periods with $\tau_0=\tau_1=\tau$ (see legend). Panels (a,b) correspond to the integrable regime, whereas panels (c,d) represent the nonintegrable regime. PBC are used in panels (a,c), and OBC in panels (b,d). The system parameters are $N=8$, $J=1$, $h_x=0$ for the integrable case or $h_x=1$ for the nonintegrable case, $h_z=1$, and $\omega=1$.}
    \label{E_NN_tau_int}
\end{figure}

\begin{figure}
    \centering
\includegraphics[width=0.49\linewidth,height=0.35\linewidth]{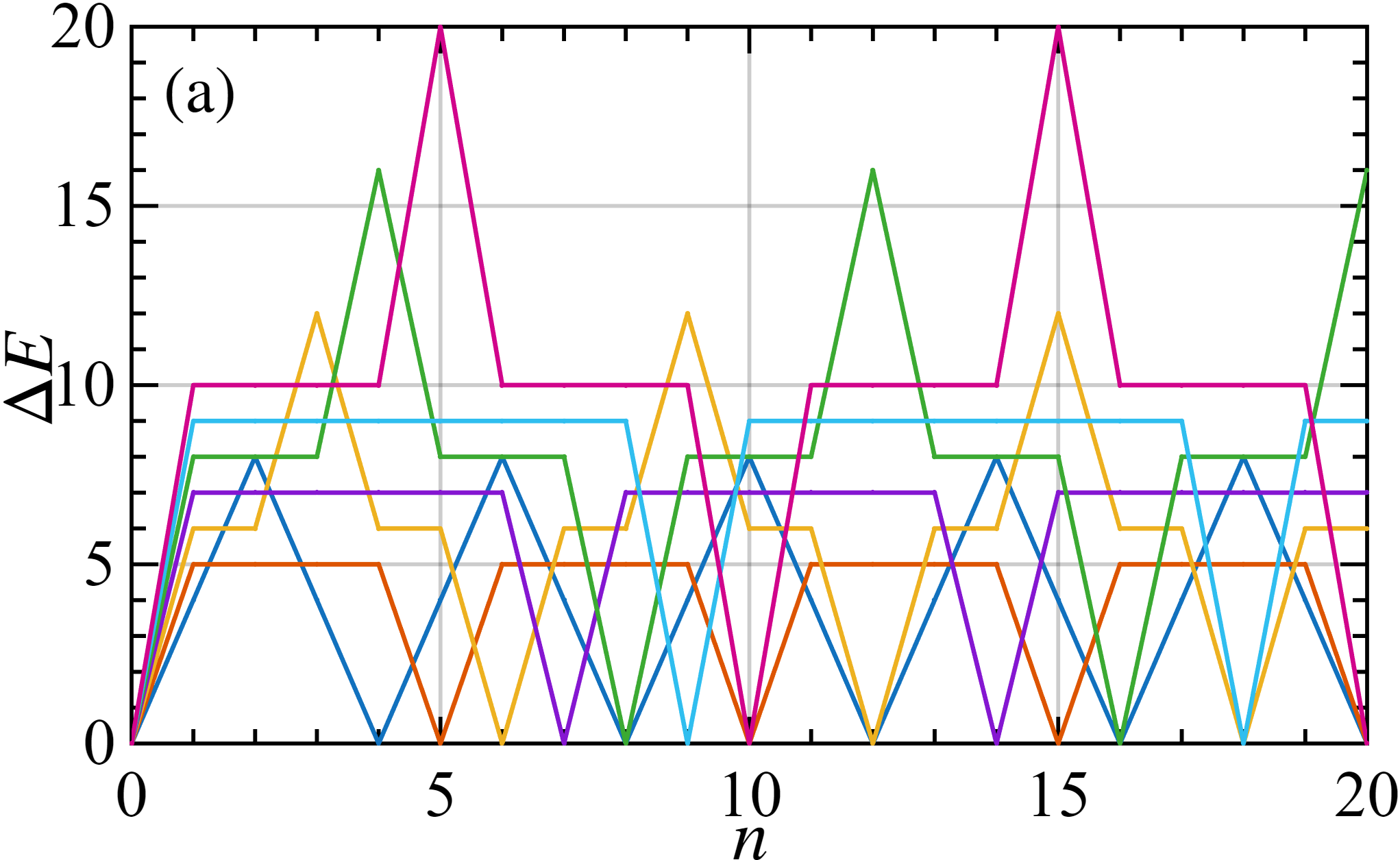}
\includegraphics[width=0.49\linewidth,height=0.35\linewidth]{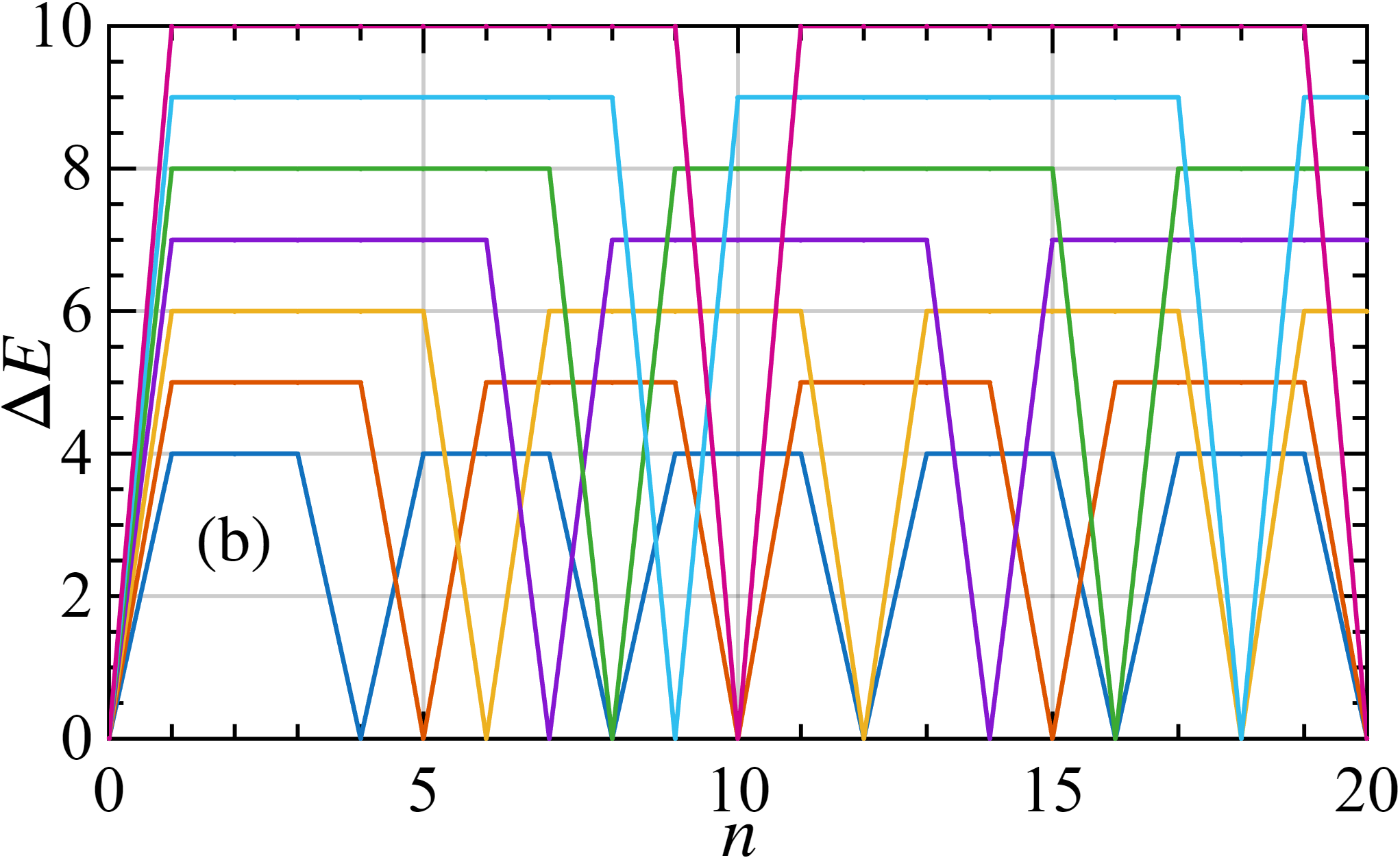}
\includegraphics[width=0.49\linewidth,height=0.35\linewidth]{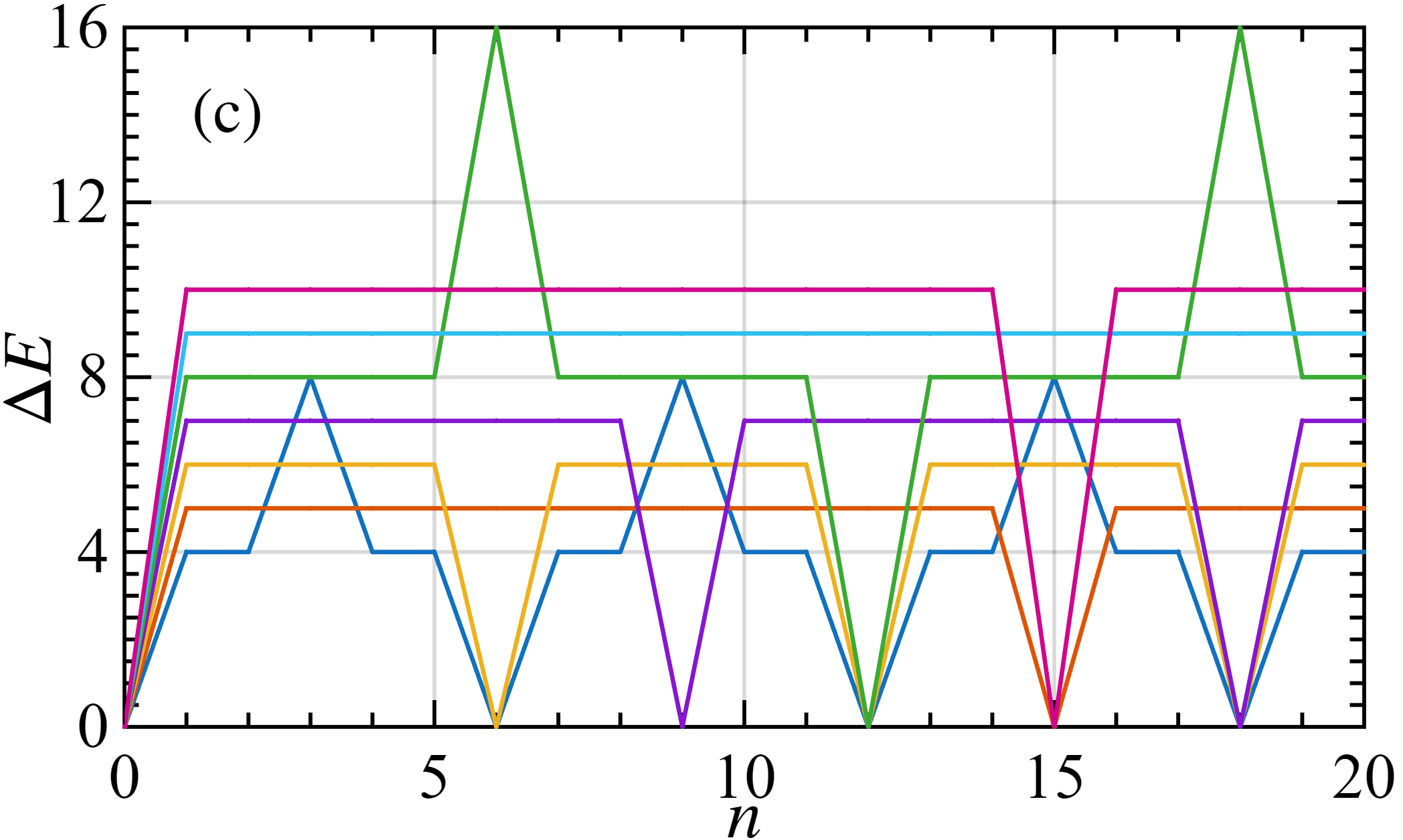}
\includegraphics[width=0.49\linewidth,height=0.35\linewidth]{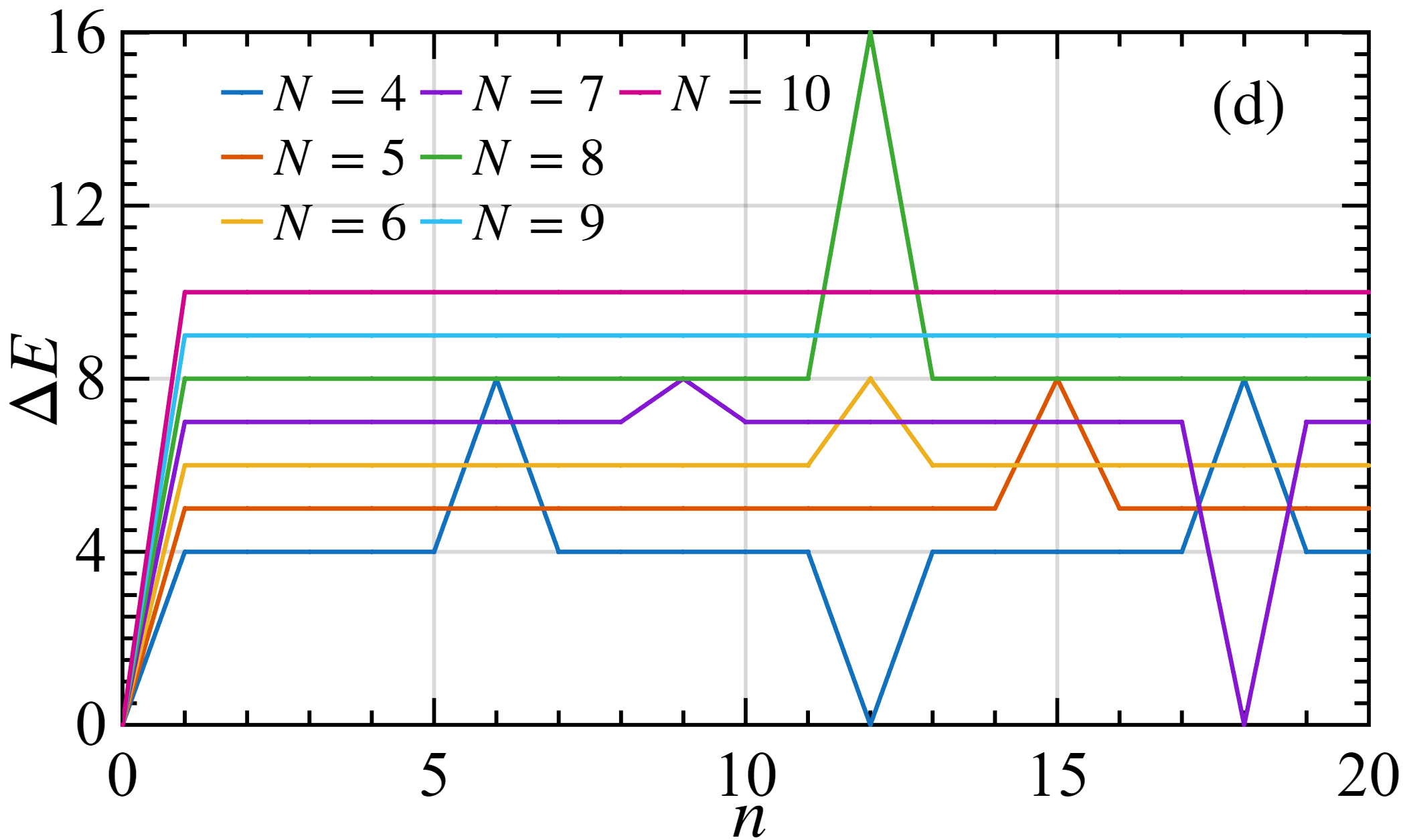}
    \caption{Stored energy $\Delta E$ as a function of the number of kicks $n$ for a noninteracting QB driven by a time-periodic nearest-neighbor Ising charger for different system sizes $N$ (see legend). Panels (a,b) correspond to the integrable regime, while panels (c,d) show the nonintegrable regime. PBC are adopted in panels (a,c), and OBC are used in panels (b,d).  The parameters are set to $\tau_0=\tau_1=\pi/4$, $J=1$, $h_x=0$ for the integrable case or $h_x=1$ for the nonintegrable case, with $h_z=1$ and $\omega=1$.}
\label{E_NN_int_N}
\end{figure}

\textit{Nearest-neighbor interacting charger.—}To reveal several distinctive features, we now turn to the nearest-neighbor interacting charger by restricting the interaction range to $k=1$ in Eq.~(\ref{Charger_H}). We first analyze the integrable regime by setting $\tau_0=\tau_1=\tau$ at a fixed system size $N=8$, under both PBC and OBC. To capture commensurability effects, we consider four representative driving periods: $\tau=\pi/8$, $\pi/4$, $3\pi/8$, and $\pi/2$.

{Figure~\ref{E_NN_tau_int}(a) shows that for the integrable charger under PBC, the dynamics at $\tau=\pi/4$ is strictly periodic, with a temporal period equal to the system size, satisfying $\Delta E(mN+n)=\Delta E(n)$ for integer $m$. In this regime, the QB reaches the optimal value $\Delta E_{\max}=2\omega N$ at $n=N/2$. In contrast, at $\tau=\pi/2$ the stored energy vanishes. For intermediate periods $\tau=\pi/8$ and $3\pi/8$, the maximum stored energy lies between half and the full optimal value, $\omega N<\Delta E_{\max}<2\omega N$, and the energy profile is symmetric under the transformation $\tau\rightarrow\pi/2-\tau$. Figure~\ref{E_NN_tau_int}(b) reveals that for the integrable charger under OBC, the periodic structure at $\tau=\pi/4$ is preserved, although the maximum stored energy is reduced to $\Delta E_{\max}=\omega N$. At $\tau=\pi/2$, the oscillation period doubles and the energy reaches only $\Delta E_{\max}=\omega N/2$. For $\tau=\pi/8$ and $3\pi/8$, the maximum stored energy lies in the range $\omega N/2 < \Delta E_{\max} < \omega N$, and the dynamics no longer exhibits a clear periodic pattern.

In Figure~\ref{E_NN_tau_int}(c,d), we further examine the nonintegrable regime by introducing a longitudinal field $h_x=1$, which leads to modified charging dynamics. Under PBC, at $\tau=\pi/4$ the system exhibits periodic behavior with period $3N/2$ and reaches the optimal energy at $n=3N/4$, whereas at $\tau=\pi/2$ the dynamics becomes periodic with period 2, attaining the optimal value at odd-numbered kicks. For $\tau=\pi/8$ and $3\pi/8$, the evolution is no longer strictly periodic; the maximum stored energy remains below half of the optimal value for $\tau=\pi/8$ and exceeds half for $\tau=3\pi/8$. Under OBC, similar trends appear, including period doubling at $\tau=\pi/4$ and $\tau=\pi/2$, with maximal energies $\Delta E_{\max}=2\omega N$ and $\Delta E_{\max}=3\omega N/2$, respectively.

To summarize, Table~\ref{tab:optimization_landscape_structured} presents the optimization landscape of the maximum stored energy as a function of the driving period $\tau$, capturing how many-body \emph{structural effects} govern energy storage, with further details provided in the SM~\cite{Structural_SM}. Notably, the commensurate point $\tau=\pi/4$ generally emerges as a distinguished regime enabling optimal charging, whereas for $\tau>\pi/4$ the maximum stored energy becomes highly sensitive to integrability and boundary conditions. This sensitivity underscores the intricate interplay among the driving period, many-body structure, and system geometry in determining the achievable energy storage. Additionally, unequal driving intervals $\tau_0\neq\tau_1$ exhibit analogous \emph{structural effects}. In the integrable regime, optimal charging arises only at the commensurate point $\tau_0=\tau_1=\pi/4$ under PBC, while deviations from this condition reduce the maximal stored energy. Under OBC, charging is further suppressed, and the optimal value is no longer attainable. By contrast, the nonintegrable charger demonstrates enhanced robustness: optimal or near-optimal energy can be realized for multiple combinations of $\tau_0$ and $\tau_1$ under PBC, and OBC still supports substantial energy storage. These results highlight that integrability imposes stringent symmetry constraints, whereas nonintegrability promotes resilience and improves charging efficiency under asymmetric driving.

{\it Short-range finite-size effects.—}As a limiting case of the long-range finite-size effects, we now present the corresponding analysis for the short-range regime. At the commensurate driving period $\tau_0=\tau_1=\tau=\pi/4$, the Floquet dynamics acquire a particularly simple structure, motivating a detailed investigation of finite-size effects at this point. We therefore analyze the dependence of the stored energy on the system size at $\tau=\pi/4$. For the integrable charger under PBC, as shown in Fig.~\ref{E_NN_int_N}(a), the stored energy displays strictly periodic behavior with a period equal to the system size. A pronounced odd--even effect is observed: even system sizes achieve optimal charging, $\Delta E_{\max}=2\omega N$, at $n=N/2$, whereas odd system sizes are limited to half of the optimal value, $\Delta E_{\max}=\omega N$. In Fig.~\ref{E_NN_int_N}(b), under OBC, the dynamics remain periodic with the same system-size-dependent period; however, the maximum stored energy is uniformly bounded by $\Delta E_{\max}=\omega N$ for all system sizes, indicating that boundary effects suppress full charging even at the commensurate point. 
Notably, a similar analysis of the kicked Ising model shows that optimal energy storage occurs when the charger interactions act along the $x$ direction under PBC, while interactions along the $z$ direction become optimal under OBC~\cite{romero2025kicked}. By contrast, when the boundary conditions are interchanged—namely, $x$-direction interactions with OBC or $z$-direction interactions with PBC—the stored energy reaches only about half of the optimal value~\cite{romero2025kicked}.

A qualitatively different behavior arises for the nonintegrable charger, as shown in Fig.~\ref{E_NN_int_N}(c,d). In this case, periodic energy growth persists under both boundary conditions, with a period $3N/2$ under PBC and $3N$ under OBC. When the system size is a multiple of four, the stored energy reaches the absolute upper bound $\Delta E_{\max}=2\omega N$ at kick number $n=3N/4$ under PBC and $n=3N/2$ under OBC. This enhancement originates from commensurate Floquet quasienergies, which enable fully constructive interference among excitation pathways generated by successive kicks. For system sizes that are not divisible by four, this commensurability is lost, leading to reduced but still substantial energy storage, with $\Delta E_{\max}\gtrsim \omega N$.  Moreover, a broader comparison of finite-size scaling at representative driving periods is presented in the SM~\cite{Structural_SM}. At $\tau=\pi/4$, integrability induces pronounced parity effects and high boundary sensitivity, whereas nonintegrable dynamics generally sustain extensive energy storage with nontrivial size dependence. At the longer period $\tau=\pi/2$, integrability results in a severe suppression of charging, which becomes complete under PBC and size-independent under OBC, whereas nonintegrable systems restore linear scaling with system size. Taken together, these findings demonstrate that integrability and boundary geometry play a decisive role in governing both finite-size effects and asymptotic energy-storage performance in periodically driven QBs.

The dependence of the stored energy on the nearest-neighbor coupling strength at $\tau=\pi/4$, discussed in detail in the SM~\cite{Structural_SM}, further underscores these \emph{structural effects}. In the integrable regime, charging exhibits a pronounced parity dependence on the coupling strength: under PBC, optimal charging occurs only for odd integer couplings, while even couplings completely suppress energy storage, with noninteger values yielding intermediate energies. Under OBC, this suppression is softened, although the maximum stored energy remains confined to fixed fractions of the optimal value. In contrast, the nonintegrable charger demonstrates significantly greater robustness, sustaining optimal or near-optimal charging across a broad range of coupling strengths under both boundary conditions. Overall, integrability enforces stringent parity- and boundary-induced constraints, whereas nonintegrability smooths the parameter dependence and stabilizes energy storage.

\begin{table}[t]
\centering
\footnotesize
\caption{Optimization landscape of the maximum stored energy governed by many-body \emph{structural effects}.}
\label{tab:optimization_landscape_structured}
\resizebox{\columnwidth}{!}{%
\begin{tabular}{|c|c|c|c|}
\hline
Interaction & Dynamics & Boundary & \multicolumn{1}{c|}{$\Delta E_{\rm max}$} \\
\hline
\multirow{2}{*}{LR}
& \multirow{2}{*}{Integrable}
& PBC
& $
\begin{cases}
0 \sim 2\omega N, & \qquad \quad \ \ 0 \leq \tau < \pi/2 \\
2\omega N\;(\text{even }N), & \qquad \quad \ \ \tau = \pi/2 \\
\omega N\;(\text{odd }N), & \qquad \quad \ \ \tau = \pi/2
\end{cases}$ \\
\cline{3-4}
&  & OBC
& $
\begin{cases}
0\sim 2\omega N, & \qquad \qquad \quad \ \ \ \ 0 \leq \tau \leq \pi/2 \\
\end{cases}$ \\
\cline{2-4}
\multirow{2}{*}{LR}
& \multirow{2}{*}{Nonintegrable}
& PBC
& $
\begin{cases}
0 \sim 2\omega N, & \qquad \quad \ \ 0 \leq \tau < \pi/2 \\
2\omega N\;(\text{even }N), & \qquad \quad \ \ \tau = \pi/2 \\
\omega N\;(\text{odd }N), & \qquad \quad \ \ \tau = \pi/2
\end{cases}$ \\
\cline{3-4}
&  & OBC
& $
\begin{cases}
0\sim 2\omega N, & \qquad \qquad \quad \ \ \ \ 0 \leq \tau \leq \pi/2 \\
\end{cases}$ \\
\hline
\multirow{2}{*}{NN}
& \multirow{2}{*}{Integrable}
& PBC
& $
\begin{cases}
0 \sim 2\omega N, & 0 \leq \tau \leq \pi/2 \\
0, & \tau = \pi/2 \\
2\omega N\;(N=4m, m \in \mathbb{Z}), & \tau = \pi/4 \\
\omega N\;(N\neq 4m, m \in \mathbb{Z}), & \tau = \pi/4
\end{cases}$ \\
\cline{3-4}
&  & OBC
& $
\begin{cases}
< 2\omega N, & \qquad \qquad \quad \ \ \ \ \ \ 0<\tau < \pi/2 \\
\omega N/2, &  \qquad \qquad \quad \ \ \ \ \ \ \tau = \pi/2 \\
\omega N, & \qquad \qquad \quad \ \ \ \ \ \ \tau = \pi/4
\end{cases}$ \\
\cline{2-4}
\multirow{2}{*}{NN}
& \multirow{2}{*}{Nonintegrable}
& PBC
& $
\begin{cases}
< 2\omega N, &\!\!\!\!\!\!\!\!\!\!\!0<\tau\neq \pi/4 < \pi/2  \\
2\omega N\;(N=4m, m \in \mathbb{Z}), & \tau = \pi/4 \\
\omega N\;(N\neq 4m, m \in \mathbb{Z}), & \tau = \pi/4\\
2\omega N\;(\forall N), & \tau = \pi/2 \\
\end{cases}$ \\
\cline{3-4}
&  & OBC
& $
\begin{cases}
< 2\omega N,  &\!\!\!\!\!\!\!\!\!\!\!0<\tau\neq \pi/4 < \pi/2 \\
2\omega N\;(N=4m, m \in \mathbb{Z}), & \tau = \pi/4 \\
\omega N\;(N\neq 4m, m \in \mathbb{Z}), & \tau = \pi/4 \\
3\omega N/2\;(\forall N) & \tau = \pi/2
\end{cases}$ \\
\hline
\end{tabular}
}
\end{table}

\textit{Summary.—}
In this Letter, we discuss how the structural properties of periodically driven Ising chargers govern the performance of QBs. Building upon the kicked-Ising QB framework introduced in~\cite{romero2025kicked} for the integrable regime, we extend the analysis to more general periodically driven settings in the nonintegrable regime. Our study shows that many-body \emph{structural effects} fundamentally shape both the achievable stored energy and the robustness of the charging protocol. In charging settings dominated by long-range interactions, the driving period $\tau$ acts as a crucial control parameter. At the commensurate point $\tau=\pi/2$, optimal or near-optimal charging is generally achieved, whereas periods satisfying $\tau>\pi/4$ exhibit pronounced sensitivity to boundary conditions and integrability. Furthermore, introducing unequal driving durations, $\tau_0\neq\tau_1$, reveals the robustness of nonintegrable dynamics, which sustain substantial energy storage even under temporal asymmetry, whereas integrable chargers are constrained by symmetry and exhibit reduced performance. Interestingly, the short-range one exhibits behavior fundamentally distinct from that of the long-range regime. In the nearest-neighbor case, commensurability and finite-size effects dominate the dynamics. Under PBC, integrable chargers exhibit pronounced odd-even and parity-dependent behavior, achieving full charging only for even system sizes or odd-integer couplings, whereas OBC reduces the maximal stored energy due to boundary-induced constraints. By contrast, nonintegrable dynamics smooth out these many-body structural sensitivities, enabling robust energy storage across a broad range of system sizes, boundary conditions, and coupling strengths.~Overall, our results demonstrate that integrability imposes strict symmetry and parity constraints, whereas nonintegrable chaotic dynamics enhance ergodicity and reduce sensitivity to boundary effects.~This interplay establishes a general framework for engineering efficient QBs, highlighting that optimal charging depends not only on interaction strength or system size but, more fundamentally, on the driving protocol and the underlying many-body \emph{structural effects}. Furthermore, the maximum charging power scales approximately linearly with system size for both long-range and nearest-neighbor chargers and decreases monotonically with increasing driving period, independent of integrability or boundary conditions, as detailed in the SM~\cite{Structural_SM}.

Finally, in the End Matter, we provide a complementary perspective on the charging dynamics of QBs based on bipartite entanglement entropy.~In particular, we present the stored energy and bipartite mutual information of a QB driven by a nonintegrable time-periodic charger for representative driving periods; see Appendix A. In addition, we briefly discuss the relationship between QB structural optimization and system dimensionality; see Appendix B.

{\textit{Acknowledgments.—}We thank Dr.~Yue Ban for valuable comments on this manuscript. We note that Ref.~\cite{romero2025kicked} is a prior work that introduced the kicked-Ising QB and established its basic Floquet framework at the self-dual point in the integrable regime. The present work builds on that recent study by considering a more general periodically driven setting, in which the interaction and field durations are independently tunable. Within this extended framework, we further investigate structural aspects of the charging dynamics, including long-range interactions, system-size effects, and the behavior in both integrable and nonintegrable regimes. Cheng Shang acknowledges the Hakubi projects of RIKEN. 

\textit{Data availability.—}The data that support the findings of this Letter are not publicly available. The data are available from the authors upon reasonable request.

\bibliography{QB}
\onecolumngrid
\begin{center}
\textbf{End Matter}
\end{center}
\twocolumngrid

\textit{Appendix A: Bipartite entanglement entropy and QB charging dynamics}.—In this additional section, for completeness, we discuss an interesting finding on the role of quantum coherence in the charging dynamics of structured many-body QBs. To characterize quantum correlations during the charging process, we analyze the bipartite entanglement entropy (BEE) between a subsystem $X$ and its complement $X^c$. The BEE quantifies the extent to which a subsystem is entangled with the rest of the QB. It is defined as the von Neumann entropy of the reduced density matrix of the subsystem~\cite{hamma2005bipartite,PhysRevLett.96.181602,calabrese2004entanglement,page1993average,jiang2025entanglement,shukla2025scrambling,kadar2010entanglement}: ${S_X} =  - {\rm{T}}{{\rm{r}}_X}\left( {{\rho _X}\log {\rho _X}} \right)$, where ${\rho _X} = {\rm{T}}{{\rm{r}}_{{X^c}}}\left[ \rho  \right]$ is obtained by tracing out all degrees of freedom outside the subsystem $X$. The full system is assumed to be in a pure state, ${\rho} = |\psi_0\rangle \langle \psi_0|$, with $|\psi_0\rangle$ representing the ground state of the QB Hamiltonian ${H}_B$.

During charging, the BEE captures the spreading of quantum correlations and provides insight into how energy is distributed across the system. Below, we compute the BEE for a $1|(N-1)$ bipartition. Specifically, we consider two representative system sizes, $N=8$ and $N=10$, in the nonintegrable regime under both PBC and OBC, as shown in Fig.~\ref{E_ENT_NN}.

For $N=8$, the entanglement entropy closely mirrors the temporal structure of the stored energy under both PBC and OBC. In Fig.~\ref{E_ENT_NN}(a,c), at instants of maximal energy storage, $S_X$ exhibits pronounced minima. This behavior indicates that when the QB reaches its extremal energy configuration, the system approaches a product state across the bipartition, with local excitations coherently aligned rather than distributed in entangled superpositions, such that maximal energy storage coincides with minimal bipartite entanglement. In contrast, for $N=10$, the BEE follows the stored energy dynamics without exhibiting sharp minima, as shown in Fig.~\ref{E_ENT_NN}(b,d). In this case, the QB does not attain its maximal energy due to the absence of perfect constructive interference, so its energy is distributed across many-body eigenstates, and substantial entanglement persists throughout the evolution.

Overall, these results delineate a clear distinction between perfect and imperfect charging regimes. For a commensurate system size such as $N=8$, where the protocol enables extremal energy storage, bipartite entanglement is strongly suppressed at the charging peaks and reaches a minimum. In contrast, for a noncommensurate system size such as $N=10$, where perfect constructive interference is absent, persistent entanglement accompanies submaximal energy accumulation.

\begin{figure}[t]
    \centering
\includegraphics[width=0.48\linewidth,height=0.34\linewidth]{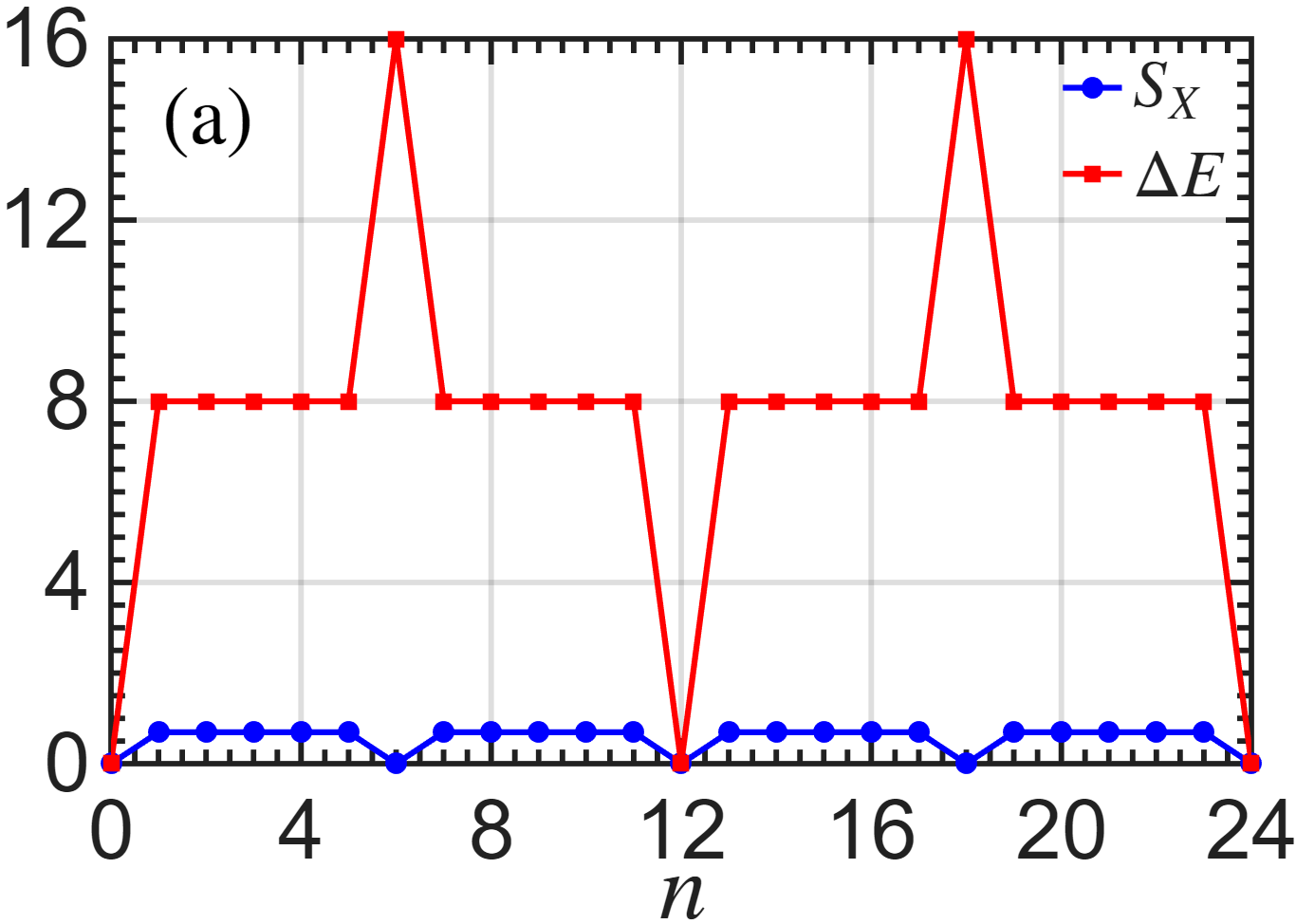}
\includegraphics[width=0.48\linewidth,height=0.34\linewidth]{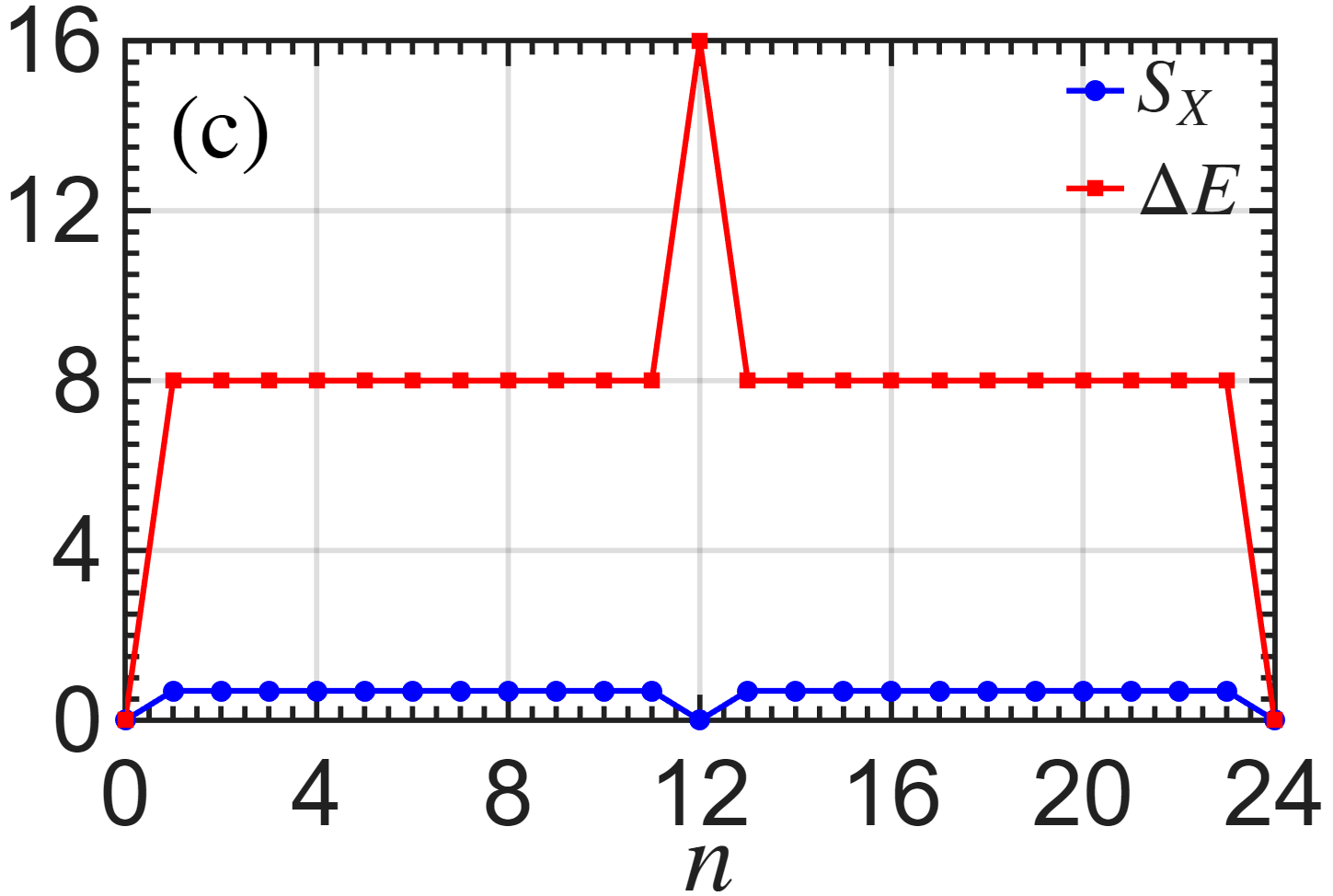}
\includegraphics[width=0.48\linewidth,height=0.34\linewidth]{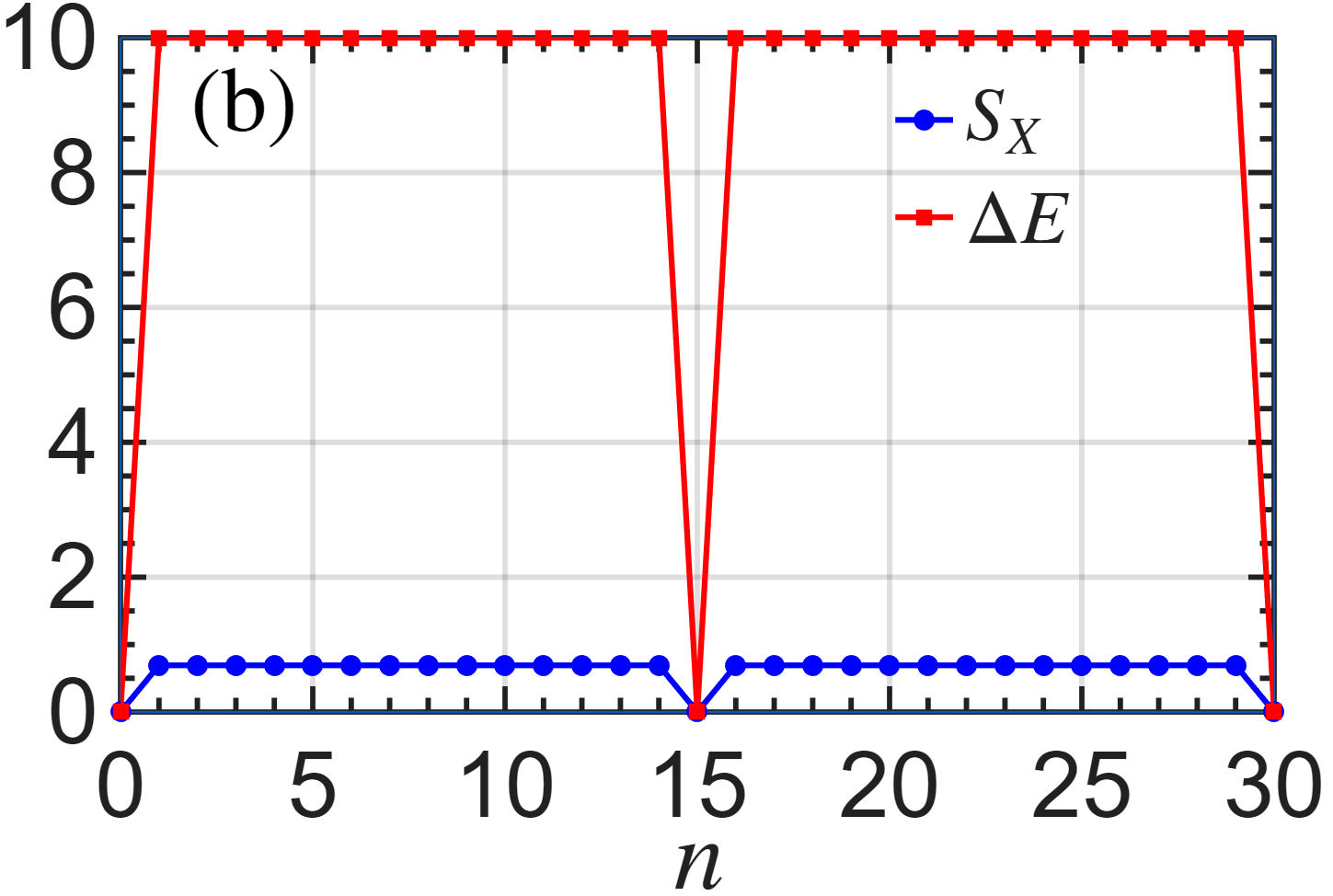}
\includegraphics[width=0.48\linewidth,height=0.34\linewidth]{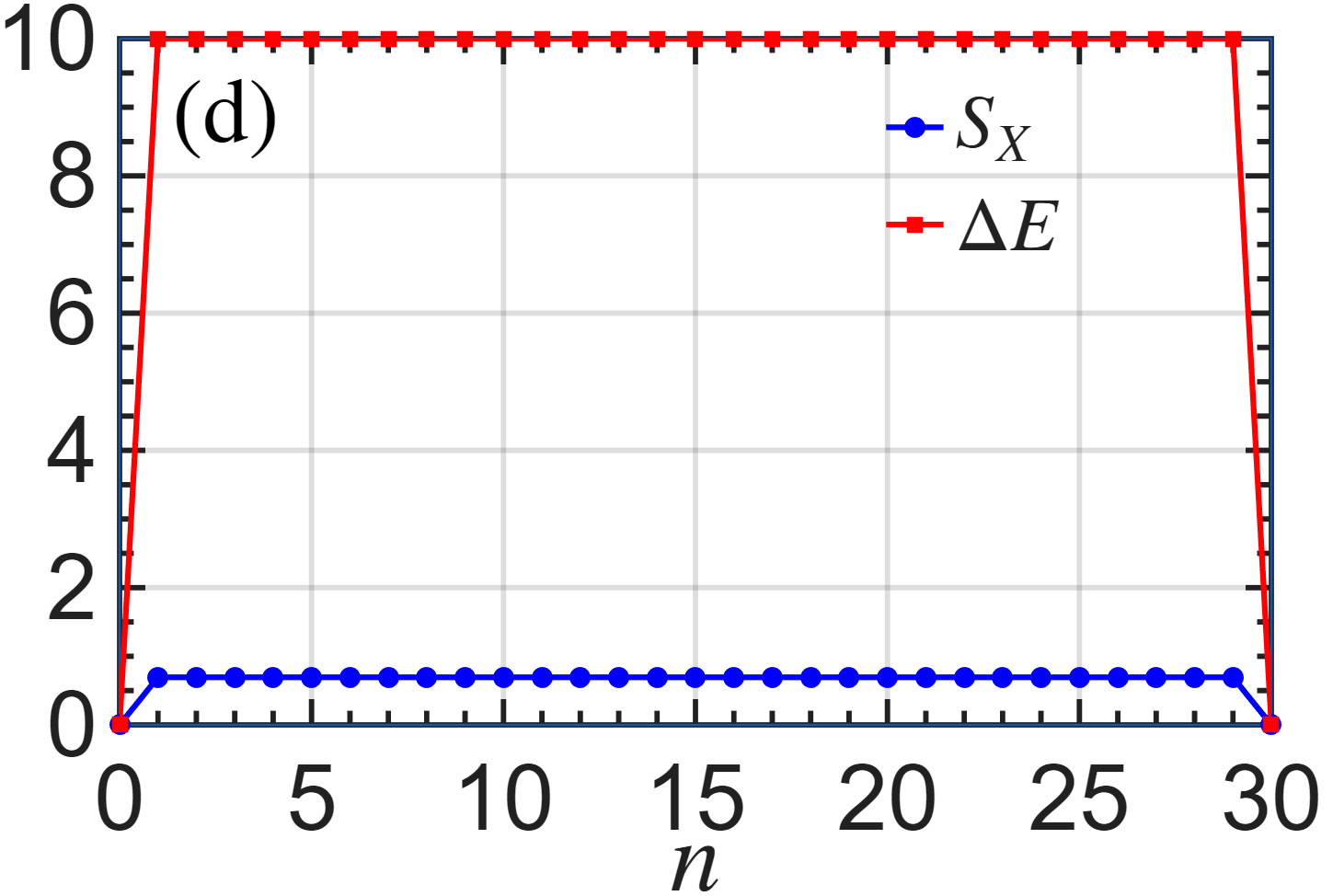}
    \caption{Stored energy and bipartite mutual information of a QB driven by a nonintegrable time-periodic charger with driving periods $\tau_0=\tau_1=\pi/4$. Panels (a,b) correspond to PBC, whereas panels (c,d) correspond to OBC. Results in panels (a,c) are shown for system size $N=8$, and those in panels (b,d) for $N=10$. The remaining parameters are $J=1$ for the interaction strength, $h_x=1$ for the transverse field strength, and $h_z=1$ for the QB field strength.}
    \label{E_ENT_NN}
\end{figure}

\textit{Appendix B: Dimensional robustness of QB.—}It is worth noting that although our analysis throughout this study has focused on a one-dimensional Ising model, the conclusions hold more generally. To substantiate the dimensional robustness of the charging performance of long-range interacting chargers, one may consider a dimensional-reduction protocol in which bosons from a higher-dimensional Mott-insulating lattice are concentrated onto a quasi-one-dimensional information path. Such a transformation can be implemented by combining the free-boson and Bose-Hubbard Hamiltonians and can be achieved in $\mathcal{O}(1)$ time.~The resulting high-density bosonic states may then be encoded into an effective qubit representation, followed by the application of CNOT gates~\cite{Kuwahara2024}. Based on the above procedure, the long-range boson-boson interactions can be mapped onto an effective long-range Ising-type interaction, with the coupling strength renormalized via bosonic enhancement as $J \to C J$, where $C \propto \bar{n}_t^{\,p}$, $\bar{n}_t$ denotes the boson density, and $p$ characterizes the order of the local $p$-body repulsion~\cite{PhysRevX.10.031010}. Although this procedure substantially amplifies the interaction strength $J$, our numerical analysis shows that the charging performance of long-range interacting chargers is remarkably robust. For both integrable and nonintegrable dynamics, and under both PBC and OBC, the maximum stored energy remains essentially unchanged over a wide range of $J$. Hence, increasing $J$ does not provide additional charging advantages within our analytical framework. Detailed discussions are provided in the SM~\cite{Structural_SM}.

These results demonstrate that charging optimization is governed by the interplay of multiple structural parameters rather than by the enhancement of a single interaction scale such as $J$. Instead, optimal QB performance arises from a coordinated interplay among interaction range, boundary conditions, integrability, and the temporal structure of the Floquet drive. Simply modifying a single ingredient without respecting this multi-parameter compatibility may not only fail to enhance performance but can even degrade charging efficiency by driving the system away from structurally optimal regimes.

\clearpage
\onecolumngrid

\begin{center}
{\large\bf Supplementary Material for\\ \vspace{0.3em}
``Many-Body \emph{Structural Effects} in Periodically Driven Quantum Batteries''}
\end{center}

\begin{center}
Rohit Kumar Shukla\,\orcidlink{0000-0003-2693-8745}$^{1,\ *}$ and Cheng Shang\,\orcidlink{0000-0001-8393-2329}$^{2,\ \dag}$
\end{center}

\begin{center}
$^{1}$\textit{Department of Chemistry, Institute of Nanotechnology and Advanced Materials,\\
Center for Quantum Entanglement Science and Technology, Bar-Ilan University, Ramat-Gan 5290002, Israel}\\


$^{2}$\textit{Analytical Quantum Complexity RIKEN Hakubi Research Team,\\
RIKEN Center for Quantum Computing (RQC), Wako, Saitama 351-0198, Japan}
\end{center}
\vspace{0.25cm}

\noindent
This Supplementary Material is organized into three main sections. In Sec.~\ref{LR_S}, we present a comprehensive study of the long-range interacting charger, beginning with the charging dynamics for equal driving periods $\tau_0=\tau_1=\pi/4$ and then extending the analysis to unequal periods $\tau_0 \neq \tau_1$, followed by an investigation of the effect of interaction strength and a detailed discussion of how the maximum stored energy depends on the driving period and system size, highlighting structural features such as parity, commensurability, integrability, and boundary conditions. In Sec.~\ref{NN_S}, we turn to the nearest-neighbor interacting charger, where we analyze energy storage under unequal driving intervals, examine the role of the coupling strength $J$ in both integrable and nonintegrable regimes under periodic and open boundary conditions, and study the scaling of the maximum stored energy with driving period and system size. Finally, in Sec.~\ref{Power_S}, we investigate the charging power for both long-range and nearest-neighbor chargers, focusing on its dependence on system size and driving period, and clarifying how interaction range, integrability, and boundary conditions collectively govern the overall charging performance of periodically driven Ising quantum batteries.

\begin{textblock*}{0.999\textwidth}(0.108\textwidth,0.999\textheight)
\raggedright
Corresponding author: \textcolor{blue}{$^{*}$\ rohitkrshukla.rs.phy17@itbhu.ac.in}\\
Corresponding author: \textcolor{blue}{$^{\dag}$\ cheng.shang@riken.jp}
\end{textblock*}

\setcounter{tocdepth}{2}
\tableofcontents
\clearpage

\setcounter{section}{0}
\renewcommand{\thesection}{S\Roman{section}} 
\renewcommand{\thesubsection}{\Alph{subsection}} 

\setcounter{secnumdepth}{2} 

\setcounter{figure}{0}
\renewcommand{\thefigure}{S\arabic{figure}}

\section{Long-range interacting Charger} 
\label{LR_S}
\subsection{Off-resonant driving at $\tau_0=\tau_1=\pi/4$}

\noindent
In the main text, optimal energy storage for the long-range interacting charger under PBC (in both integrable and nonintegrable cases) was obtained at the resonant period $\tau_0=\tau_1=\pi/2$, whereas for the nearest-neighbor case the optimum occurs at $\tau_0=\tau_1=\pi/4$. To enable a consistent comparison, we analyze here the long-range interacting charger at the shorter period $\tau_0=\tau_1=\pi/4$, focusing on the effects of integrability and boundary conditions for system sizes $5 \leq N \leq 10$.

We consider both integrable ($h_x=0$) and nonintegrable ($h_x \neq 0$) regimes under PBC and OBC. In contrast to the nearest-neighbor case at the same driving period, the storage energy does not exhibit periodic revivals, nor does it reach the optimal value observed at resonance. After the first kick, the battery acquires a finite amount of energy, followed by oscillatory dynamics whose amplitude and frequency content depend sensitively on boundary conditions, system size, and integrability. For the integrable system with PBC, the energy displays multi-frequency oscillations across all system sizes. Under OBC, oscillations are pronounced for smaller $N$ but become progressively suppressed as the system size increases. In the nonintegrable regime with PBC, oscillations persist with an approximately fixed dominant frequency that shows weak dependence on $N$. Under OBC, larger systems exhibit a small hump around the fourth kick, after which the stored energy approaches a near-saturation value.

Despite these dynamical differences, the maximum stored energy shows qualitatively similar behavior in all cases. At this intermediate driving period, the energy remains extensive and satisfies $\omega N \leq \Delta E_{\max} < 2\omega N,$
as illustrated in Fig.~\ref{E_ATA_N_taupi4_int}(a–d). The close agreement between integrable and nonintegrable regimes indicates that, at $\tau=\pi/4$, the charging dynamics are largely insensitive to the presence of chaos. Instead, the off-resonant Floquet driving governs the behavior: constructive interference between successive kicks is incomplete, leading only to partial coherent energy accumulation. As a result, the battery does not reach full charge, and the stored energy remains bounded below its optimal resonant value. These findings highlight that, away from resonance, the structure of the driving protocol plays a more dominant role than the underlying dynamical character of the charger.
\vspace{0.25cm}
\begin{figure}[H]
    \centering
\includegraphics[width=0.246\linewidth,height=0.18\linewidth]{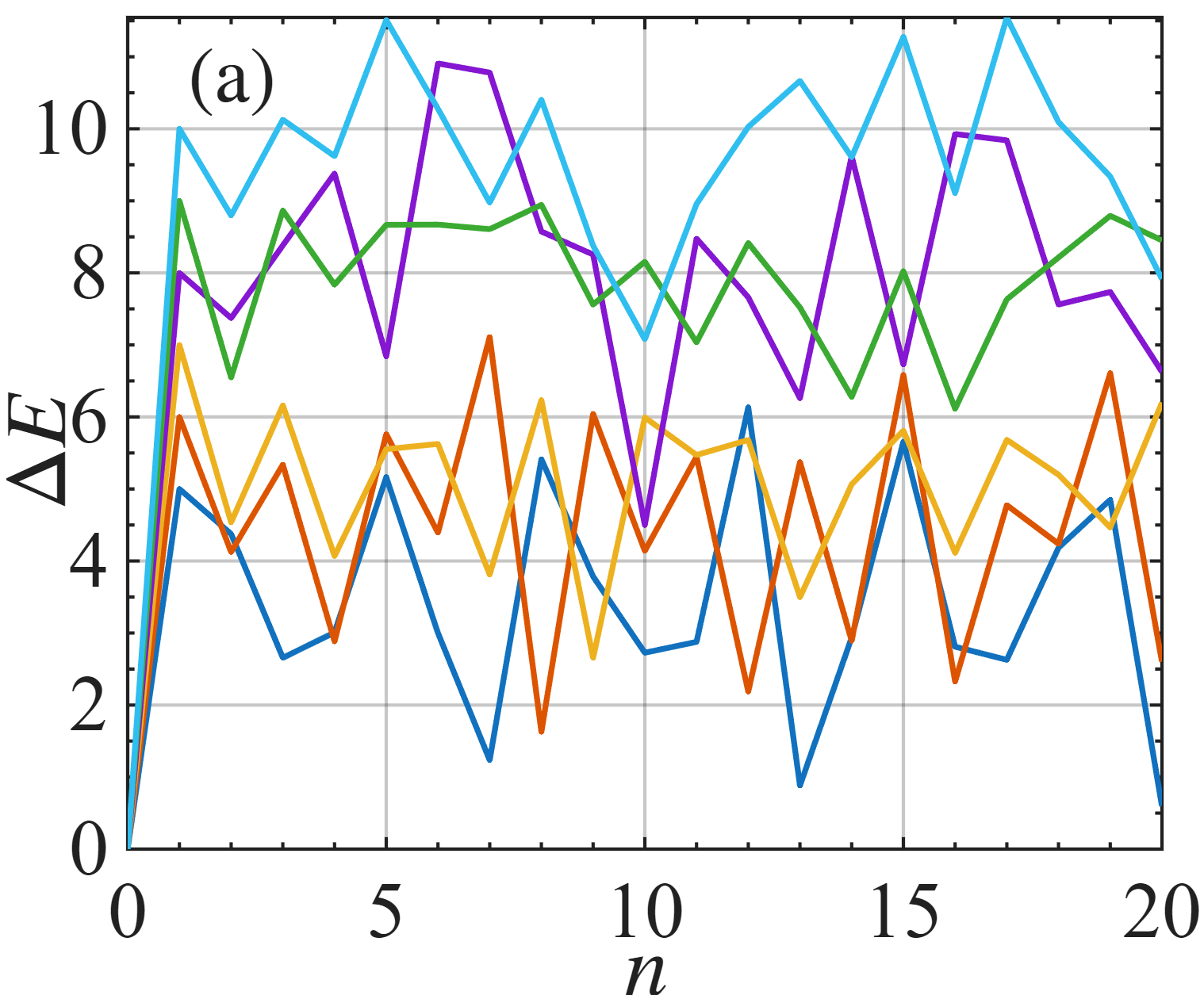}
\includegraphics[width=0.246\linewidth,height=0.18\linewidth]{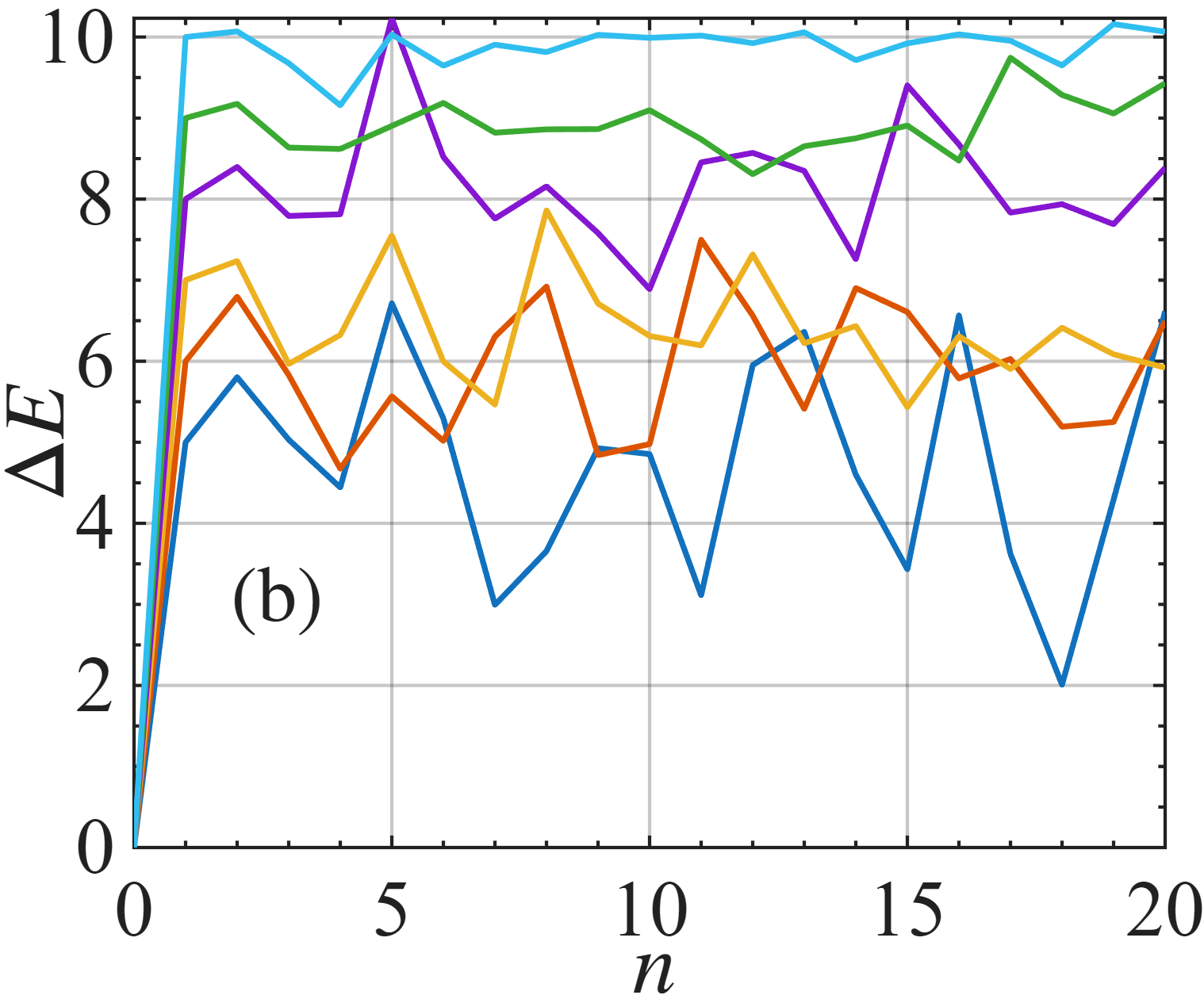}
\includegraphics[width=0.246\linewidth,height=0.18\linewidth]{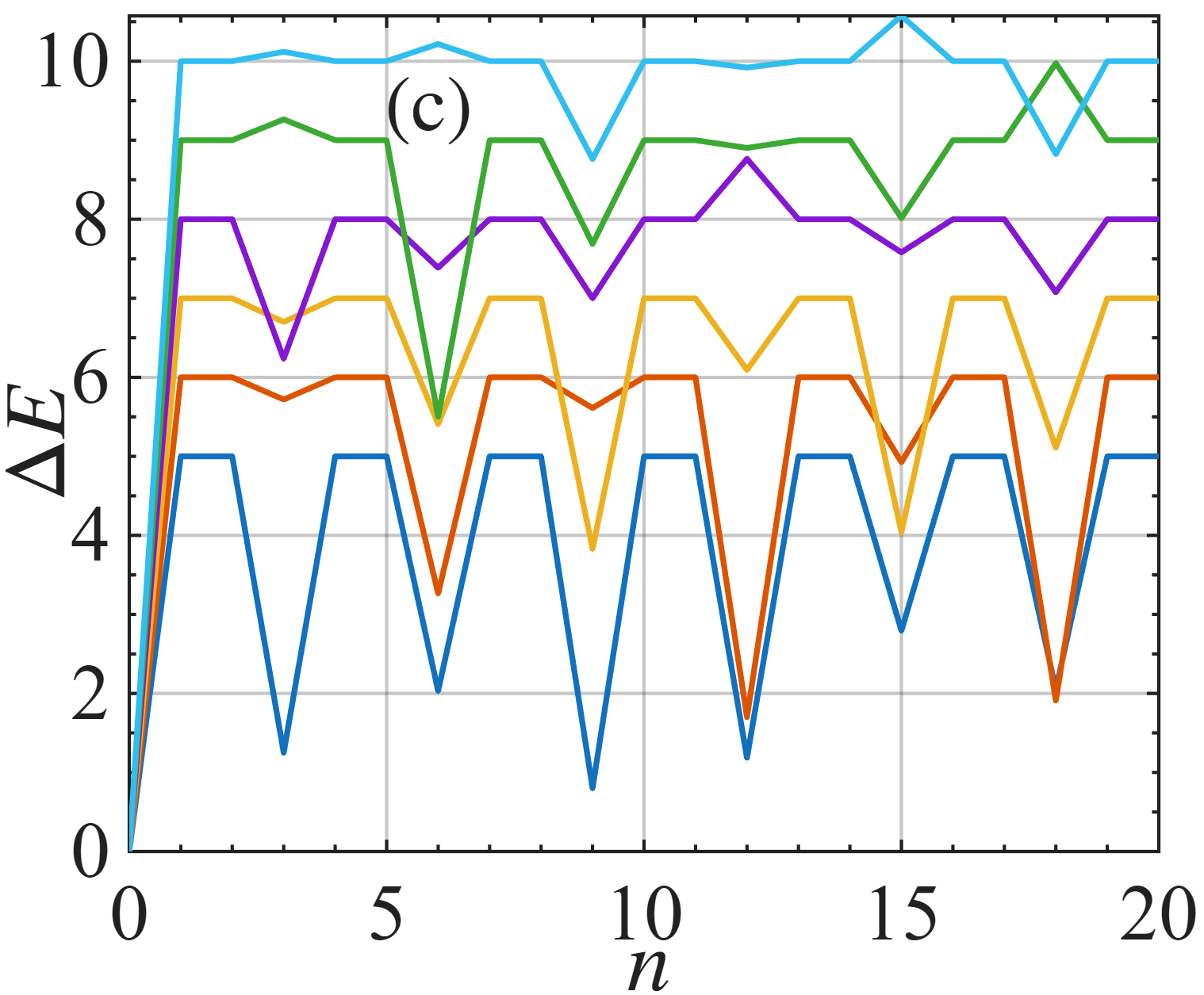}
\includegraphics[width=0.246\linewidth,height=0.18\linewidth]{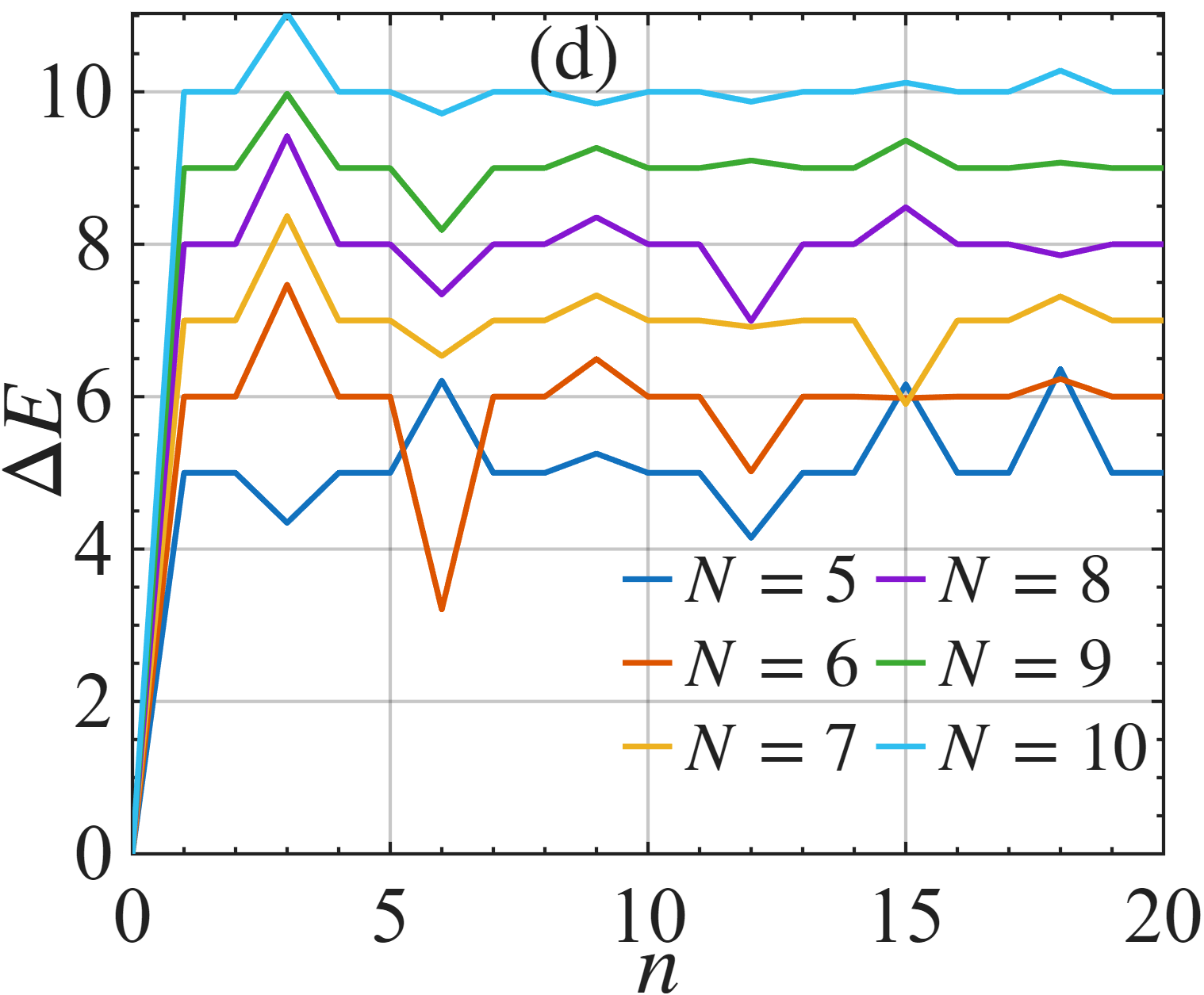}
    \caption{Stored energy of a noninteracting quantum battery driven by a long-range interacting charger for different system sizes (as indicated in the legend). Panels (a,b) correspond to the integrable case, while panels (c,d) show the nonintegrable case. Periodic and open boundary conditions are shown in panels (a,c) and (b,d), respectively. The driving periods are fixed to $\tau_0=\tau_1=\pi/4$. Parameters used are $J=1$, $h_z=1$, $\omega=1$, and $h_x=0$ ($h_x=1$) for the integrable (nonintegrable) regime.}
 \label{E_ATA_N_taupi4_int}
\end{figure}

\subsection{Energy storage beyond symmetric Floquet protocols}

To obtain a more comprehensive understanding of the charging dynamics beyond the symmetric protocol discussed in the main text, we analyze the energy storage for unequal driving periods, $\tau_0 \neq \tau_1$. While the main text establishes the resonant behavior for $\tau_0=\tau_1=\tau$, here we explicitly examine how temporal asymmetry modifies the optimization landscape. We consider both integrable ($h_x=0$) and nonintegrable ($h_x=1$) long-range Ising chargers under PBC and OBC, with system size fixed at $N=8$.

We first fix $\tau_1=\pi/4$ and vary $\tau_0=\pi/8,\ \pi/4,\ 3\pi/8,$ and $\pi/2$. For both integrable and nonintegrable chargers, and under both PBC and OBC, the stored energy exhibits a smooth but nontrivial dependence on $\tau_0$. Although the perfect resonance identified in the symmetric case is generally weakened by the asymmetry, the maximum stored energy remains above half of the optimal value and less than the optimum value, i.e., $\omega N<\Delta E_{\max} <2\omega N $, for all considered values of $\tau_0$, as shown in Fig.~\ref{long_int_E_tau1_tau2}. This demonstrates that efficient charging is robust against moderate temporal asymmetry. In particular, even when $\tau_0$ deviates significantly from $\pi/4$, the charger retains the ability to explore a substantial portion of the Hilbert space, preventing strong suppression of energy absorption.

\begin{figure}[t]
    \centering
\includegraphics[width=0.246\linewidth,height=0.18\linewidth]{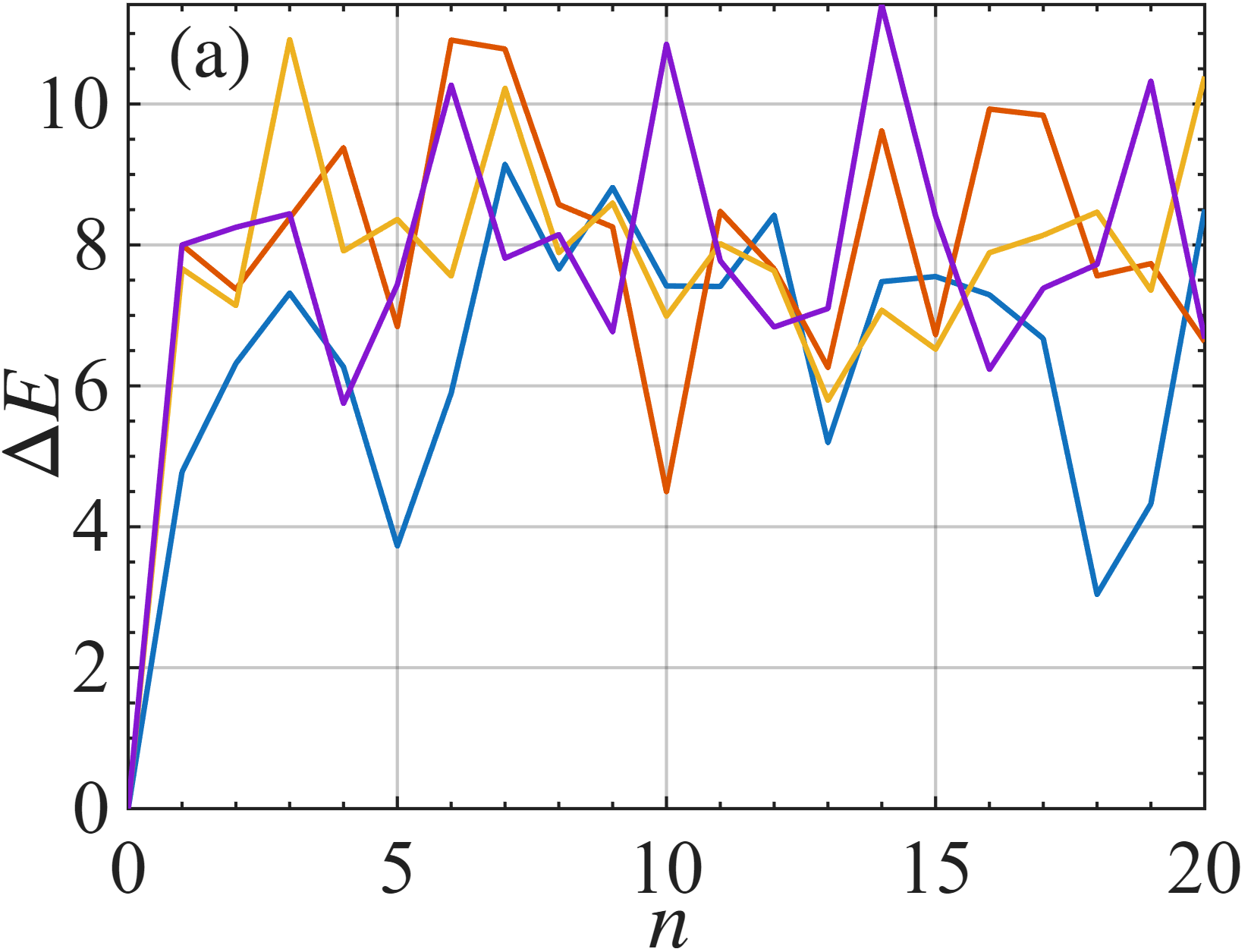}
\includegraphics[width=0.246\linewidth,height=0.18\linewidth]{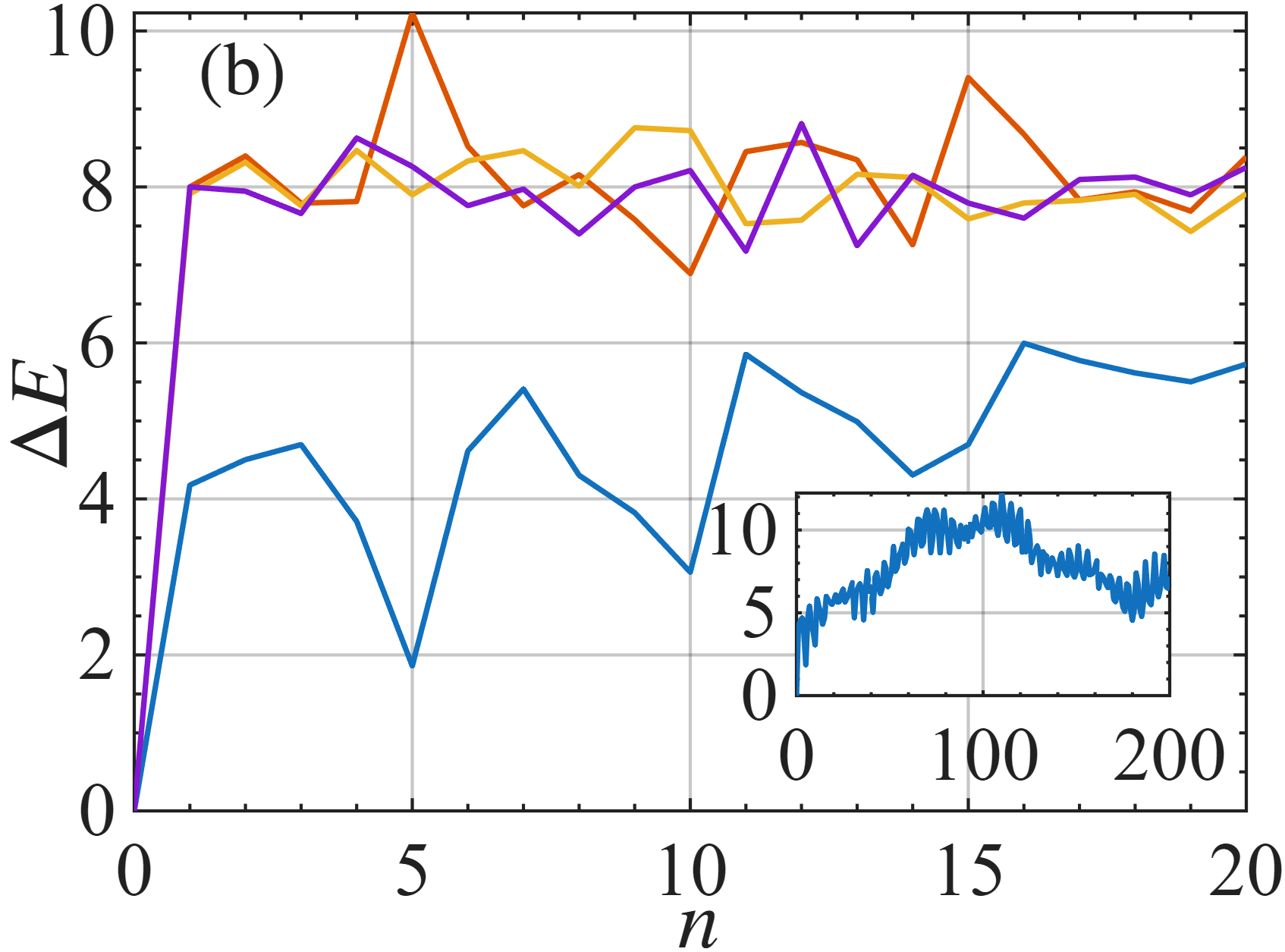}
\includegraphics[width=0.246\linewidth,height=0.18\linewidth]{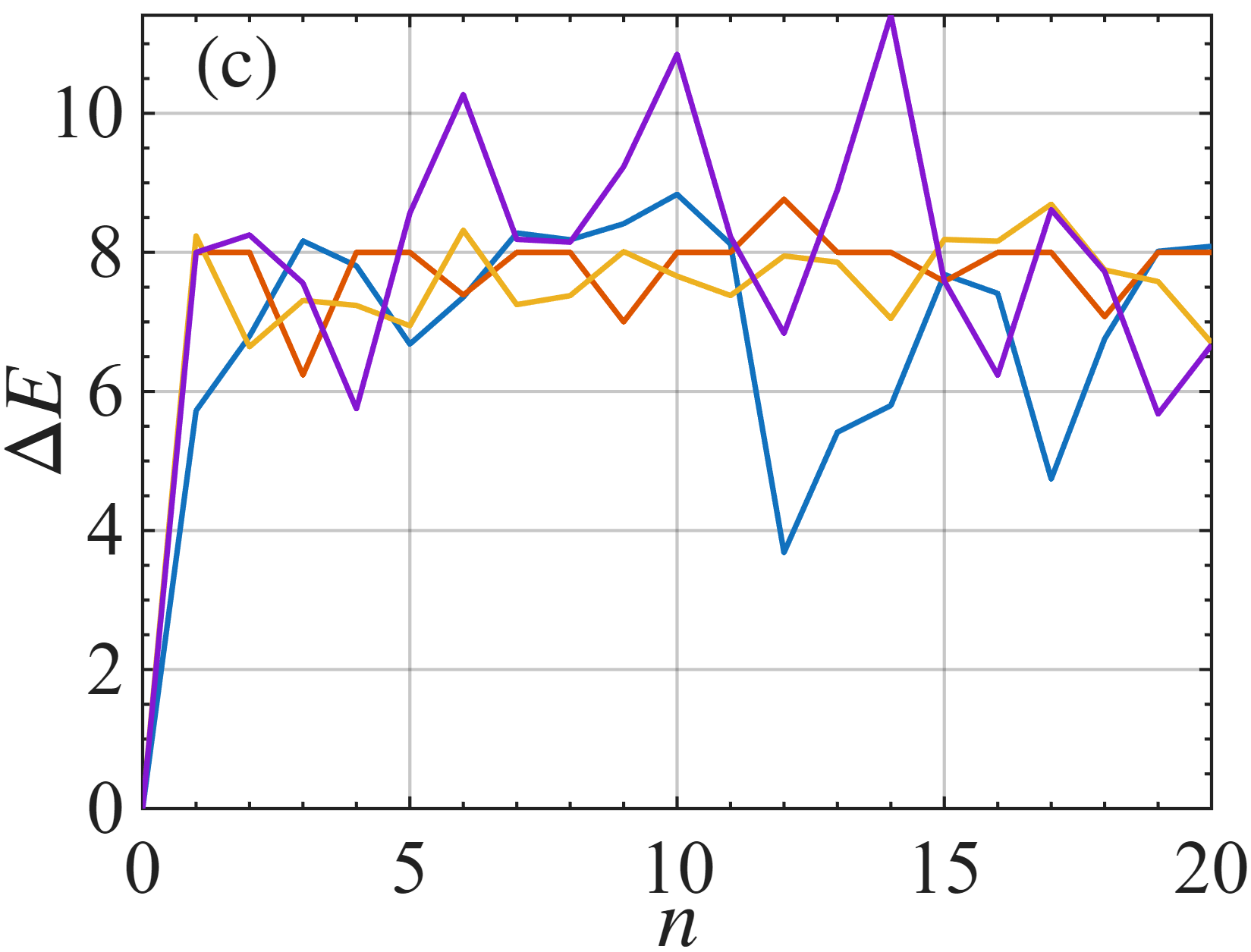}
\includegraphics[width=0.246\linewidth,height=0.18\linewidth]{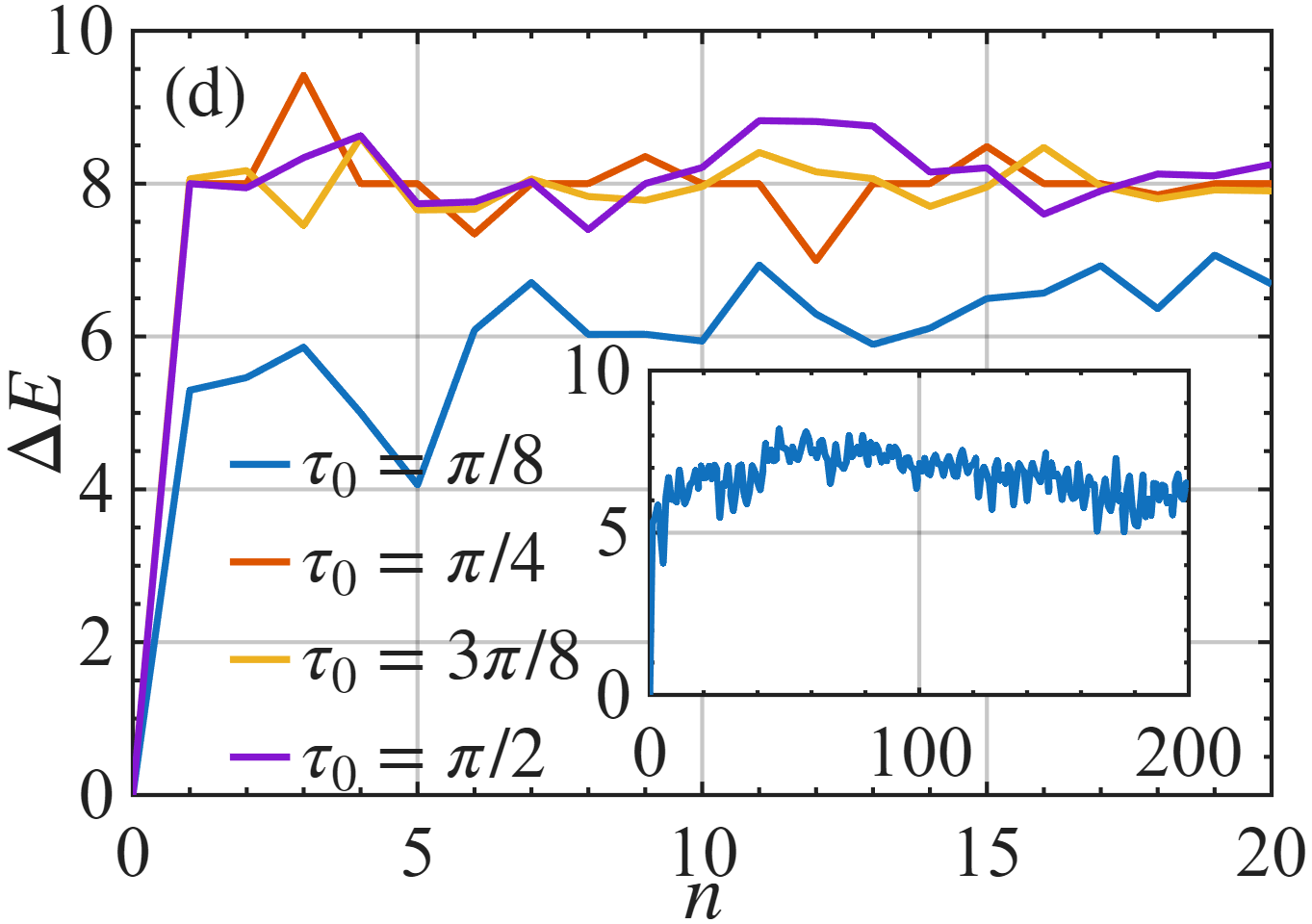}
 \caption{Storage energy $\Delta E$ as a function of the number of kicks $n$ for a non-interacting battery driven by a long-range interacting charger with fixed period $\tau_1 = \pi/4$, while $\tau_0$ is varied as $\pi/8$, $\pi/4$, $3\pi/8$, and $\pi/2$. Panels (a) and (b) correspond to the integrable case ($h_x = 0$), and panels (c) and (d) correspond to the non-integrable case ($h_x = 1$). Panels (a) and (c) show results for PBC, while panels (b) and (d) correspond to OBC. The insets in panels (b) and (d) show the long-time behavior for $\tau_1 = \pi/8$ for the corresponding cases. The system parameters are $N=8$, $J = 1$, $h_z = 1$, and $\omega = 1$.}
    \label{long_int_E_tau1_tau2}
\end{figure}

Next, we fix the driving period $\tau_0=\pi/4$ and vary $\tau_1$ over $\pi/8$, $\pi/4$, $3\pi/8$, and $\pi/2$ to examine the effects of temporal asymmetry on energy storage. For the integrable charger under PBC [Fig.~\ref{long_nint_E_tau1_tau2}(a)], the stored energy reaches its optimal value, $\Delta E_{\rm max}=2\omega N$, at $\tau_1=\pi/2$, restoring the resonant enhancement seen in the symmetric protocol, while for other values of $\tau_1$, the maximum stored energy remains above half of the optimal value, $\Delta E_{\rm max}>\omega N$, though full charging is not achieved. Under OBC [Fig.~\ref{long_nint_E_tau1_tau2}(b)], the maximum stored energy also exceeds half of the optimal value for all $\tau_1$, $\Delta E_{\rm max}>\omega N$, but remains below the PBC bound, reflecting boundary-induced suppression. Considering the nonintegrable charger under both PBC and OBC [Fig.~\ref{long_nint_E_tau1_tau2}(c,d)], the behavior is qualitatively similar: the resonant point $\tau_1=\pi/2$ under PBC yields optimal charging, while away from resonance the stored energy remains robustly above $\omega N$ and less than the optimum value {\it i.e.}, $\omega N<\Delta E_{\rm max} <2\omega N$. These observations indicate that efficient energy storage persists under unequal driving intervals regardless of integrability, with boundary conditions primarily determining whether full optimal charging can be achieved.

Overall, unequal driving periods generally reduce the sharpness of the resonant enhancement observed for $\tau_0=\tau_1=\pi/2$, but they do not destroy efficient energy storage. Under PBC, optimal charging can still be achieved when one of the driving periods satisfies the resonant condition $\pi/2$, while under OBC the stored energy remains robustly above half of the optimal value throughout the explored $(\tau_0,\tau_1)$ parameter space. These findings confirm that the long-range interacting charger exhibits strong resilience against temporal asymmetry, and that many-body \emph{structural effects} continue to govern the global features of the charging landscape.

\begin{figure}[t]
    \centering
    \includegraphics[width=0.246\linewidth,height=0.18\linewidth]{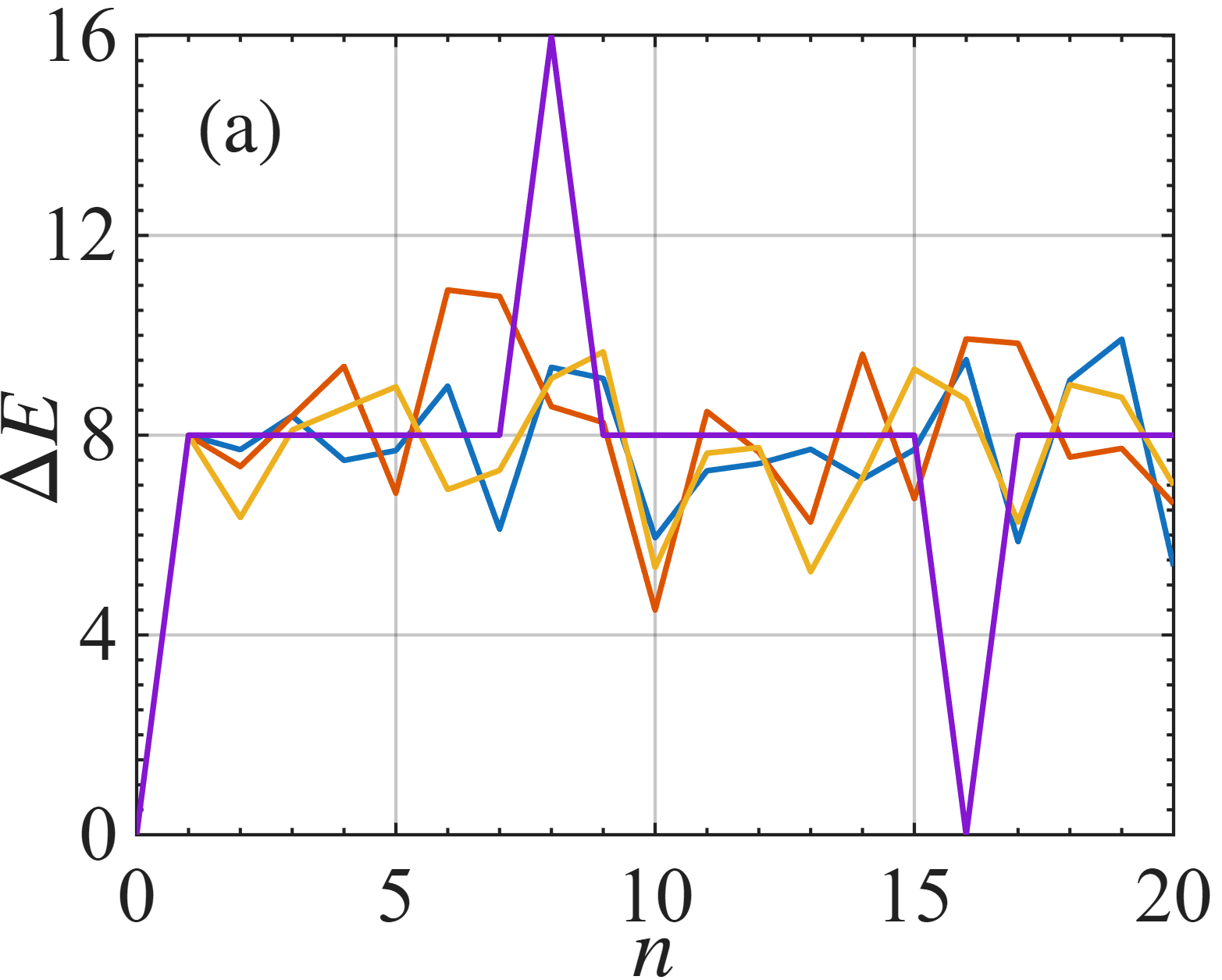}
 \includegraphics[width=0.246\linewidth,height=0.18\linewidth]{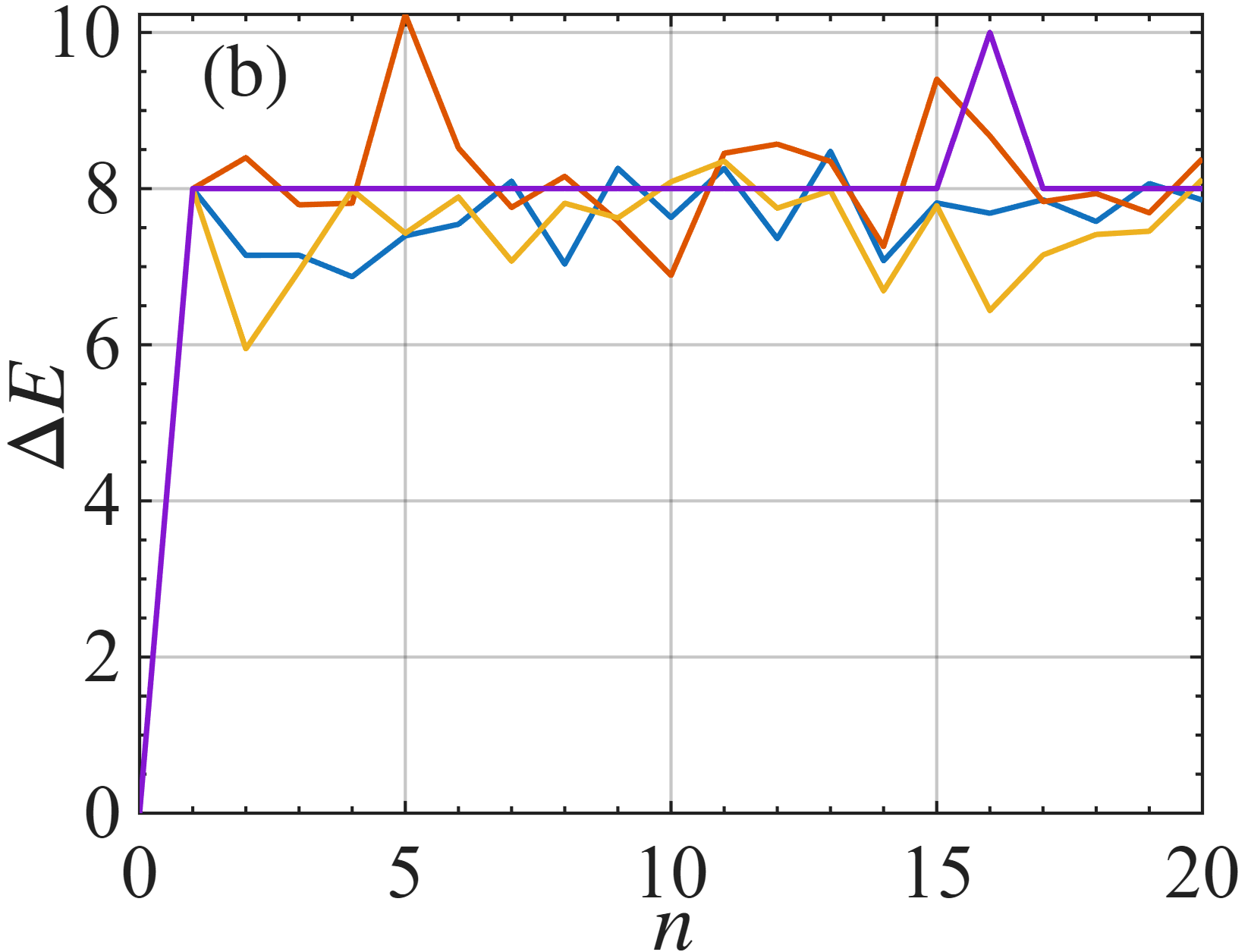}
  \includegraphics[width=0.246\linewidth,height=0.18\linewidth]{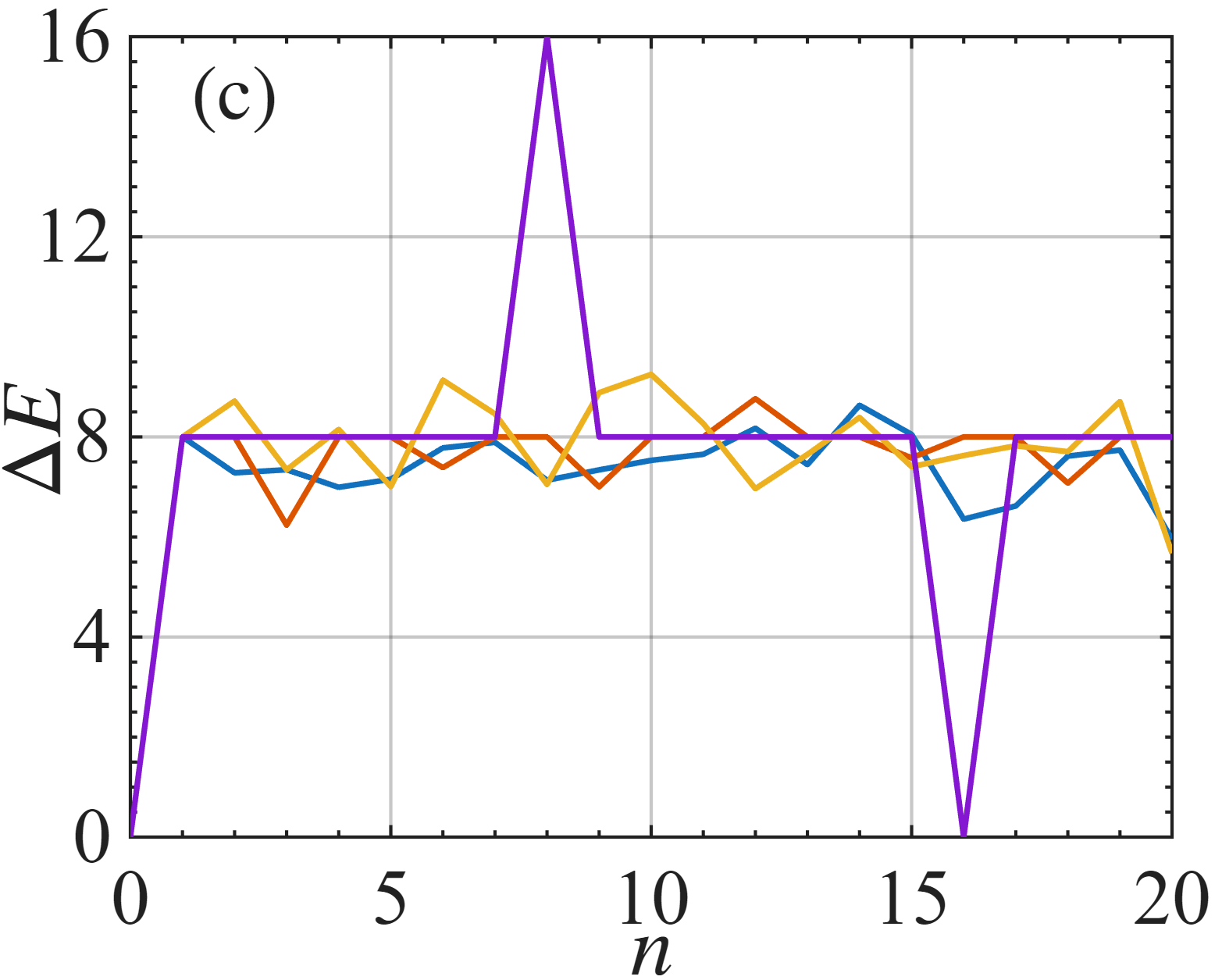}
   \includegraphics[width=0.246\linewidth,height=0.18\linewidth]{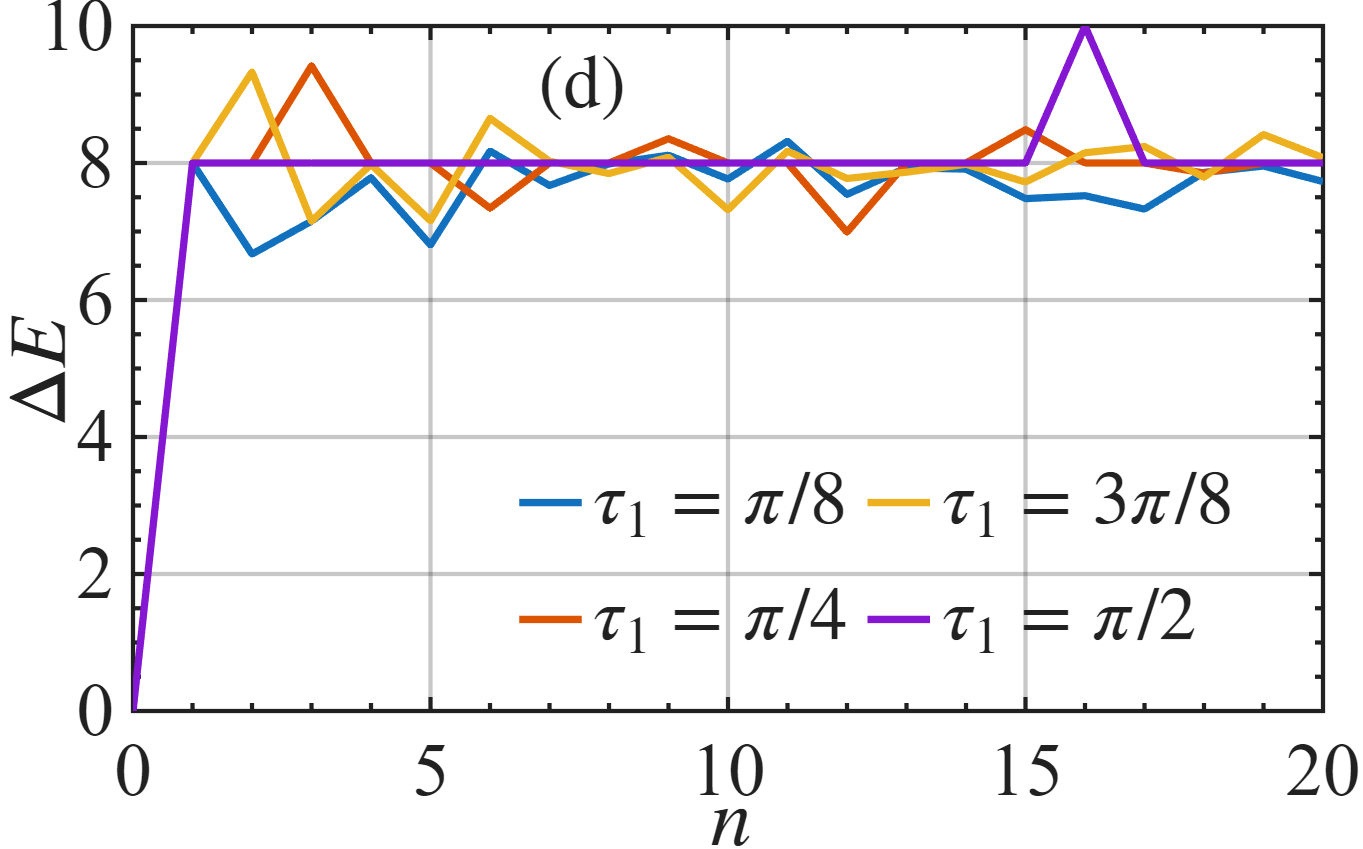}
\caption{Same as Fig.~\ref{long_int_E_tau1_tau2}, here $\tau_0$ fixed at $\pi/4$ and $\tau_1$ varied as $\pi/8$, $\pi/4$, $3\pi/8$, and $\pi/2$.}
    \label{long_nint_E_tau1_tau2}
\end{figure}

\subsection{Interaction-strength dependence and structural robustness}

Although the main analysis is carried out for a one-dimensional Ising charger, it is crucial to establish that the observed charging behavior is not an artifact of dimensionality or of a particular choice of interaction strength. In experimentally relevant platforms, higher-dimensional Mott-insulating bosonic lattices can be effectively reduced to a quasi-one-dimensional information channel by concentrating bosons along a selected path through a dimensional-reduction protocol that combines free-boson and Bose–Hubbard Hamiltonians, achievable in $\mathcal{O}(1)$ time. The resulting high-density bosonic states may then be encoded into an effective qubit representation and mapped onto a long-range Ising-type interaction, where the coupling is renormalized via bosonic enhancement as $J \rightarrow C J$ with $C \propto \bar{n}_t^{,p}$, $\bar{n}_t$ being the boson density and $p$ the order of the local $p$-body repulsion. This construction demonstrates that the effective interaction scale can be significantly amplified in realistic implementations. Consequently, it becomes necessary to determine whether such an enhancement of $J$ can further improve the charging performance, or whether the optimal behavior identified in the main text is intrinsically protected by the Floquet-engineered resonance mechanism. This consideration motivates a systematic study of the interaction-strength dependence within the resonant driving protocol.

We therefore extend this reasoning to a systematic analysis of the resonant protocol $\tau_0=\tau_1=\pi/2$ by varying the interaction strength over $J\in[0.5,2.0]$ in steps of $0.5$, considering both integrable and nonintegrable chargers under PBC and  OBC. This exploration allows us to assess whether amplifying the effective coupling scale translates into enhanced charging capability. For PBC, we observe that the battery attains the optimal storage $\Delta E_{\rm max}=2\omega N$ across the entire parameter window, except at special commensurate points where $J$ is an integer multiple of four (it will be discussed later), as depicted in Fig.~\ref{E_long pi2_J}(a). Outside these isolated values, the resonant Floquet structure secures maximal energy absorption, while variations in $J$ primarily influence the accumulated many-body phase and hence the stroboscopic recurrence time of the storage dynamics, leading to a progressive reduction of the oscillation period with increasing interaction strength. In contrast, for the integrable charger with OBC, the attainable storage saturates at $\Delta E_{\rm max}=\omega N+2$ and remains entirely insensitive to $J$, as shown in Fig.~\ref{E_long pi2_J}(b), indicating that boundary-induced spectral constraints fix the charging ceiling. The nonintegrable charger, at period $\tau_0=\tau_1=\pi/2$, follows the same qualitative trend as in the integrable case under both PBC and OBC for all noncommensurate values of $J$, where resonance ensures optimal or near-optimal charging and the interaction strength reshapes only the fine temporal features rather than the maximum energy reached.

\begin{figure}[t]
    \centering
\includegraphics[width=0.45\linewidth,height=0.32\linewidth]{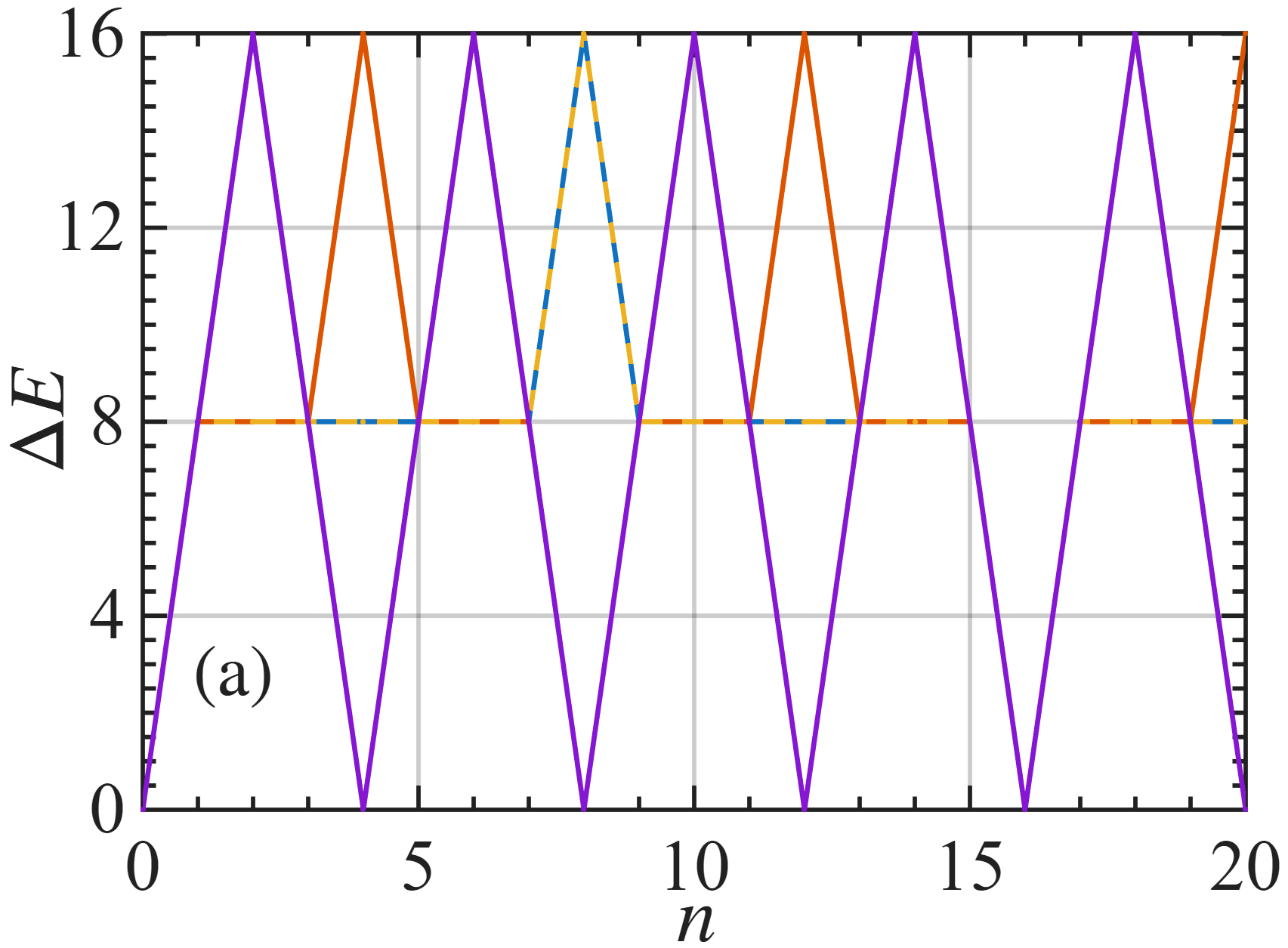}
\includegraphics[width=0.45\linewidth,height=0.32\linewidth]{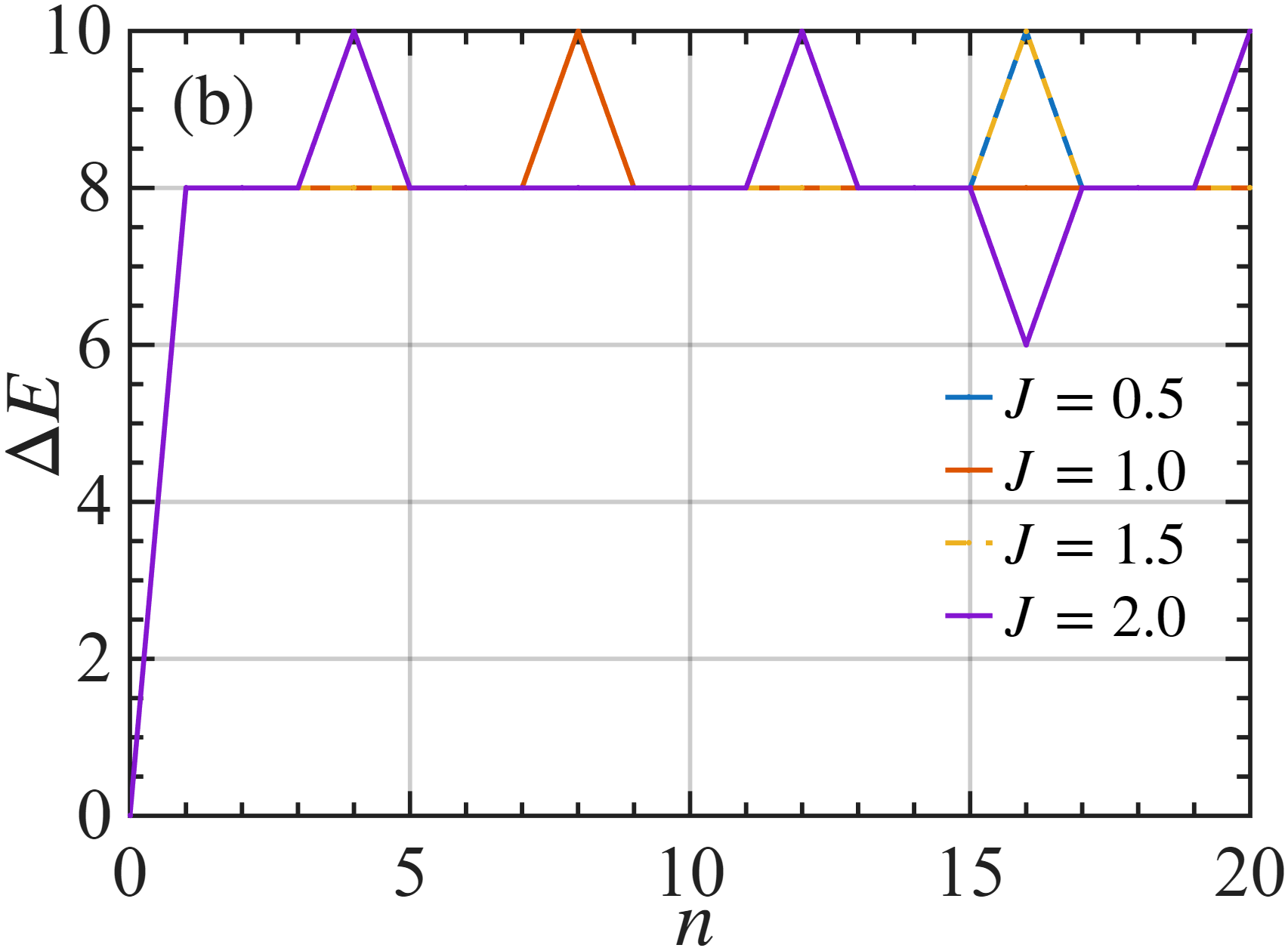}
\caption{Stored energy $\Delta E$ as a function of the number of kicks $n$ for a long-range interacting integrable charger: (a) PBC, (b) OBC, with the interaction strength $J$ varied from $0.5$ to $2.0$ in steps of $0.5$. Parameters are: $N=8$, $\tau_0=\tau_1 = \pi/2$, $h_z = \omega = 1$, and $h_x = 0$.}
\label{E_long pi2_J}
\end{figure}

\begin{figure}[b]
    \centering
\includegraphics[width=0.246\linewidth,height=0.18\linewidth]{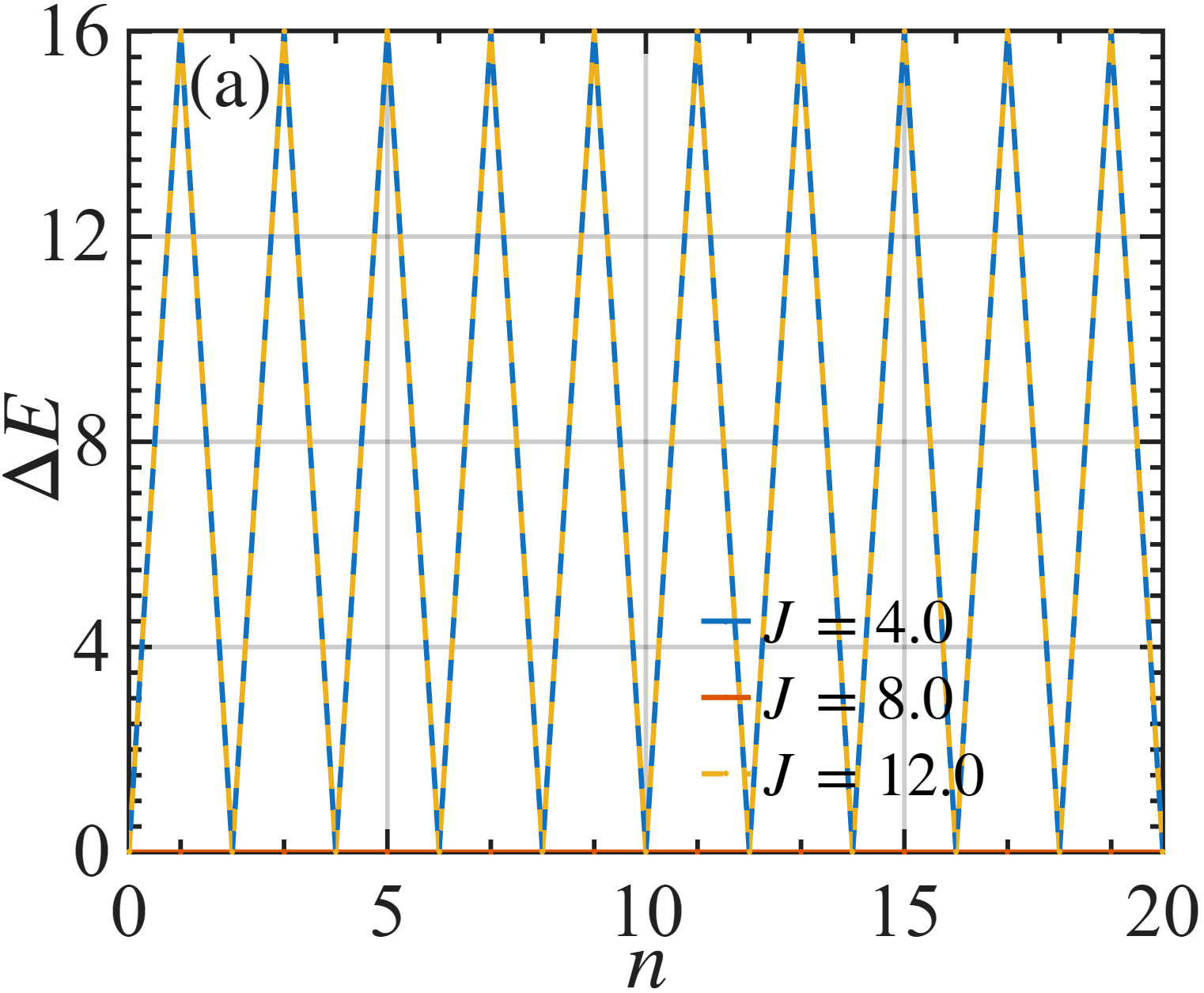}
\includegraphics[width=0.246\linewidth,height=0.18\linewidth]{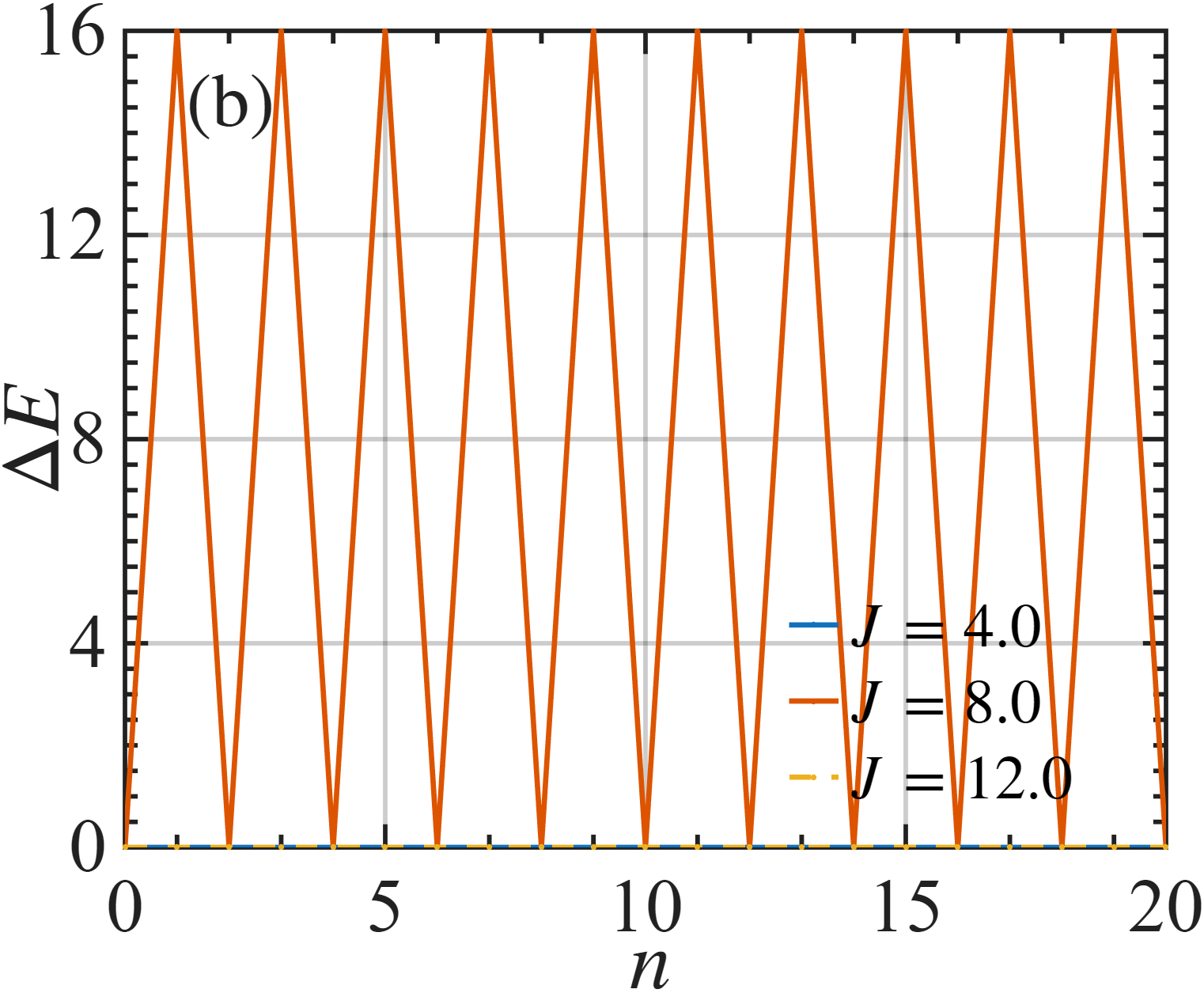}
\includegraphics[width=0.246\linewidth,height=0.18\linewidth]{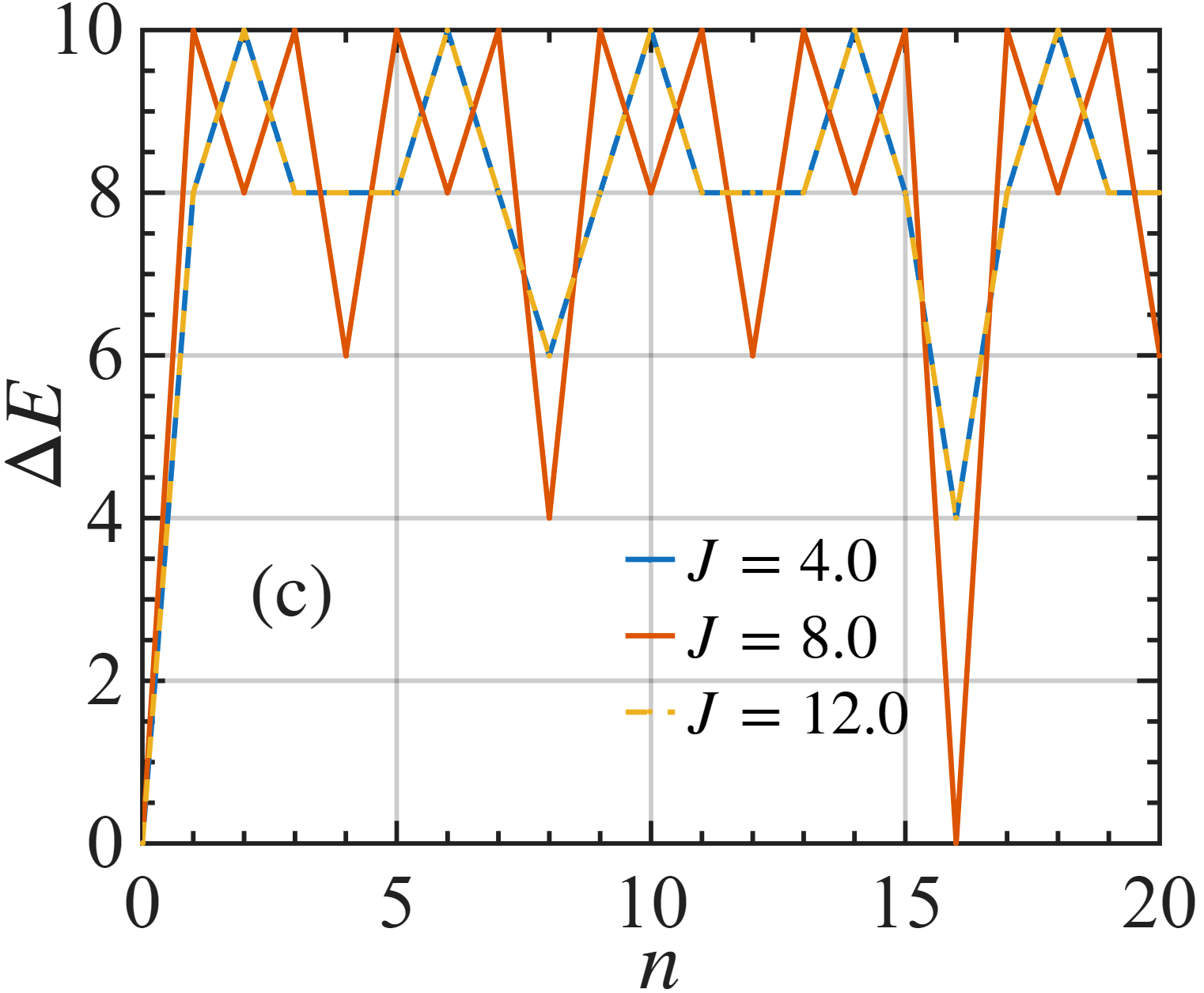}
\includegraphics[width=0.246\linewidth,height=0.18\linewidth]{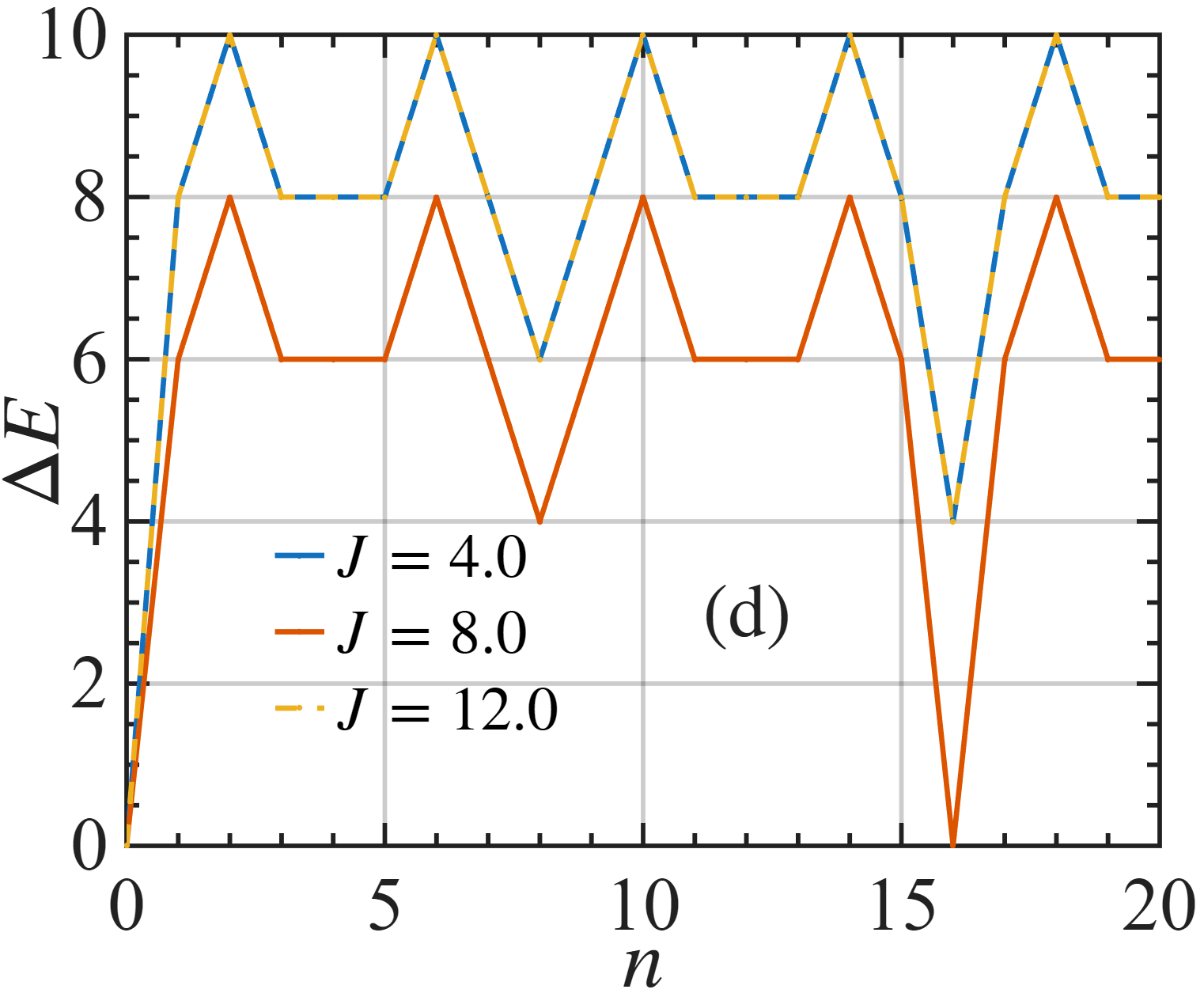}
\caption{Stored energy $\Delta E$ as a function of the number of kicks $n$ for a long-range interacting charger: integrable ($h_x = 0$) in panels (a,c) and nonintegrable ($h_x = 1$) in panels (b,d). Panels (a,b) correspond to PBC, and panels (c,d) correspond to OBC. The interaction strength $J$ is varied in multiples of four. Other parameters are $N=8$, $\tau_0=\tau_1 = \pi/2$, $h_z = \omega = 1$.}
\label{E_long pi2_J4}
\end{figure}

A qualitatively different situation arises when the commensurability condition $J=4m$ ($m \in \mathbb{Z})$ is fulfilled, as summarized in Fig.~\ref{E_long pi2_J4}. In this regime, the interplay between Floquet phase accumulation and interaction-induced level structure produces pronounced parity-dependent interference effects. Under PBCs, the integrable charger alternates between complete suppression of charging for odd multiples, $\Delta E_{\rm max}=0$, and full restoration of optimal storage for even multiples, $\Delta E_{\rm max}=2\omega N$ [Fig.~\ref{E_long pi2_J4}(a)]. The nonintegrable charger under PBC exhibits the complementary pattern, achieving optimal storage for odd multiples and vanishing energy absorption for even multiples [Fig.~\ref{E_long pi2_J4}(b)]. Open boundary conditions substantially mitigate this resonance-like sensitivity in the integrable case, preserving the constant value $\Delta E_{\rm max}=\omega N+2$ irrespective of parity [Fig.~\ref{E_long pi2_J4}(c)], whereas the nonintegrable system displays only partial suppression at even multiples, where the storage capacity is reduced to $\Delta E_{\rm max}=\omega N$ [Fig.~\ref{E_long pi2_J4}(d)]. Collectively, these observations confirm that even when the effective interaction scale is significantly enhanced, the maximal charging performance is largely protected by the Floquet-engineered resonance mechanism, with deviations appearing only at special commensurate interaction strengths determined by the combined role of integrability and boundary conditions.

\subsection{Dependence of maximum stored energy on driving period and system size}

\begin{figure}[t]
    \centering
\includegraphics[width=0.32\linewidth,height=0.22\linewidth]{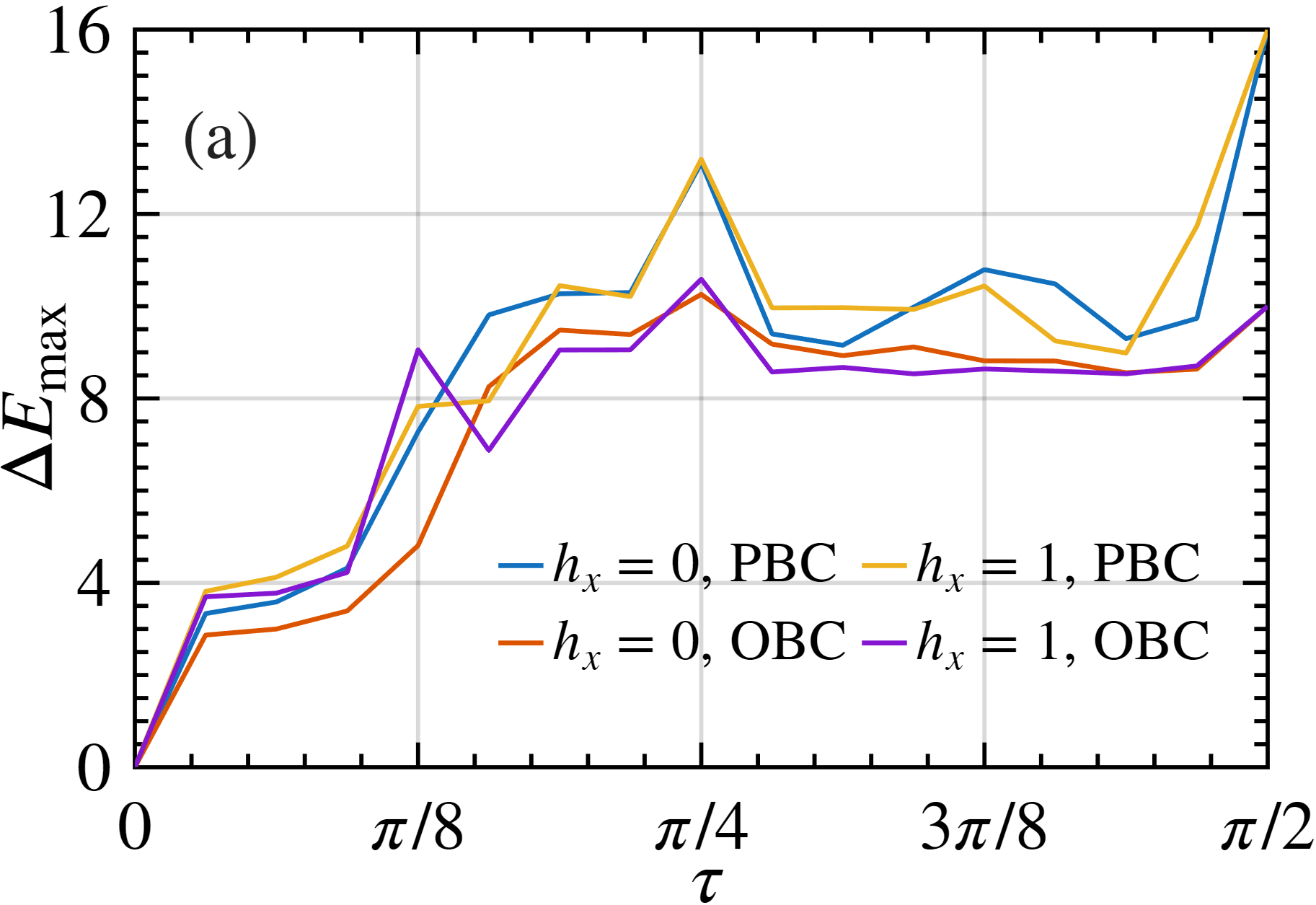}
\includegraphics[width=0.32\linewidth,height=0.22\linewidth]{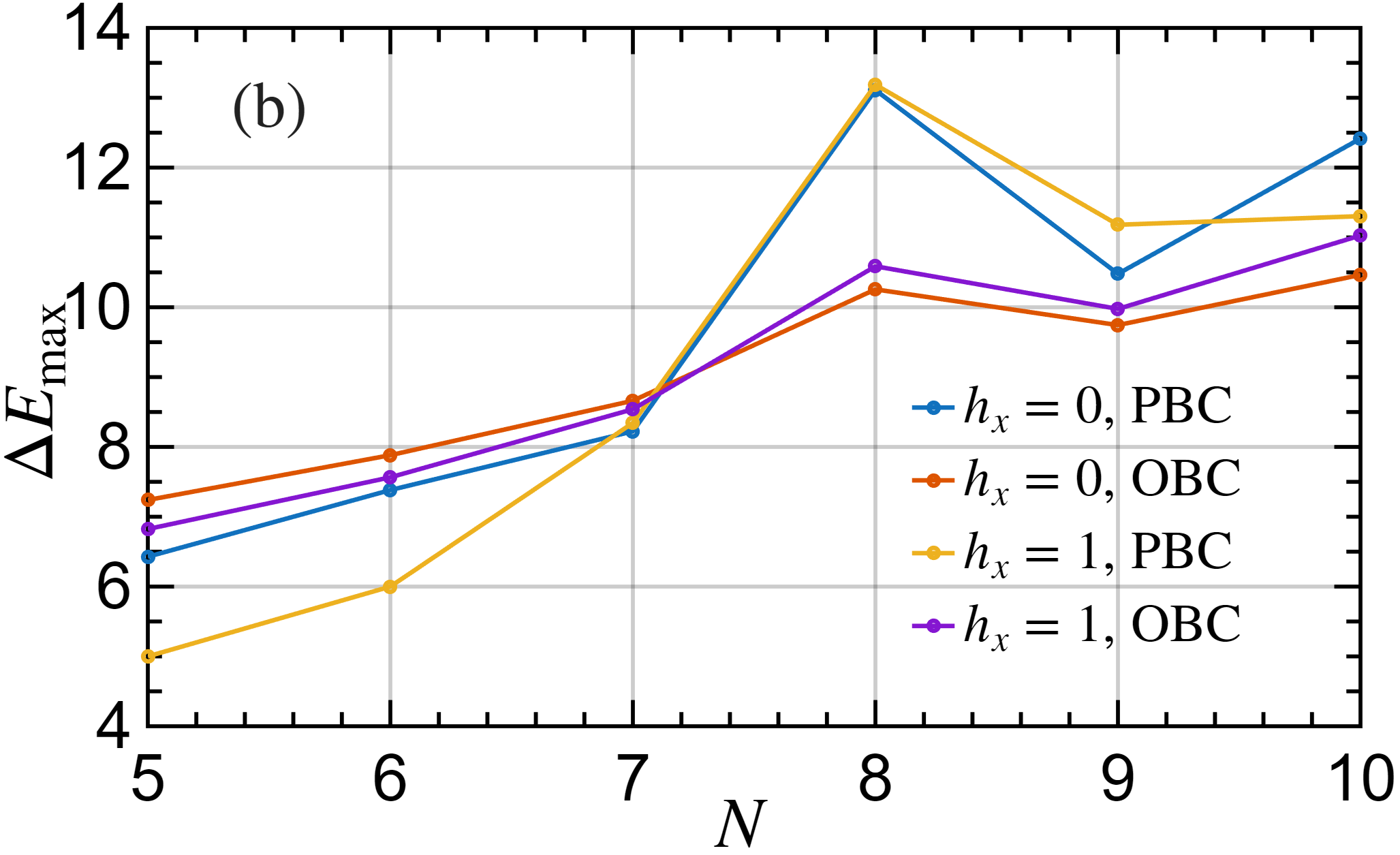}
\includegraphics[width=0.32\linewidth,height=0.22\linewidth]{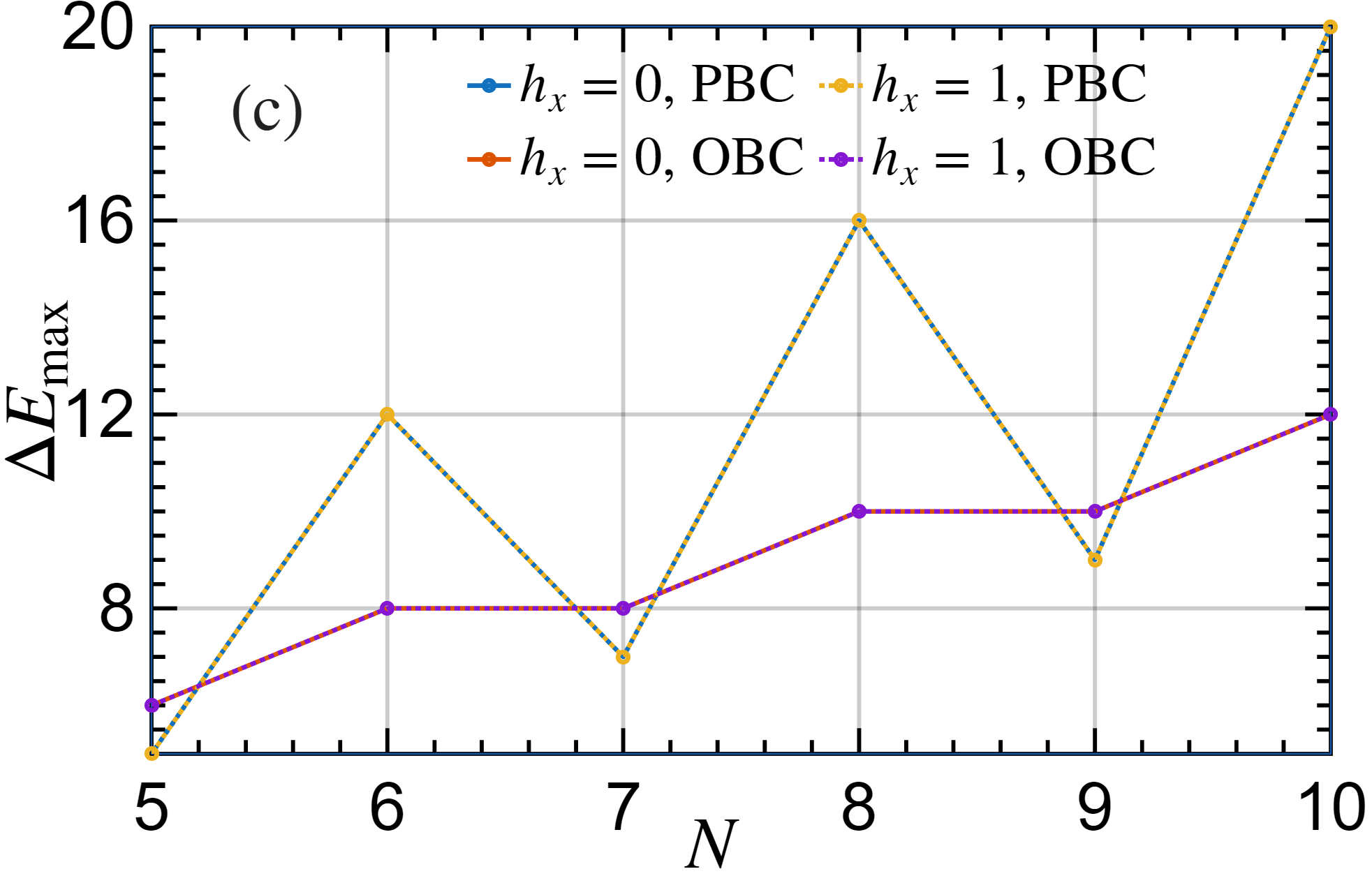}
  \caption{(a) Maximum stored energy $\Delta E$ as a function of the Floquet period $\tau_0=\tau_1 \equiv \tau \in [0,\pi/2]$ for a long-range interacting charger with system size $N=8$. The system is integrable for $h_x = 0$ (absence of longitudinal field) and nonintegrable for $h_x = 1$ (presence of longitudinal field).
(b,c) Maximum stored energy as a function of system size $N$ for fixed driving periods: (b) $\tau_0=\tau_1 = \pi/4$ and (c) $\tau_0=\tau_1 = \pi/2$, for all cases considered. Parameters: $J=1$, $h_z=1$, and $\omega=1$.}
\label{Emax_long_tau}
\end{figure}

To further clarify the role of many-body \emph{structural effects} in periodically driven Ising quantum batteries, we analyze the maximum stored energy of a long-range interacting charger as a function of the symmetric driving period $\tau_0=\tau_1\equiv\tau$. We consider both integrable ($h_x=0$) and nonintegrable ($h_x=1$) regimes under PBC and OBC, fixing the system size to $N=8$. The maximum stored energy $\Delta E_{\rm max}$ is extracted from the stroboscopic evolution over $n=0$ to $500$ kicks, while $\tau$ is varied in the range $\tau\in[0,\pi/2]$ with resolution $\pi/32$, as shown in Fig.~\ref{Emax_long_tau}(a).

For all structural configurations, the limit $\tau=0$ corresponds to identity evolution, yielding vanishing storage, $\Delta E_{\rm max}=0$. In the weak-driving regime, exemplified by $\tau=\pi/32$, the stored energy remains nearly constant and below $\omega N/2$, indicating that short-period driving suppresses energy absorption and largely masks structural distinctions. A similar trend persists at $\tau=\pi/16$, except for the nonintegrable system with PBC, where $\Delta E_{\rm max}$ slightly exceeds $\omega N/2$, signaling the onset of enhanced absorption enabled by the interplay between chaotic dynamics and translational symmetry. As $\tau$ increases toward $\tau\simeq 3\pi/32$, \emph{structural effects} become more pronounced: most configurations exhibit $\Delta E_{\rm max}>\omega N/2$, whereas the integrable system under OBC remains below this threshold, reflecting the restrictive influence of open boundaries in the presence of conserved quantities. At $\tau=\pi/8$, the integrable systems under both PBC and OBC and the nonintegrable system under PBC satisfy $\omega N/2<\Delta E_{\rm max}<\omega N$, while the nonintegrable system under OBC shows a marked enhancement with $\Delta E_{\rm max}>\omega N$, demonstrating a strong synergy between boundary-induced mode mixing and chaotic dynamics. For larger driving periods up to $\tau=\pi/2$, all configurations enter a regime of efficient charging with $\omega N<\Delta E_{\rm max}<2\omega N$, governed by Floquet resonances. A notable enhancement appears at $\tau=\pi/4$, where both integrable and nonintegrable systems under PBC achieve $\Delta E_{\rm max}>3\omega N/4$, outperforming their OBC counterparts. At the resonant point $\tau=\pi/2$, the stored energy reaches its optimal value: under PBC the battery attains $\Delta E_{\rm max}=2\omega N$, whereas under OBC and for the even system size considered here it reaches $\Delta E_{\rm max}=\omega N+2$, consistent with the resonance mechanism discussed in the main text. These results confirm that integrability and boundary conditions act as key structural controls, strongly shaping the charging landscape [Fig.~\ref{Emax_long_tau}(a)].

We next investigate the system-size dependence of the maximum stored energy for fixed driving periods $\tau=\pi/4$ [Fig.~\ref{Emax_long_tau}(b)] and $\tau=\pi/2$ [Fig.~\ref{Emax_long_tau}(c)], considering both integrable and nonintegrable dynamics under PBC and OBC. For $\tau=\pi/4$, $\Delta E_{\rm max}$ scales approximately linearly with $N$ in all cases, demonstrating extensive energy storage. In addition, a clear structural enhancement appears at commensurate sizes $N=4\mathcal{N}$, where the stored energy is systematically higher across both dynamical regimes and boundary geometries, highlighting the role of lattice commensurability in periodically driven Ising batteries. At the resonant point $\tau=\pi/2$, the behavior under PBC and OBC remains qualitatively similar for integrable and nonintegrable dynamics, but a pronounced odd–even effect emerges. Under PBC, even system sizes achieve optimal charging with $\Delta E_{\rm max}=2\omega N$, whereas odd sizes are restricted to $\Delta E_{\rm max}=\omega N$. Under OBC, the parity dependence persists in a softened form: for odd $N$, $\Delta E_{\rm max}=\omega N+1$, while for even $N$ it increases to $\Delta E_{\rm max}=\omega N+2$, with $\Delta E_{\rm max}>\omega N$ in all cases. Together, these findings underscore that charging efficiency in periodically driven Ising quantum batteries is governed not by a single control parameter but by the coordinated interplay among driving period, system size, boundary structure, and integrability.

\section{nearest-neighbor interacting charger}
\label{NN_S}

We now examine the short-range limit of the long-range charger discussed in the main text by restricting the interaction range to nearest neighbors ($k=1$). This regime provides a complementary perspective on how spatial interaction structure influences charging dynamics. In particular, we investigate (i) asymmetric driving protocols with $\tau_0 \neq \tau_1$, (ii) the role of interaction strength, and (iii) the dependence of the maximum stored energy on the symmetric driving period $\tau_0=\tau_1\equiv\tau$. The analysis highlights how short-range interactions reshape the interplay among integrability, boundary conditions, and Floquet resonances.

\subsection{Asymmetric driving protocol: $\tau_0 \neq \tau_1$}

\begin{figure}[t]
    \centering
\includegraphics[width=0.246\linewidth,height=0.18\linewidth]{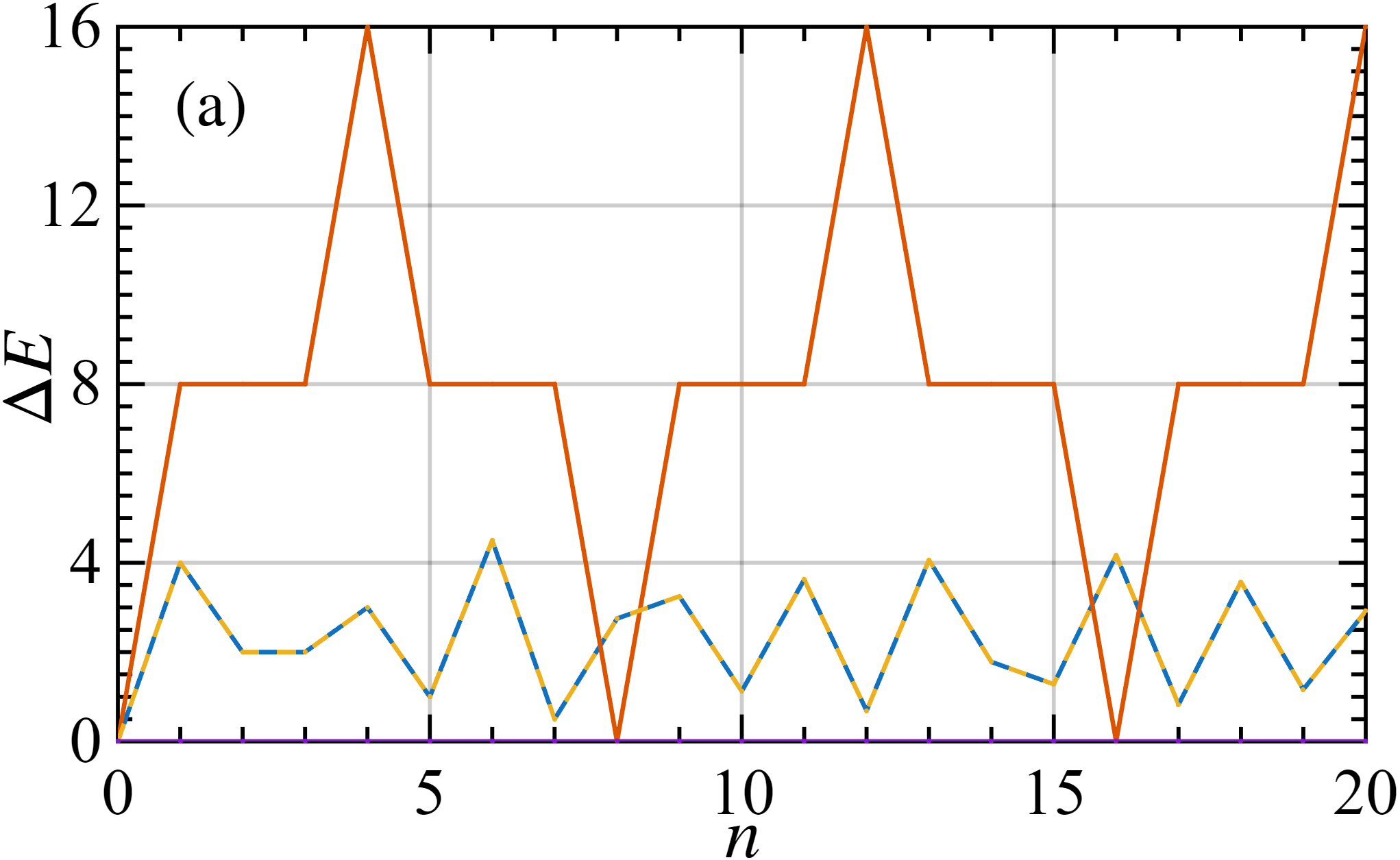}
\includegraphics[width=0.246\linewidth,height=0.18\linewidth]{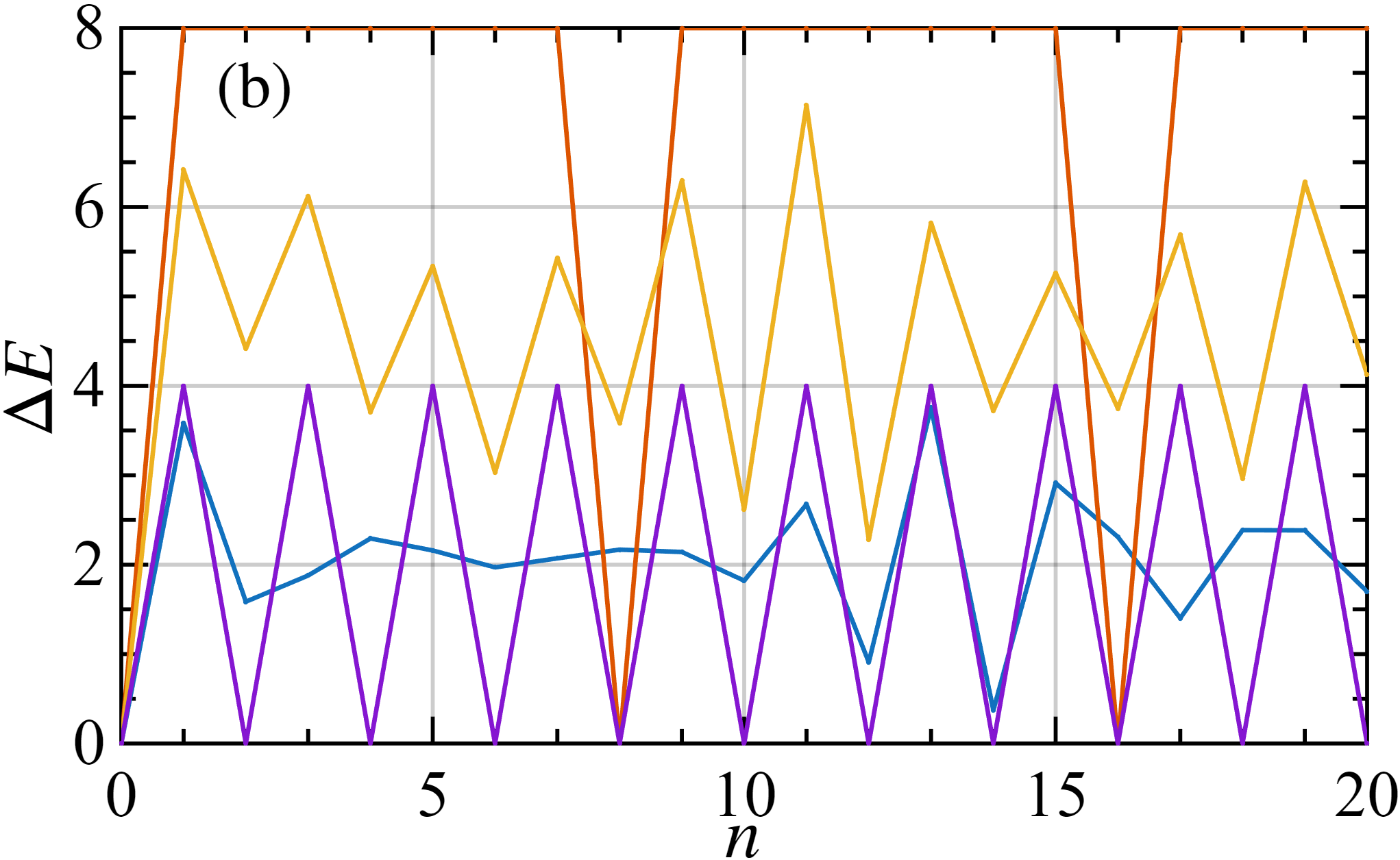}
\includegraphics[width=0.246\linewidth,height=0.18\linewidth]{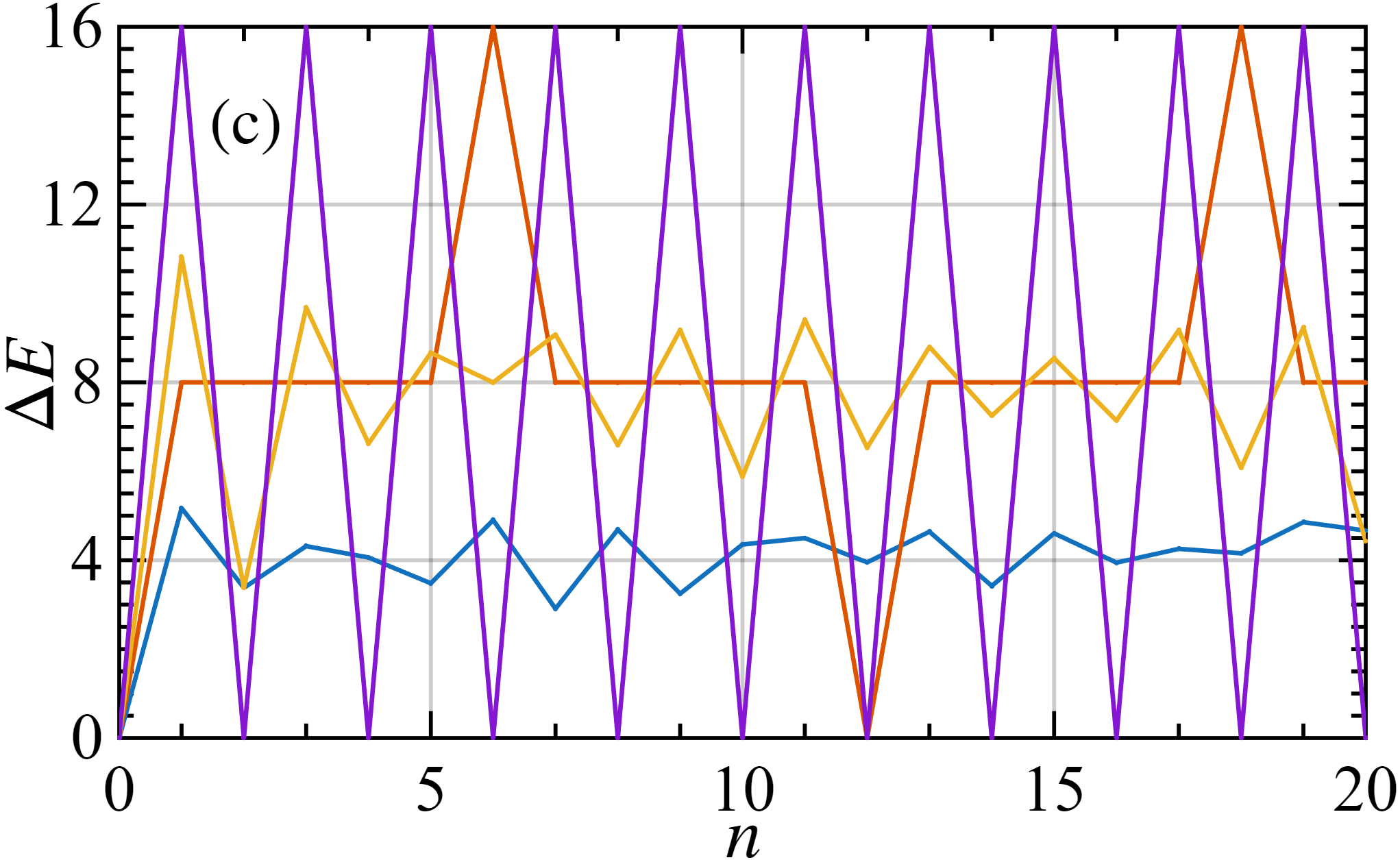}
\includegraphics[width=0.246\linewidth,height=0.18\linewidth]{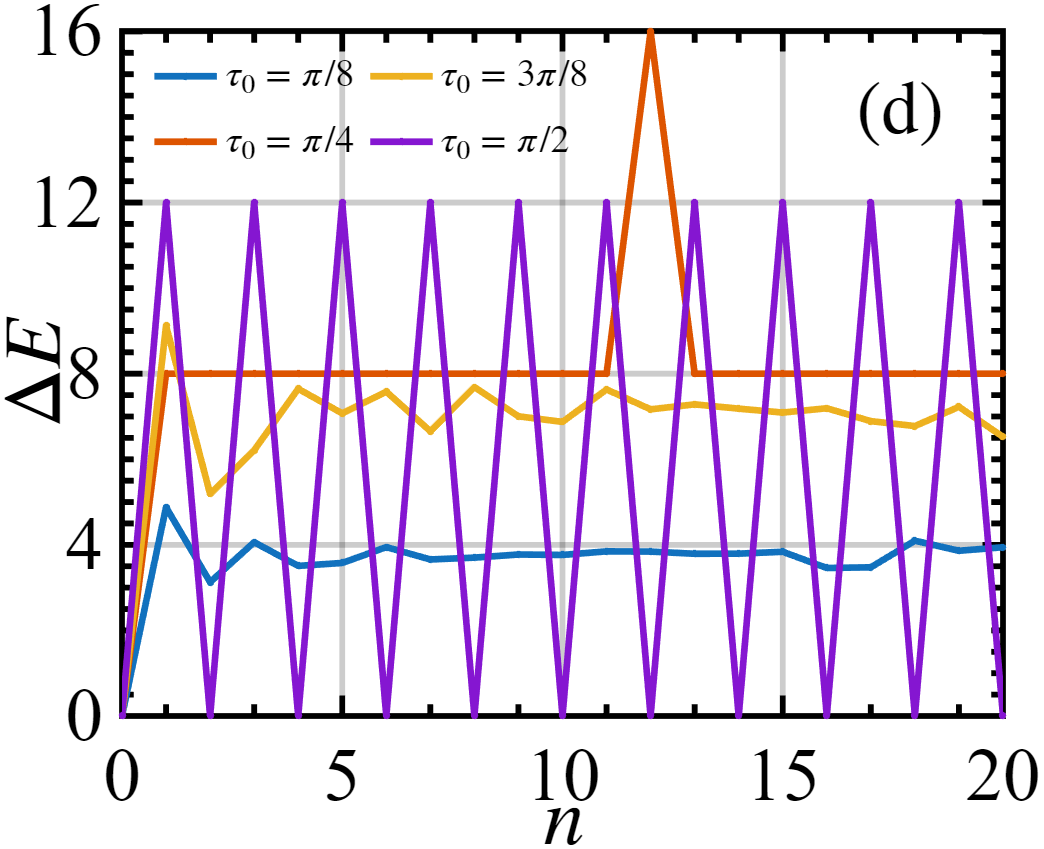}
   \caption{ Storage energy $\Delta E$ as a function of the number of kicks $n$ for a non-interacting battery driven by a NN interacting charger with fixed period $\tau_1 = \pi/4$, while $\tau_0 $ is varied as $\pi/8$, $\pi/4$, $3\pi/8$, and $\pi/2$. Panels (a) and (b) correspond to the integrable case ($h_x = 0$), and panels (c) and (d) correspond to the non-integrable case ($h_x = 1$). Panels (a) and (c) show results for PBC, while panels (b) and (d) correspond to OBC.  The system parameters are $N=8$, $J = 1$, $h_z = 1$, and $\omega = 1$. }
    \label{NN_int_E_tau1_tau2}
\end{figure}

\begin{figure}[b]
    \centering
\includegraphics[width=0.246\linewidth,height=0.18\linewidth]{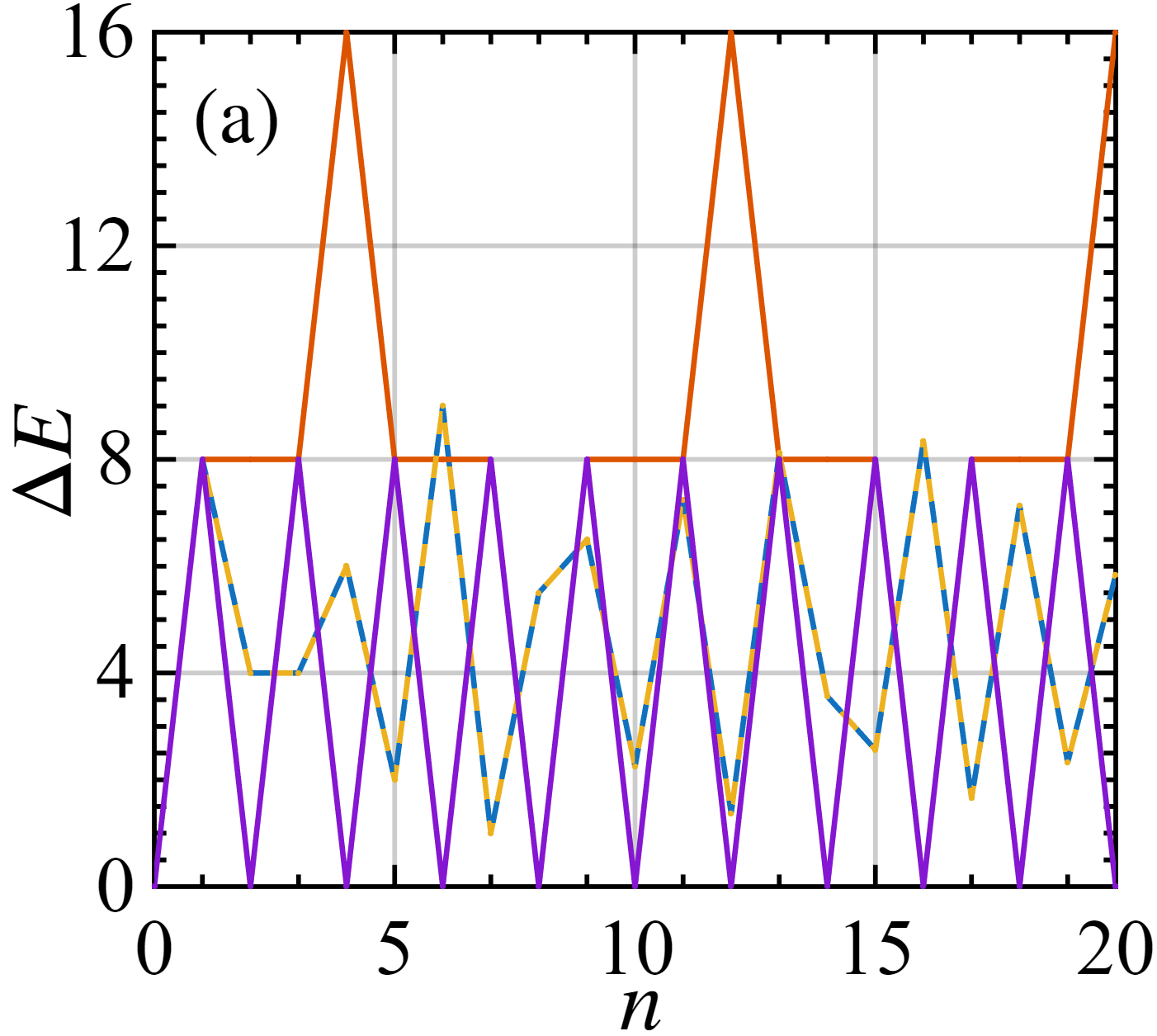}
\includegraphics[width=0.246\linewidth,height=0.18\linewidth]{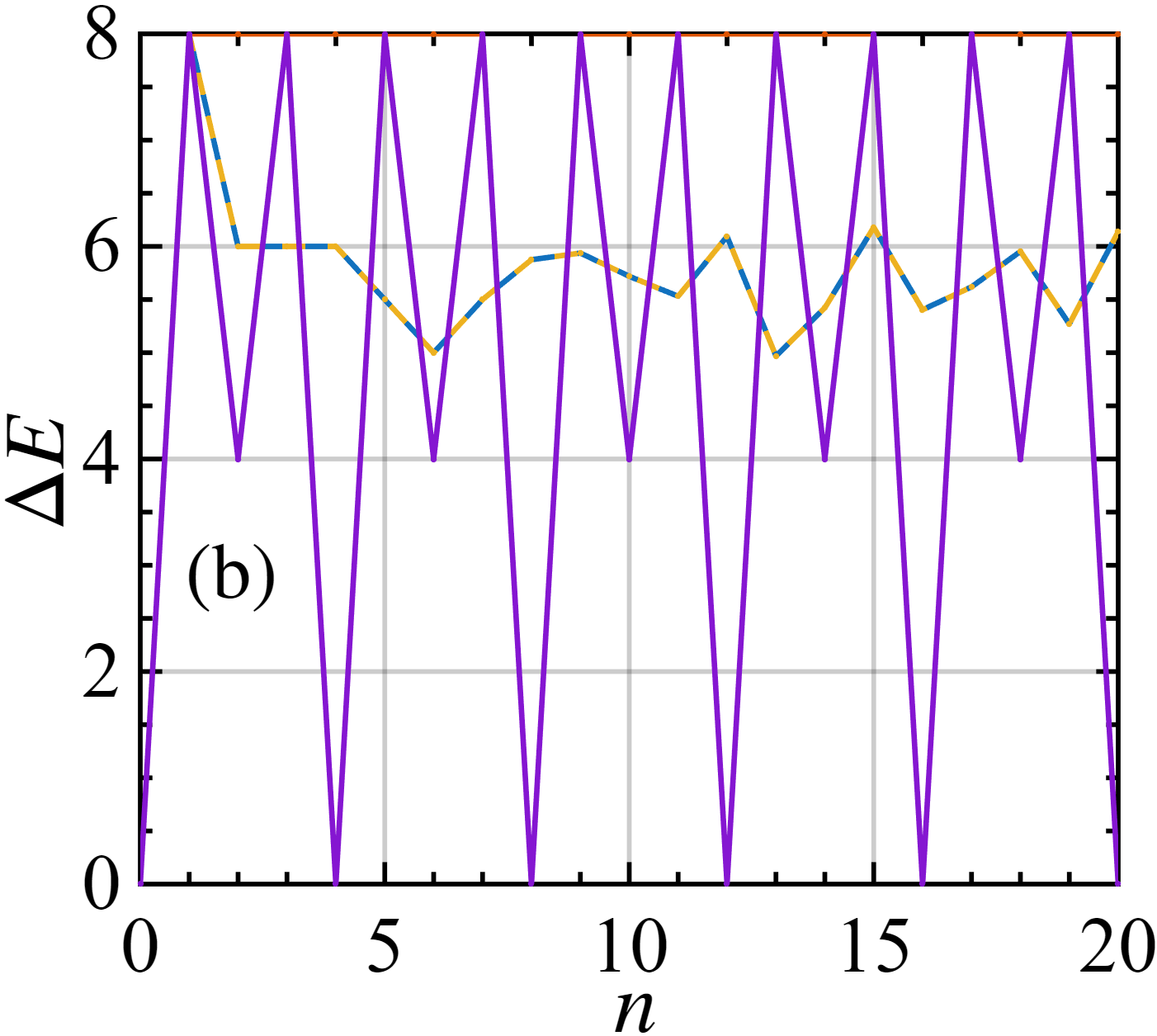}
\includegraphics[width=0.246\linewidth,height=0.18\linewidth]{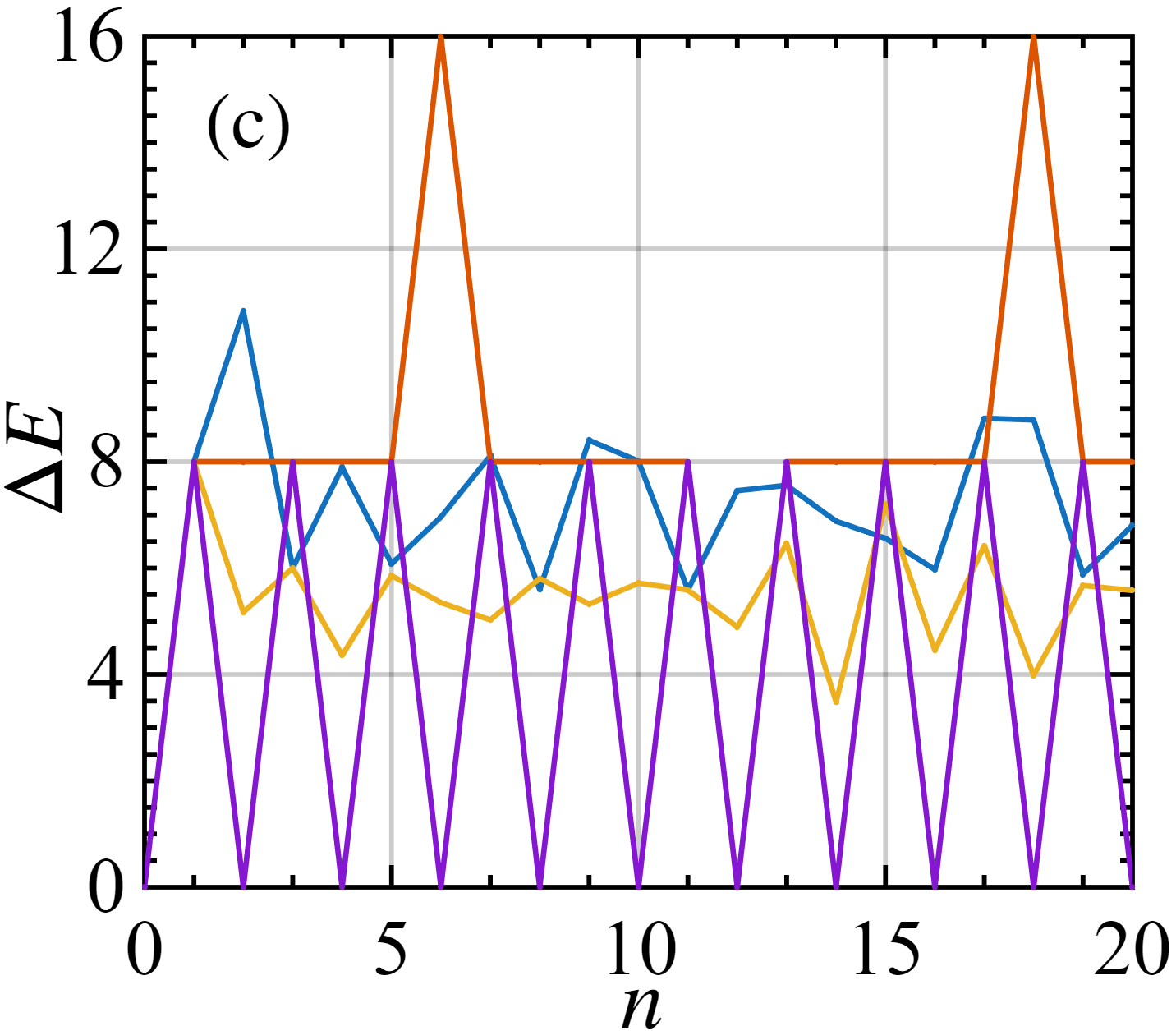}
\includegraphics[width=0.246\linewidth,height=0.18\linewidth]{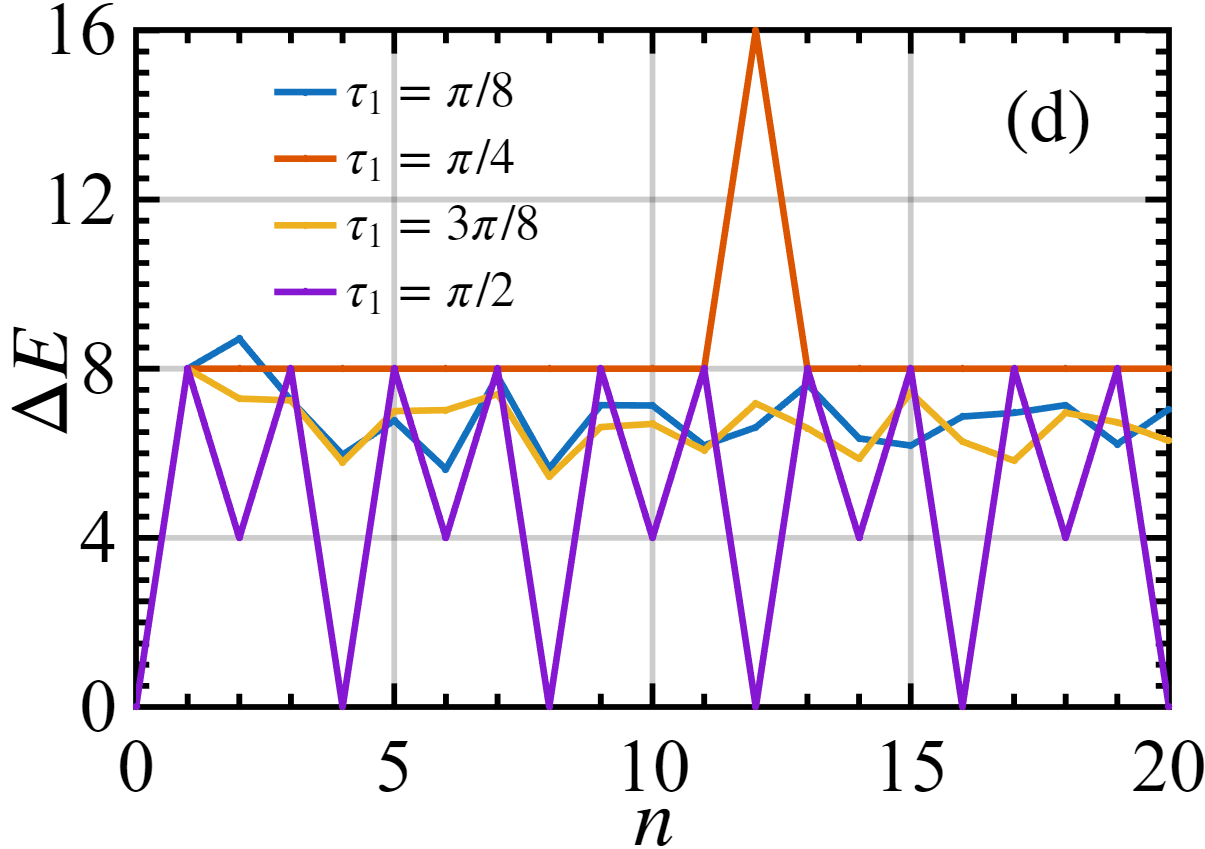}
    \caption{Same as Fig.~\ref{NN_int_E_tau1_tau2}, here $\tau_0 $ fixed at $\pi/4$ and $\tau_1$ varied as $\pi/8$, $\pi/4$, $3\pi/8$, and $\pi/2$.}
    \label{NN_nint_E_tau1_tau2}
\end{figure}

We first consider an asymmetric Floquet sequence by fixing $\tau_1=\pi/4$ and varying $\tau_0=\pi/8,\ \pi/4,\ 3\pi/8,$ and $\pi/2$. For the integrable charger under PBC [Fig.~\ref{NN_int_E_tau1_tau2}(a)], optimal energy storage is achieved exclusively at the commensurate point $\tau_0=\pi/4$, where $\Delta E_{\rm max}=2\omega N$. At $\tau_0=\pi/2$, however, the evolution suppresses charging entirely, leading to $\Delta E_{\rm max}=0$, consistent with destructive interference in the Floquet cycle. For intermediate values $\tau_0=\pi/8$ and $3\pi/8$, the stored energy remains below $\omega N$, and the charging curves exhibit symmetry about $\tau_0=\pi/4$, reflecting the invariance $\Delta E(\tau_0)=\Delta E(\pi/2-\tau_0)$. Under OBC [Fig.~\ref{NN_int_E_tau1_tau2}(b)], the integrable system displays an overall reduction in charging capacity. Even at $\tau_0=\pi/4$, the maximum stored energy decreases to $\Delta E_{\rm max}=\omega N$, while at $\tau_0=\pi/2$ it is further reduced to $\omega N/2$. For the remaining values of $\tau_0$, the stored energy remains below $\omega N$, underscoring the restrictive influence of open boundaries in the short-range integrable regime. Turning to the nonintegrable charger [Fig.~\ref{NN_int_E_tau1_tau2}(c,d)], the response becomes more robust against asymmetry. Under PBC, optimal storage is recovered at both $\tau_0=\pi/4$ and $\pi/2$, accompanied by periodic stroboscopic dynamics with periods scaling as $3N/2$ and $2$, respectively. For $\tau_0=3\pi/8$, the maximum stored energy exceeds $\omega N$, while for $\tau_0=\pi/8$ it remains below this threshold. Under OBC, the stored energy is nearly uniform across different $\tau_0$, except at $\tau_0=\pi/2$, where it increases to $\Delta E_{\rm max}=3\omega N/2$, indicating boundary-assisted enhancement in the presence of chaotic dynamics.

We next reverse the protocol, fixing $\tau_0=\pi/4$ and varying $\tau_1=\pi/8,\ \pi/4,\ 3\pi/8,$ and $\pi/2$. For the integrable charger under PBC [Fig.~\ref{NN_nint_E_tau1_tau2}(a)], optimal charging again occurs at $\tau_1=\pi/4$, yielding $\Delta E_{\rm max}=2\omega N$, with stroboscopic periodicity scaling linearly with system size. At $\tau_1=\pi/2$, the maximum stored energy decreases to $\omega N$. For $\tau_1=\pi/8$ and $3\pi/8$, the stored energy exceeds $\omega N$, and the symmetry under $\tau_1\rightarrow\pi/2-\tau_1$ is preserved. Under OBC [Fig.~\ref{NN_nint_E_tau1_tau2}(b)], the maximum stored energy remains bounded by $\omega N$ for all $\tau_1$, and the dynamics remain periodic at $\tau_1=\pi/4$ and $\pi/2$, with periods scaling as $2N$ and $N/2$, respectively. In the nonintegrable case [Fig.~\ref{NN_nint_E_tau1_tau2}(c,d)], optimal charging is obtained at $\tau_1=\pi/4$ for both PBC and OBC. For $\tau_1=\pi/8$, the stored energy exceeds $\omega N$, whereas for $\tau_1=3\pi/8$ and $\pi/2$ it reduces to $\omega N$. The stroboscopic periods depend sensitively on boundary geometry, increasing under OBC relative to PBC.

\subsection{Interaction-strength dependence and structural selectivity}

\begin{figure}[t]
    \centering
\includegraphics[width=0.32\linewidth,height=.22\linewidth]{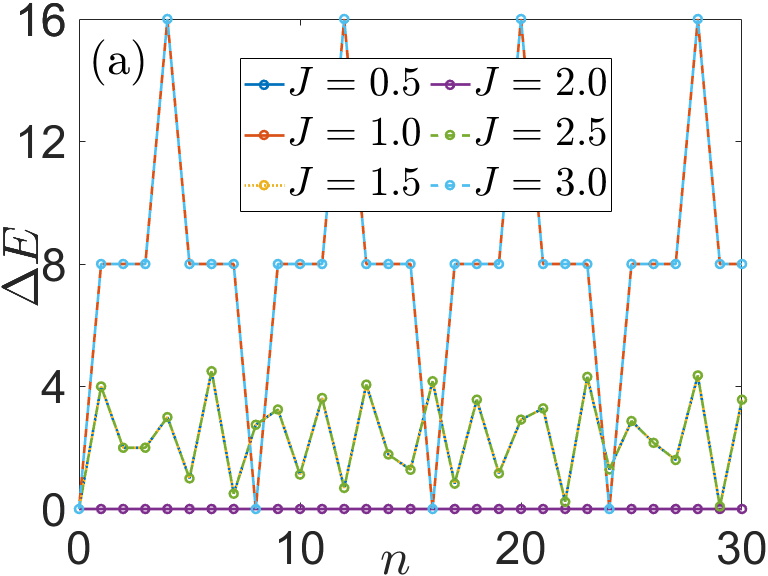}
\includegraphics[width=0.32\linewidth,height=.22\linewidth]{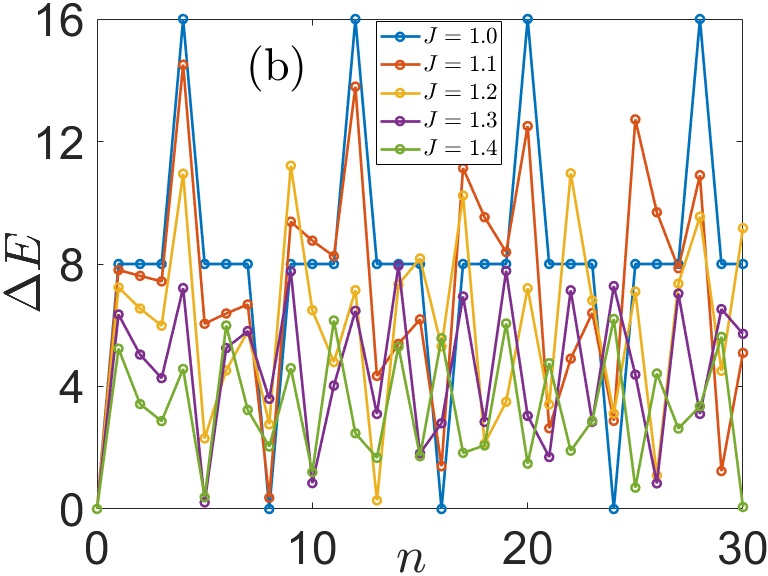}
\includegraphics[width=0.32\linewidth,height=.22\linewidth]{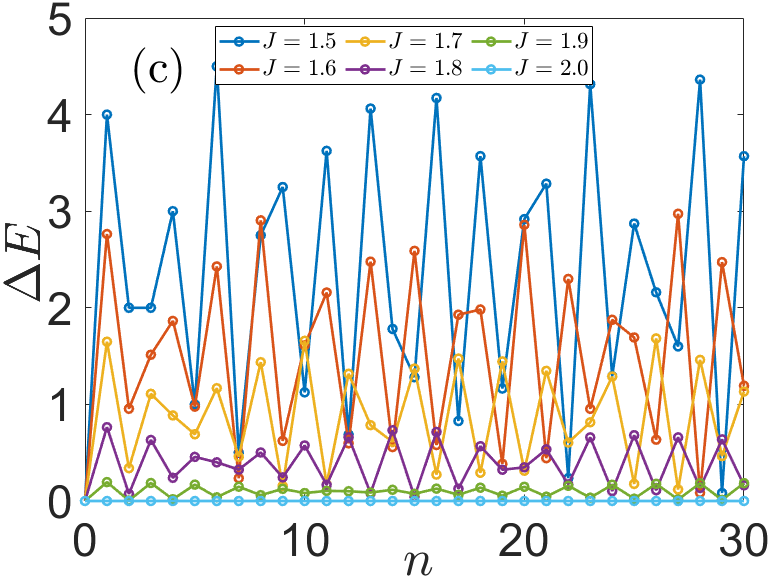}
\caption{Stored energy as a function of the number of kicks for an integrable NN charger under PBC. Panels (a)–(c) correspond to different interaction-strength ranges: (a) $J = 0.5\!-\!3.0$ in steps of $0.5$; (b) $J=1.0\!-\!1.4$ in steps of $0.1$; and (c) $J=1.5\!-\!2.0$ in steps of $0.1$. Unless stated otherwise, the parameters are $\tau_0=\tau_1=\pi/4$, $h_z=\omega=1$, $h_x=0$, and $N=8$.}
 \label{E_NN_int_p1_J}
\end{figure}

For completeness, we extend our analysis to the nearest-neighbor interacting charger in order to confirm that the conclusions obtained for long-range couplings are not intrinsically dependent on the interaction topology. When correlations are restricted to short distances, collective enhancement mechanisms are suppressed, and it becomes essential to determine whether varying the interaction scale can still qualitatively influence the resonant charging behavior. This comparison allows us to assess whether the robustness of the maximum stored energy, established in the long-range case, persists when interactions are purely local, or whether short-range correlations instead generate a more selective and structurally sensitive response.

To address this question, we systematically investigate the dependence of the charging performance on the nearest-neighbor interaction strength $J$ for both integrable and nonintegrable chargers under PBC and OBC. The symmetric driving protocol is fixed at $\tau_0=\tau_1\equiv\tau=\pi/4$, corresponding to the resonant regime previously identified for the short-range model, thereby ensuring a consistent framework for comparison. This complementary study isolates the role of interaction range and clarifies whether the observed charging characteristics arise from generic features of resonance, integrability, and boundary effects, or whether they are fundamentally linked to the presence of long-range couplings.

We first consider the integrable charger under PBC, as shown in Fig.~\ref{E_NN_int_p1_J}. In sharp contrast to the long-range case, where charging performance remains largely insensitive to variations of $J$, the nearest-neighbor integrable system exhibits a pronounced parity dependence on the coupling strength. For odd integer values of $J$, the charger reaches the optimal stored energy during its evolution. Conversely, for even integer values of $J$, energy storage is completely suppressed, with $\Delta E=0$ at all times. For noninteger values of $J$, the maximum stored energy assumes intermediate values, interpolating continuously between the optimal odd-$J$ regime and the zero-storage even-$J$ regime.

A central observation is that the charging dynamics depends only on the fractional part of $J$, leading to a periodic dependence of the stored energy on the interaction strength, $E(J)=E(J+m)$ for any integer $m$. Consequently, values such as $J=0.5$, $1.5$, and $2.5$ produce identical charging behavior, as illustrated in Fig.~\ref{E_NN_int_p1_J}(a). This periodicity reflects the underlying Floquet commensurability and highlights that increasing the absolute magnitude of $J$ does not necessarily enhance performance.

To further clarify this structural selectivity, we vary $J$ continuously from $1$ to $2$ in increments of $0.1$, thereby interpolating between an odd and an even integer value. As shown in Fig.~\ref{E_NN_int_p1_J}(b,c), the maximum stored energy decreases monotonically as $J$ approaches the even-integer point, eventually vanishing near $J=2$. Thus, rather than exhibiting monotonic enhancement with increasing interaction strength, the system transitions between resonant and nonresonant charging regimes determined by discrete structural compatibility conditions.

The integrable charger under OBC, shown in Fig.~\ref{E_NN_int_p0_J}, exhibits a distinct interaction-strength dependence compared to the PBC case. For odd integer values of the coupling strength $J$, the maximum stored energy reaches half of the optimal value, $\Delta E_{\rm max}=\omega N$. In contrast, for even integer values of $J$, the stored energy is further reduced to one-fourth of the optimal value, $\Delta E_{\rm max}=\omega N/2$. This behavior differs qualitatively from the integrable system under PBC, where even integer couplings completely suppress energy storage. For noninteger values of $J$, the stored energy assumes intermediate values between these limiting cases, reflecting a continuous crossover between distinct commensurate regimes.

To characterize this crossover more precisely, we vary the coupling strength continuously from $J=1$ to $J=2$ in increments of $0.1$. As shown in Fig.~\ref{E_NN_int_p0_J}(b), the maximum stored energy remains above half of the optimal value as $J$ increases from $1$ to approximately $1.4$. For larger values, $1.5\le J<2$, the stored energy decreases progressively, approaching $\omega N/2$ as $J\rightarrow2$, as illustrated in Fig.~\ref{E_NN_int_p0_J}(c). Thus, unlike the sharp suppression observed under PBC, the OBC case displays a smoother reduction of charging performance as the coupling strength approaches an even integer value, further emphasizing the role of boundary structure in shaping interaction-induced selectivity.

\begin{figure}[t]
    \centering
\includegraphics[width=0.32\linewidth,height=.22\linewidth]{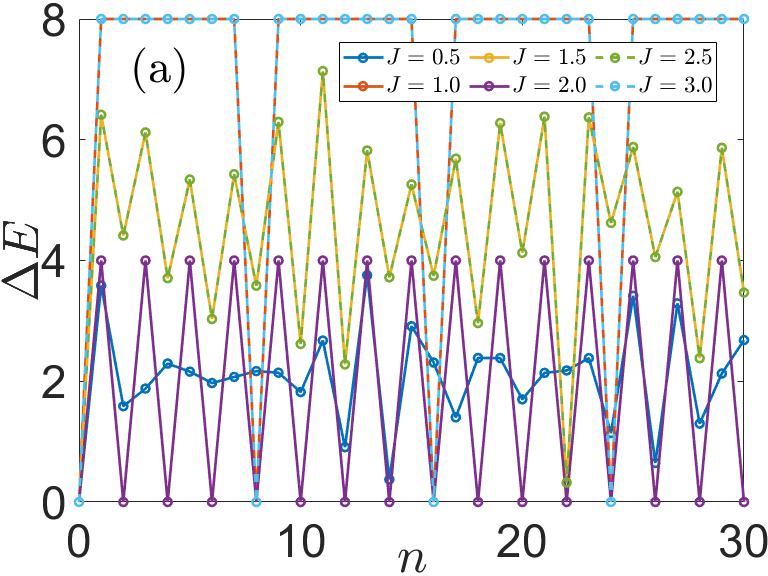}
\includegraphics[width=0.32\linewidth,height=.22\linewidth]{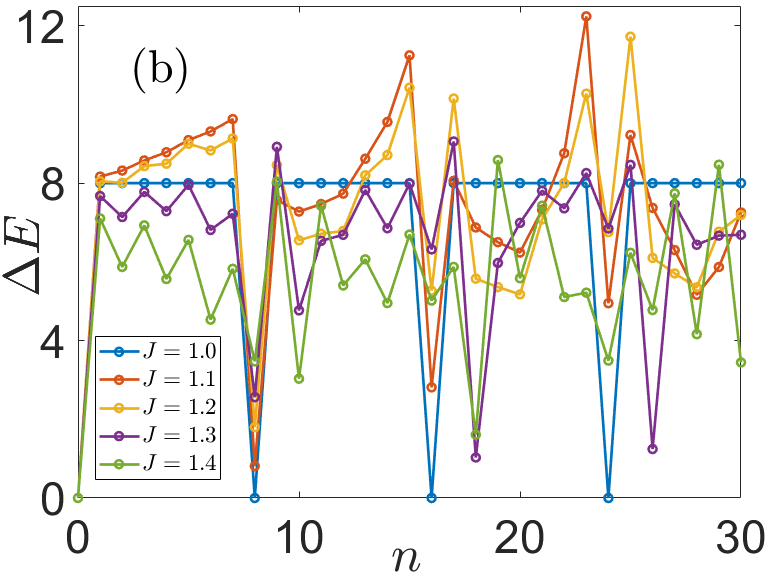}
\includegraphics[width=0.32\linewidth,height=.22\linewidth]{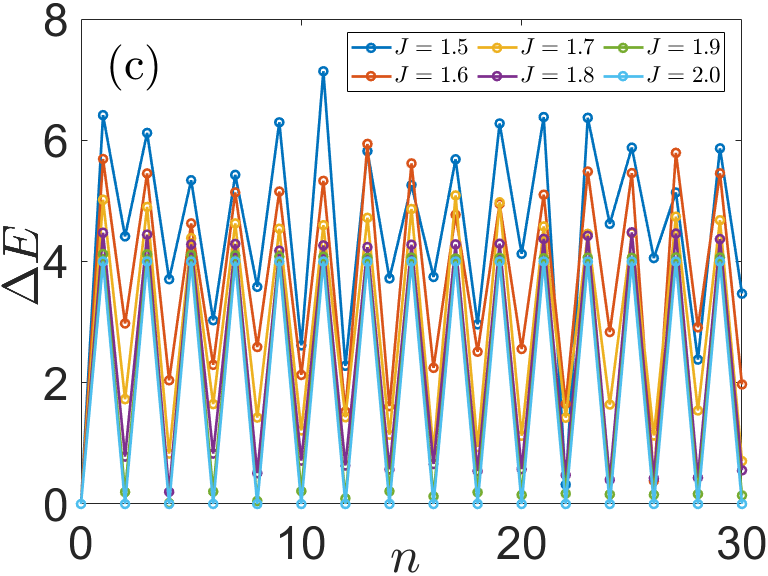}
\caption{Same as Fig.~\ref{E_NN_int_p1_J}, but for OBC.}
 \label{E_NN_int_p0_J}
\end{figure}

The nonintegrable charger under PBC, presented in Fig.~\ref{E_NN_nint_p1_J}, displays a qualitatively different interaction-strength dependence compared to the integrable case. For odd integer values of the coupling strength $J$, the system attains the optimal stored energy during its evolution. For even integer values, however, the maximum stored energy is reduced to half of the optimal value, $\Delta E_{\rm max}=\omega N$. In contrast to the integrable PBC case, where even couplings completely suppress storage, the nonintegrable dynamics retains a finite and substantial charging capacity. For noninteger values of $J$, the stored energy remains greater than or equal to half of the optimal value, as shown in Fig.~\ref{E_NN_nint_p1_J}(a), indicating enhanced robustness against variations in interaction strength.

To further resolve this crossover, we vary $J$ continuously from $1$ to $2$ in steps of $0.1$. As $J$ increases from $1$ to approximately $1.5$, the maximum stored energy decreases gradually from the optimal value but remains above $\omega N$, i.e., above half of the optimal storage. For $1.5 \le J \le 2$, the stored energy continues to decrease and eventually approaches $\Delta E_{\rm max}=\omega N$ at $J=2$, as illustrated in Fig.~\ref{E_NN_nint_p1_J}(b,c). Thus, unlike the sharp parity-induced suppression observed in the integrable regime, the nonintegrable PBC charger exhibits a smoother and more robust response to changes in the interaction strength.

\begin{figure}[b]
    \centering
\includegraphics[width=0.32\linewidth,height=.22\linewidth]{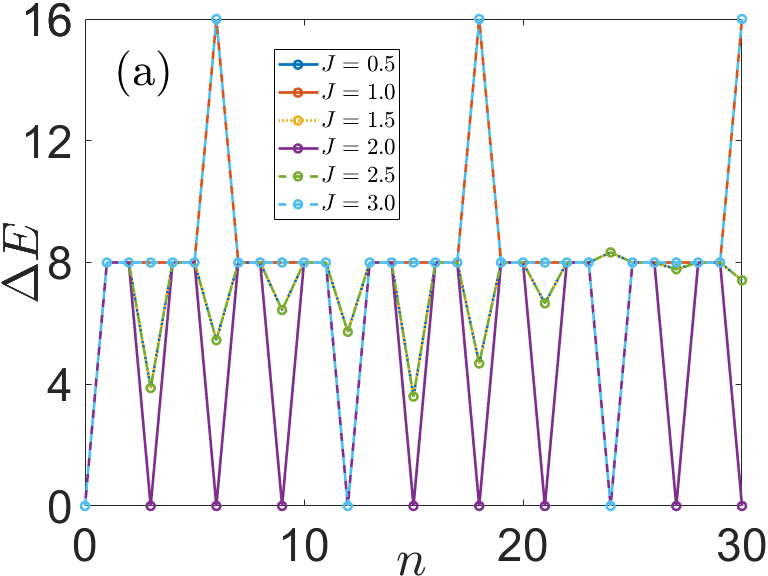}
\includegraphics[width=0.32\linewidth,height=.22\linewidth]{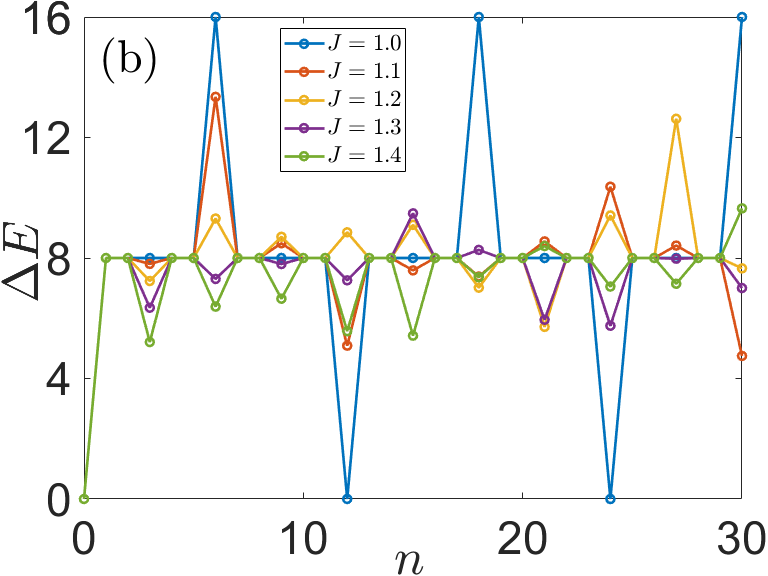}
\includegraphics[width=0.32\linewidth,height=.22\linewidth]{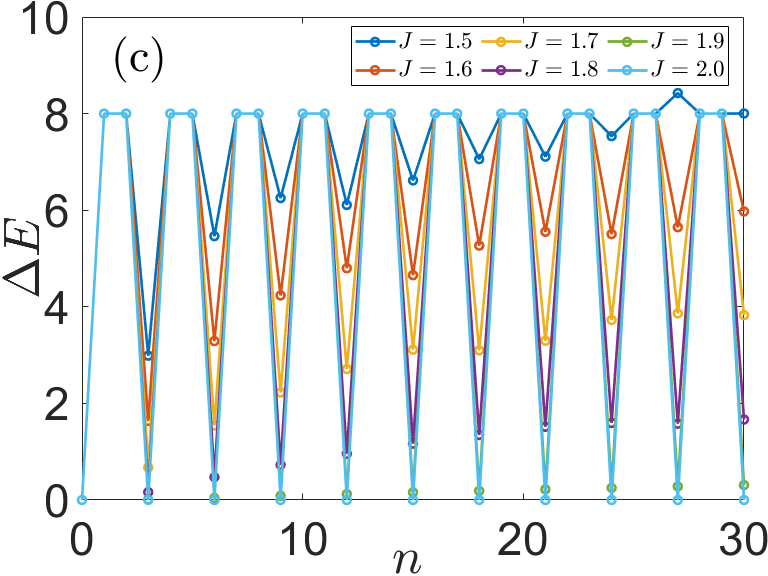}
    \caption{Same as Fig.~\ref{E_NN_int_p1_J}, but for the nonintegrable charger with $h_x = 1$.}
 \label{E_NN_nint_p1_J}
\end{figure}

The charging behavior for the nonintegrable charger under OBC, shown in Fig.~\ref{E_NN_nint_p0_J}, is also examined. We find that the dependence of the stored energy on the coupling strength closely mirrors that observed under PBC. In particular, the overall magnitude and variation of the maximum stored energy follow a quantitatively similar trend, indicating that, in the nonintegrable regime, boundary geometry plays a less restrictive role compared to the integrable case.

\begin{figure}[t]
    \centering
\includegraphics[width=0.32\linewidth,height=.22\linewidth]{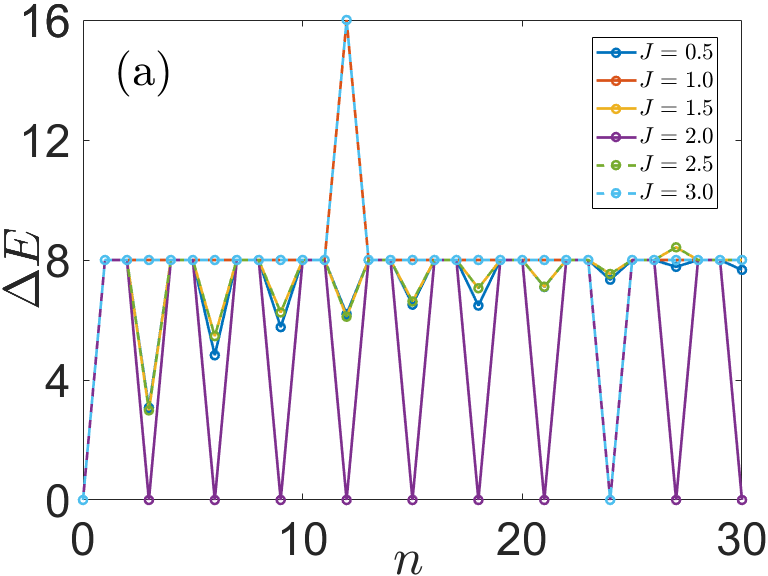}
\includegraphics[width=0.32\linewidth,height=.22\linewidth]{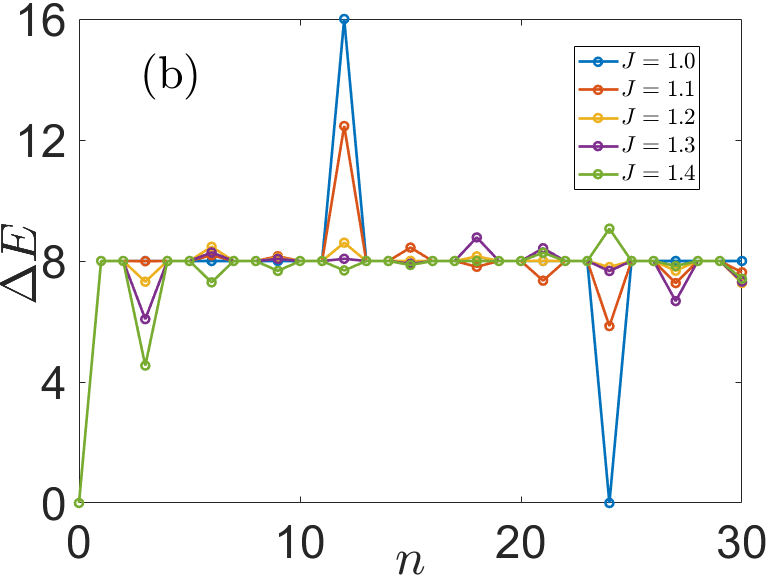}
\includegraphics[width=0.32\linewidth,height=.22\linewidth]{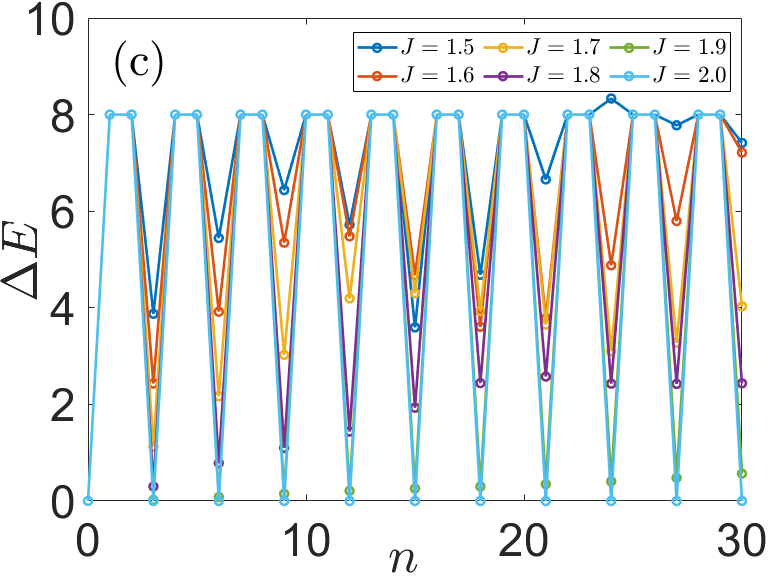}
    \caption{Same as Fig.~\ref{E_NN_int_p0_J}, but for the nonintegrable charger with $h_x=1$.}
 \label{E_NN_nint_p0_J}
\end{figure}

\subsection{Maximum stored energy: driving-period and system-size dependence}

We now investigate how the maximum stored energy of the nearest-neighbor interacting charger depends on the symmetric driving period $\tau_0=\tau_1\equiv\tau$. All structural configurations are considered, including integrable and nonintegrable dynamics under PBC and OBC, as shown in Fig.~\ref{Emax_NN_tau}(a). The driving period is varied within $\tau\in[0,\pi/2]$ in increments of $\pi/32$, while the system size is fixed at $N=8$. For each value of $\tau$, the maximum stored energy $\Delta E_{\rm max}$ is determined by optimizing the stroboscopic evolution over $n=0$ to $500$ kicks.

For all cases, the trivial limit $\tau=0$ corresponds to identity evolution, and therefore no energy is stored, $\Delta E_{\rm max}=0$. As $\tau$ increases to $\pi/32$, a finite amount of energy is accumulated, with $\Delta E_{\rm max}$ lying between one-quarter and one-half of the optimal value. Upon further increasing $\tau$, the stored energy grows steadily and, at $\tau=\pi/4$, reaches the optimal value in all configurations except for the integrable charger under OBC. In this exceptional case, the maximum stored energy remains restricted to half of the optimal value, reflecting boundary-induced constraints in the integrable short-range regime. 

Beyond $\tau=\pi/4$, the charging response becomes strongly dependent on the underlying structural properties. For the integrable charger under OBC, $\Delta E_{\rm max}$ decreases monotonically as $\tau$ increases and eventually vanishes at $\tau=\pi/2$, indicating complete suppression of charging at this commensurate point. In contrast, the integrable system under PBC exhibits a nonmonotonic behavior: $\Delta E_{\rm max}$ increases slightly past $\pi/4$, reaching a local maximum near $\tau=9\pi/32$, and then decreases toward $\Delta E_{\rm max}=3\omega N/4$ at $\tau=\pi/2$. This case also displays symmetry around $\tau=\pi/4$, consistent with Floquet commensurability. For the nonintegrable charger, both PBC and OBC show an initial reduction of $\Delta E_{\rm max}$ beyond $\pi/4$, followed by a recovery at larger $\tau$. Under PBC, the optimal value is restored at $\tau=\pi/2$, whereas under OBC the maximum stored energy reaches $3\omega N/4$, as illustrated in Fig.~\ref{Emax_NN_tau}(a).

We next analyze the scaling of $\Delta E_{\rm max}$ with system size for two representative driving periods, $\tau=\pi/4$ [Fig.~\ref{Emax_NN_tau}(b)] and $\tau=\pi/2$ [Fig.~\ref{Emax_NN_tau}(c)]. At $\tau=\pi/4$, the integrable charger under PBC exhibits a pronounced parity dependence: for odd $N$, the maximum stored energy scales as $\Delta E_{\rm max}=\omega N$, whereas even system sizes achieve the optimal scaling $\Delta E_{\rm max}=2\omega N$. By contrast, the integrable charger under OBC and the nonintegrable charger under PBC both display linear scaling with $N$, consistently attaining $\Delta E_{\rm max}=\omega N$, i.e., half of the optimal value. A distinct behavior arises for the nonintegrable charger under OBC. In this case, the scaling is nonmonotonic: small system sizes yield $\Delta E_{\rm max}=\omega N$, the optimal value $\Delta E_{\rm max}=2\omega N$ is reached at $N=8$, and for larger sizes ($N>8$) the stored energy initially drops to $\omega N$ before recovering a linear growth trend with increasing $N$.

At the longer driving period $\tau=\pi/2$, the structural contrast becomes even more pronounced. The integrable charger under PBC exhibits complete quenching of energy storage, with $\Delta E_{\rm max}=0$ for all system sizes. Under OBC, however, the integrable system maintains a finite but size-independent stored energy, $\Delta E_{\rm max}=4\omega$. In sharp contrast, the nonintegrable charger under both boundary conditions displays extensive scaling with system size. Specifically, $\Delta E_{\rm max}=2\omega N$ under PBC and $\Delta E_{\rm max}=3\omega N/4$ under OBC. 

These results clearly demonstrate that, in the nearest-neighbor regime, the charging capacity is highly sensitive to the combined effects of integrability, boundary geometry, and system-size parity. The emergence of suppression, parity selection, and nonmonotonic scaling behaviors highlights the strong \emph{structural effects} governing periodically driven Ising quantum batteries with short-range interactions.

  \begin{figure}[H]
\includegraphics[width=0.32\linewidth,height=0.22\linewidth]{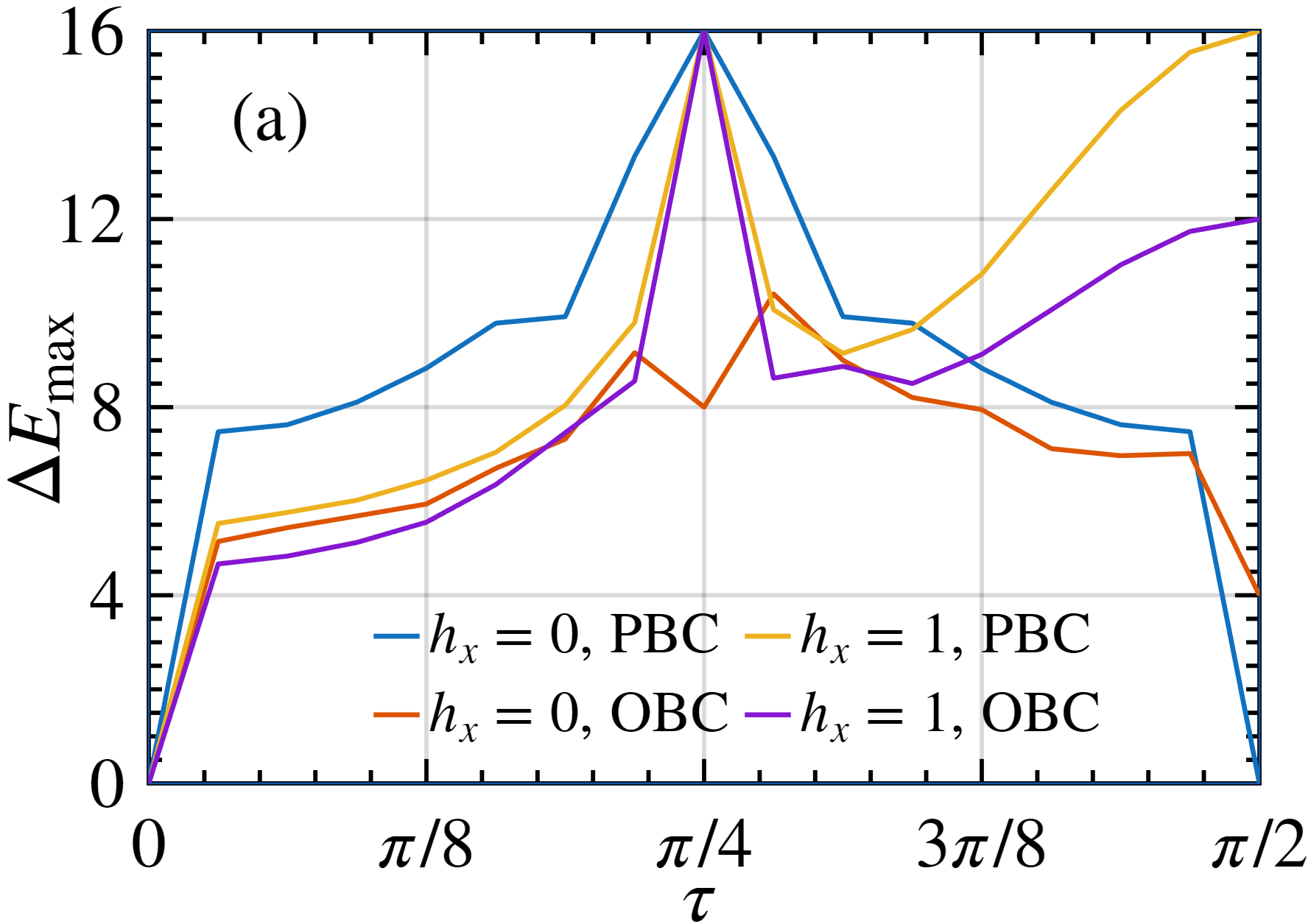}
\includegraphics[width=0.32\linewidth,height=0.22\linewidth]{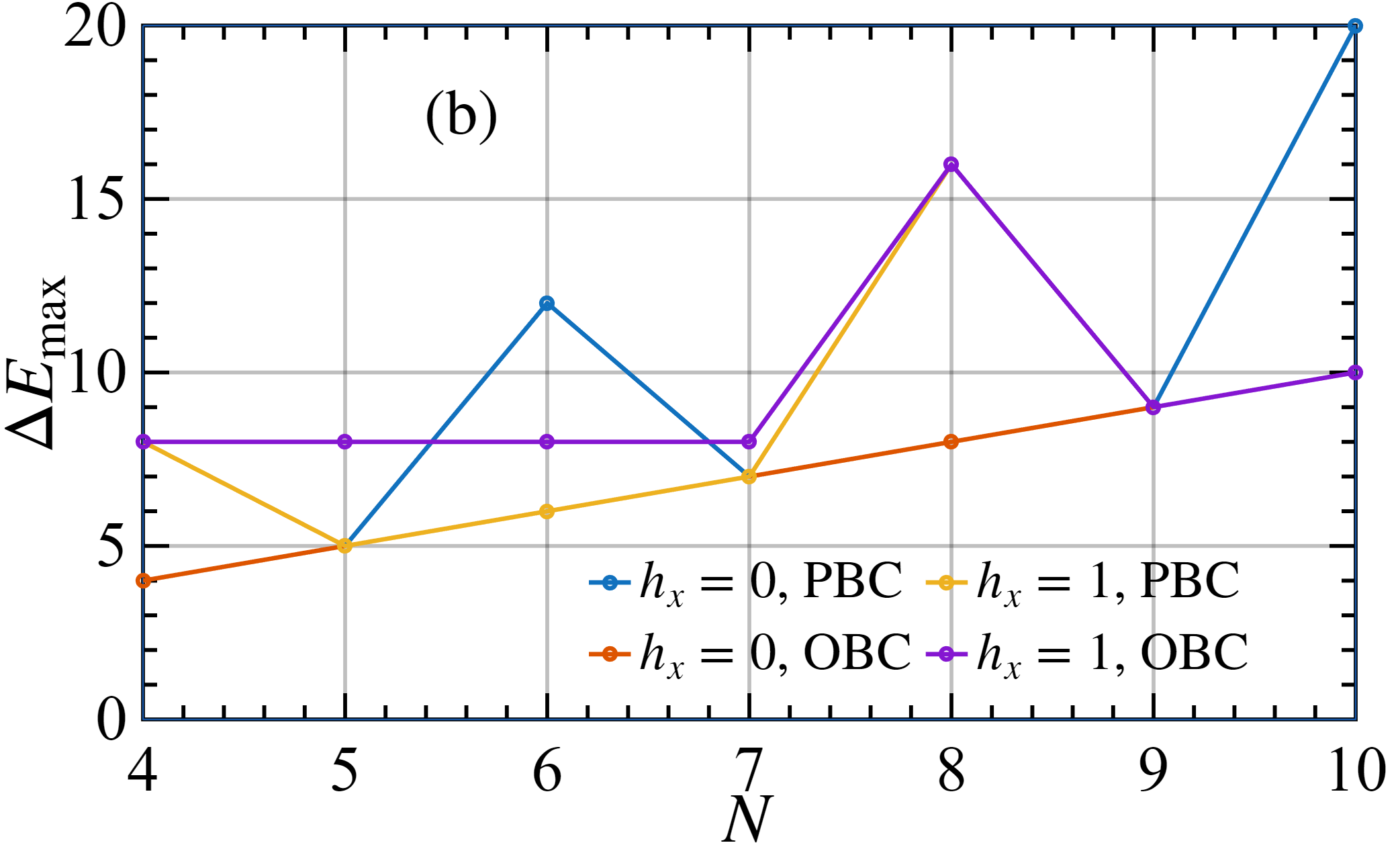}
\includegraphics[width=0.32\linewidth,height=0.22\linewidth]{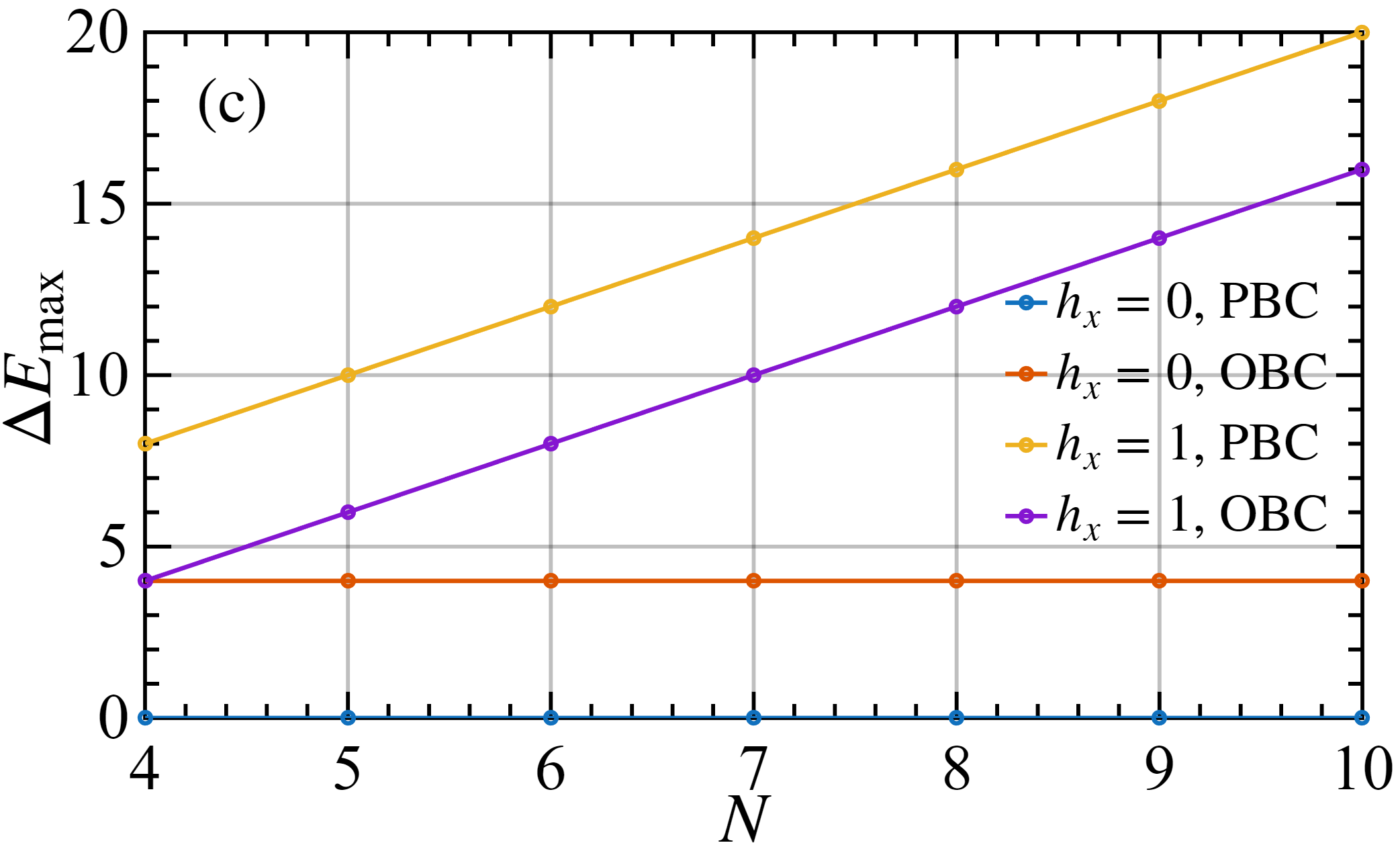}
    \caption{(a) Maximum stored energy as a function of the Floquet period $\tau_0=\tau_1=\tau$ for Nearest neighbour interacting charger with system size $N=8$.
The parameter $h_x=0$ ($h_x\neq 0$) corresponds to the absence (presence) of a longitudinal field, for which the system is integrable (nonintegrable).
(b,c) Maximum stored energy as a function of system size $N$ for fixed driving periods
(b) $\tau_0=\tau_1=\pi/4$ and
(c) $\tau_0=\tau_1=\pi/2$, for all cases considered.
}

    \label{Emax_NN_tau}
\end{figure}

\section{Charging Power}
\label{Power_S}

We now turn to the charging power of the noninteracting quantum battery subjected to a time-periodically driven interacting Ising charger. Our objective is to clarify how key structural ingredients—interaction range, boundary geometry, integrability, and system size govern the rate of energy transfer. Throughout this section, we systematically compare OBC and PBC, as well as integrable and nonintegrable driving regimes, in order to identify the dominant mechanisms controlling power enhancement or suppression.

\subsection{Long-range interacting charger}

We begin with the long-range interacting charger at fixed system size $N=8$, examining both integrable and nonintegrable dynamics under OBC and PBC. By varying the symmetric driving period $\tau$, we observe that the charging power decreases monotonically as $\tau$ increases, irrespective of boundary condition or dynamical regime. As shown in Figs.~\ref{long_E_tau}, shorter driving periods consistently yield higher charging power, indicating that rapid Floquet modulation enhances the rate of energy transfer. This trend demonstrates that, in the long-range case, the dominant control of charging power arises from the temporal structure of the drive rather than from integrability or boundary-induced structural differences.

\begin{figure}[b] 
\centering 
\includegraphics[width=0.246\linewidth,height=0.20\linewidth]{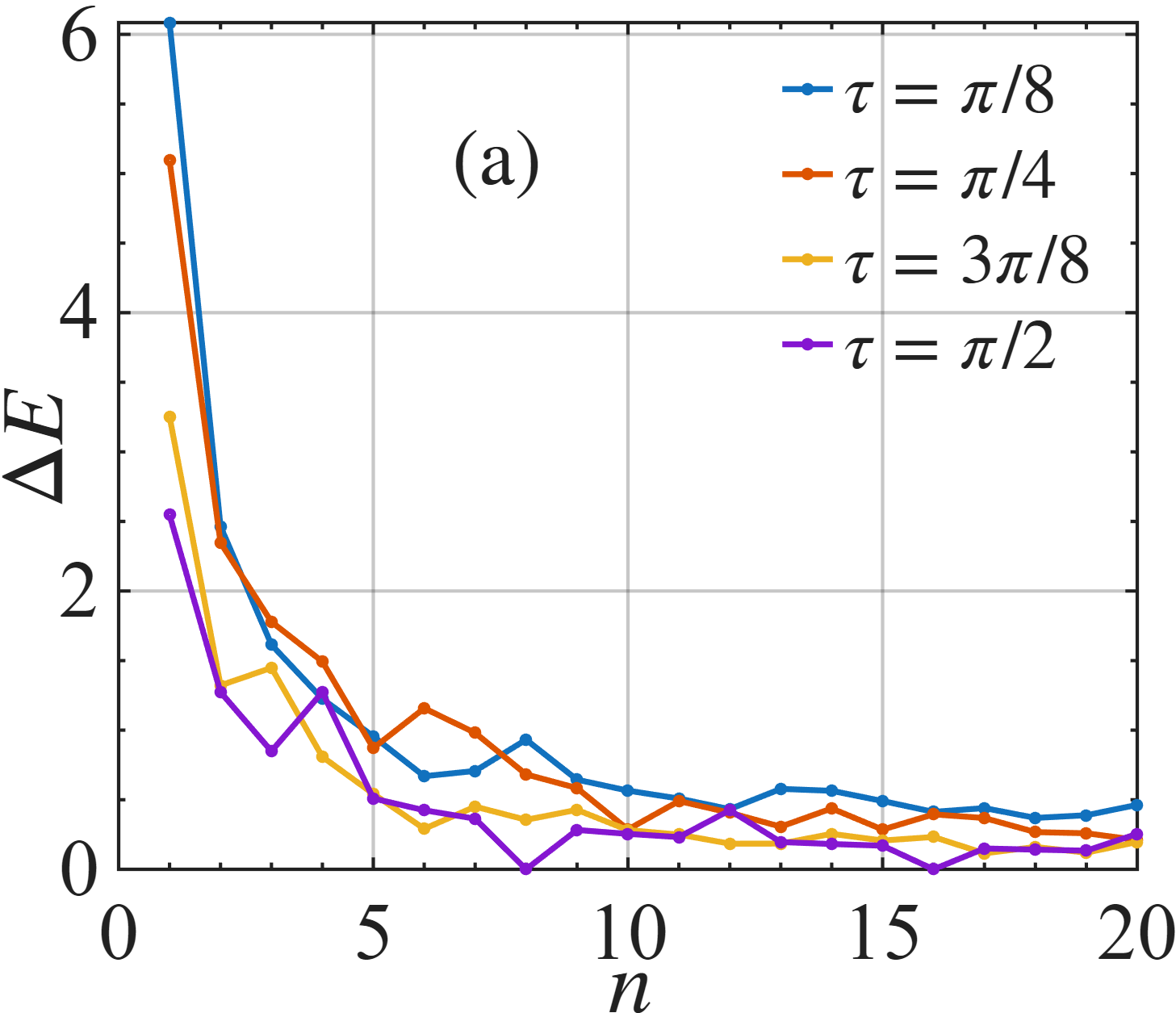}
\includegraphics[width=0.246\linewidth,height=0.20\linewidth]{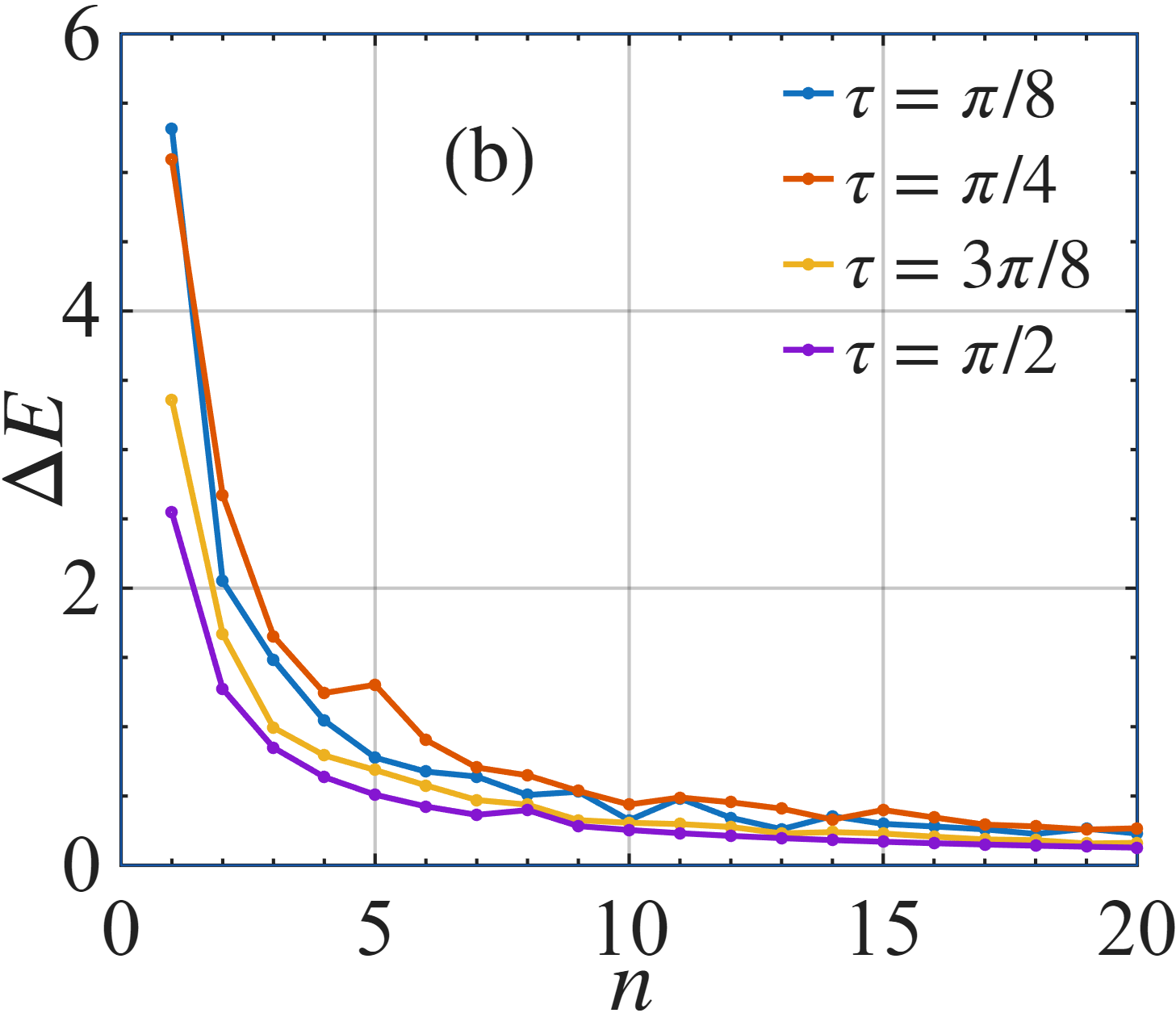}
\includegraphics[width=0.246\linewidth,height=0.20\linewidth]{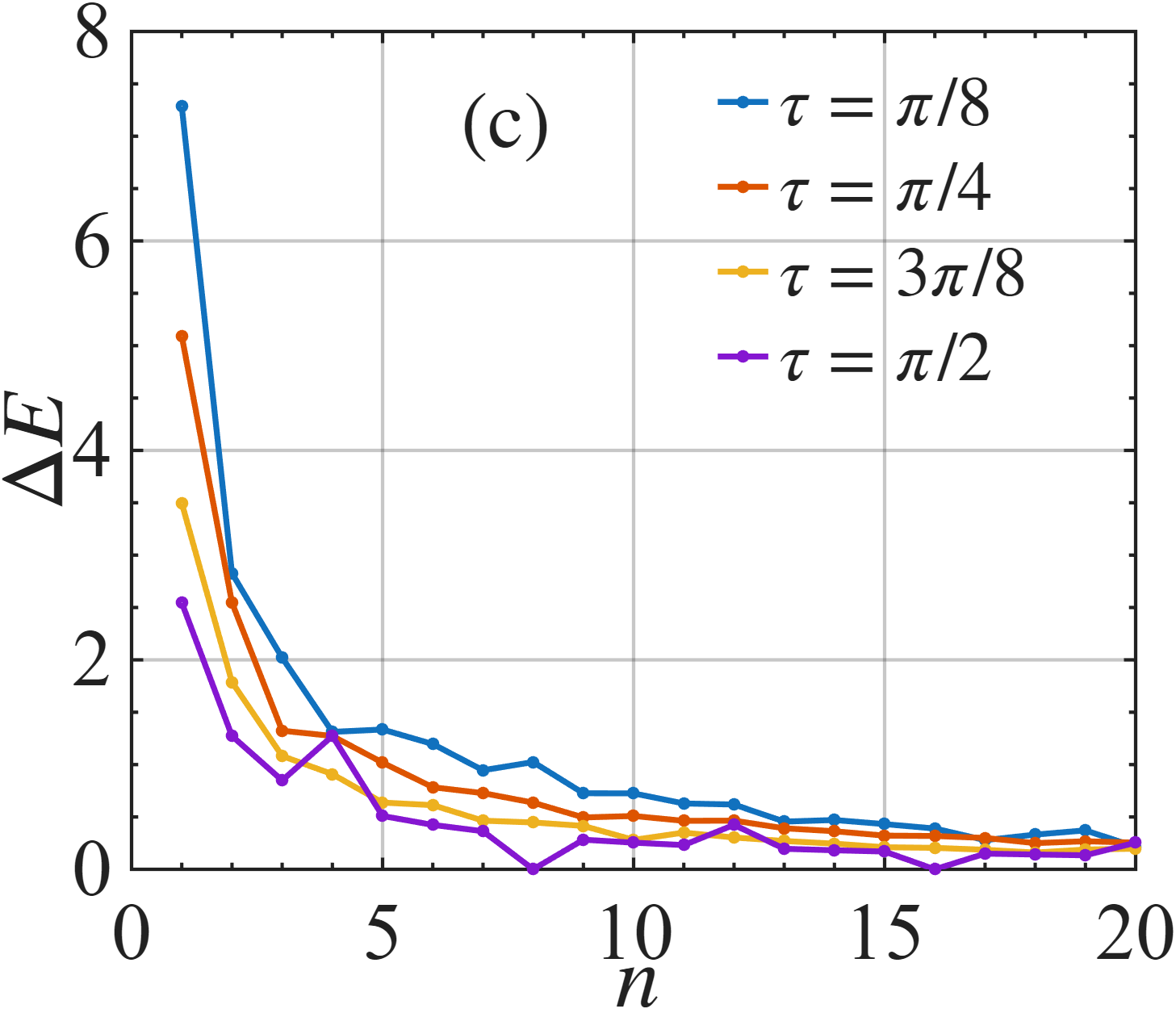}
\includegraphics[width=0.246\linewidth,height=0.20\linewidth]{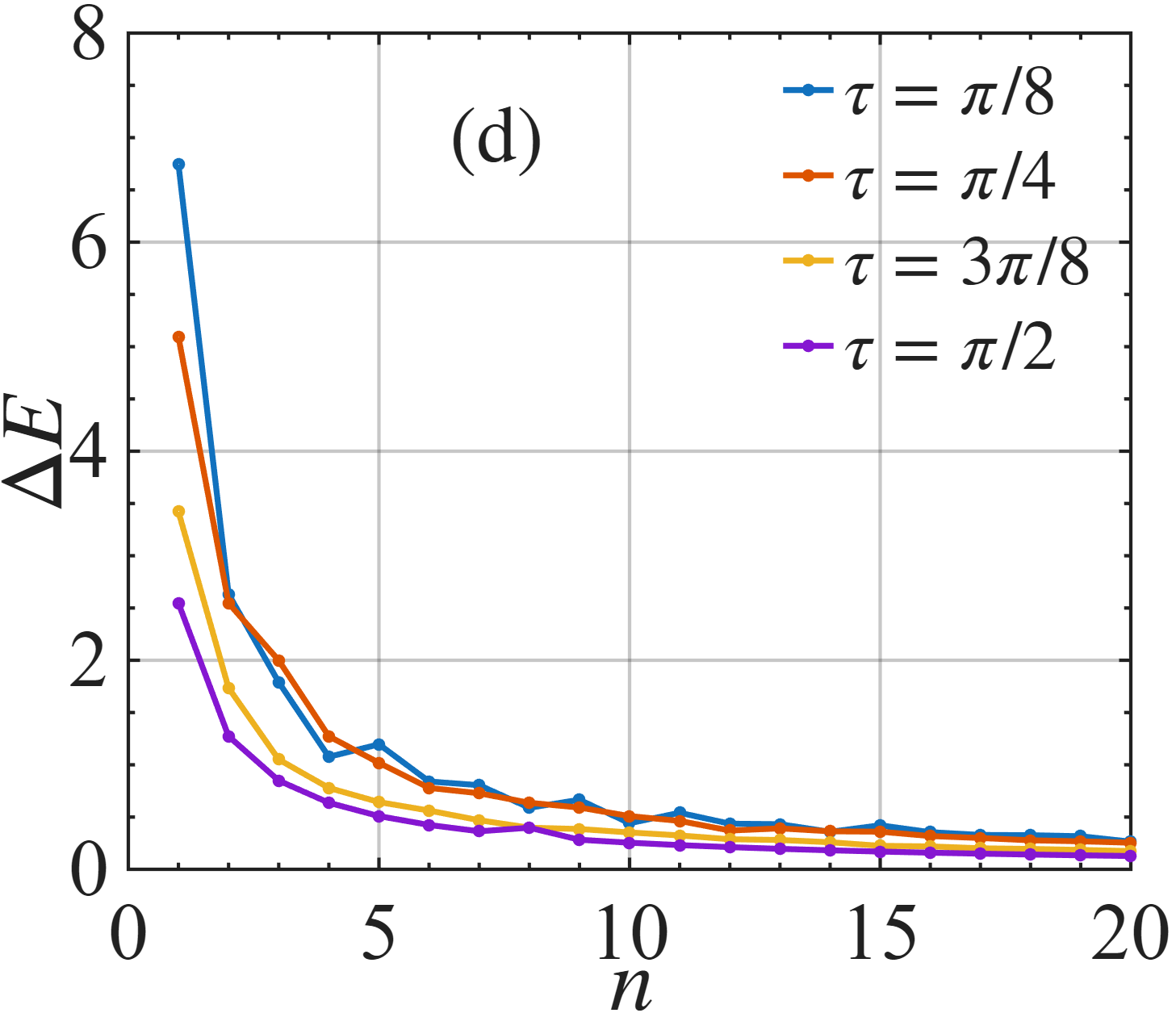}
\caption{Charging power of a noninteracting quantum battery driven by a time-periodic Long Range Ising charger for different driving periods $\tau_0=\tau_1\equiv\tau$, as indicated in the legend. Panels (a,b) correspond to the integrable charger, while panels (c,d) show the nonintegrable case. PBC are used in panels (a,c), and OBC in panels (b,d). Parameters are $N=8$, $J=1$, $h_x=0$ (integrable) or $h_x=1$ (nonintegrable), $h_z=1$, and $\omega=1$.} 
\label{long_E_tau} 
\end{figure}

To further clarify the impact of system size, we fix the symmetric driving protocol to $\tau_0=\tau_1=\pi/4$ and analyze the scaling of the maximum charging power with $N$. For all long-range configurations, independent of boundary geometry or integrability, the peak charging power exhibits an approximately linear growth with system size, as shown in Figs.~\ref{E_ATA_N}. This extensive scaling reflects the collective contribution of long-range interactions to the energy-transfer process and indicates that, in this regime, the charging protocol remains structurally robust across different dynamical and boundary settings.

\begin{figure}[t]
\centering 
\includegraphics[width=0.246\linewidth,height=0.18\linewidth]{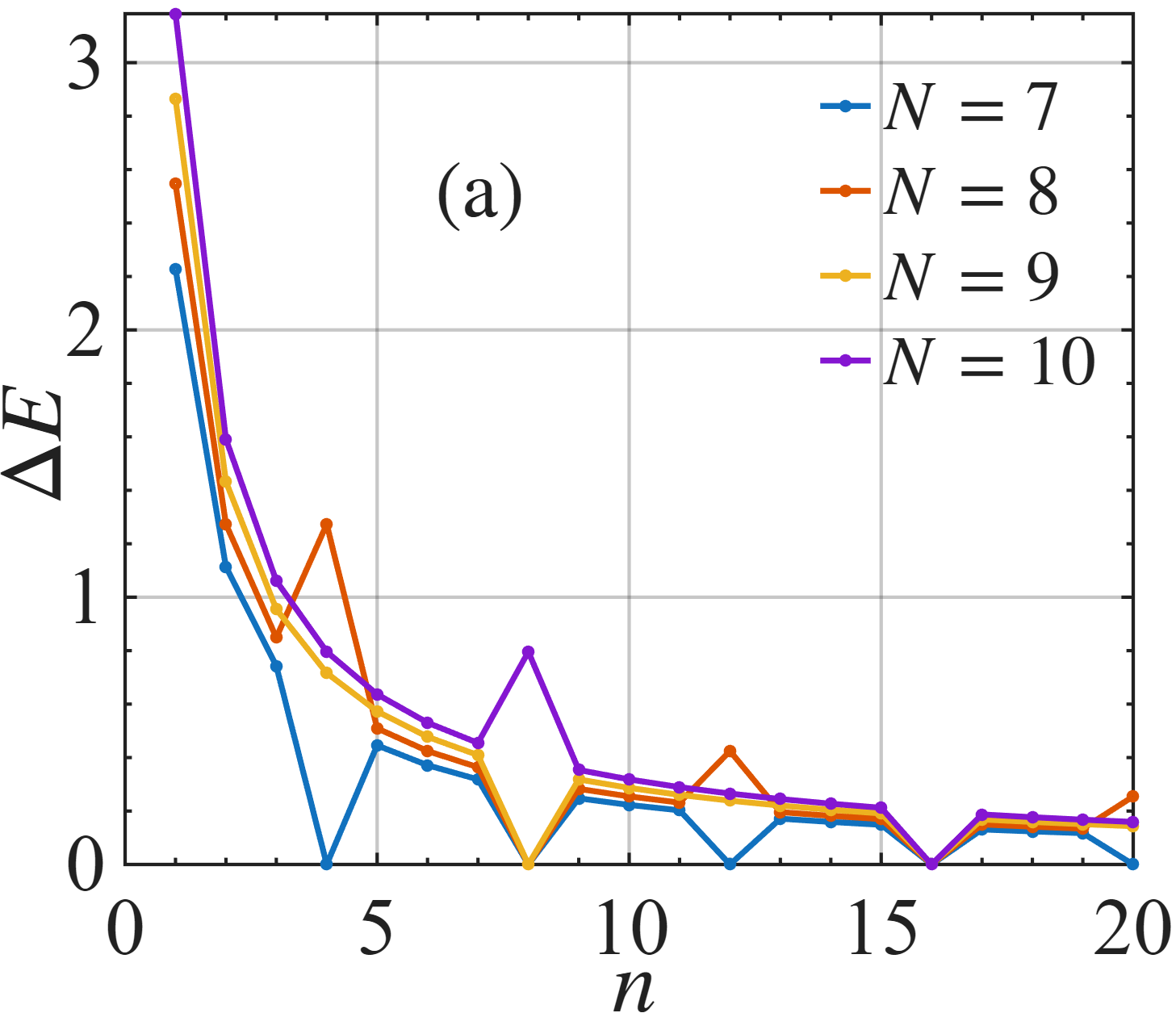}
\includegraphics[width=0.246\linewidth,height=0.18\linewidth]{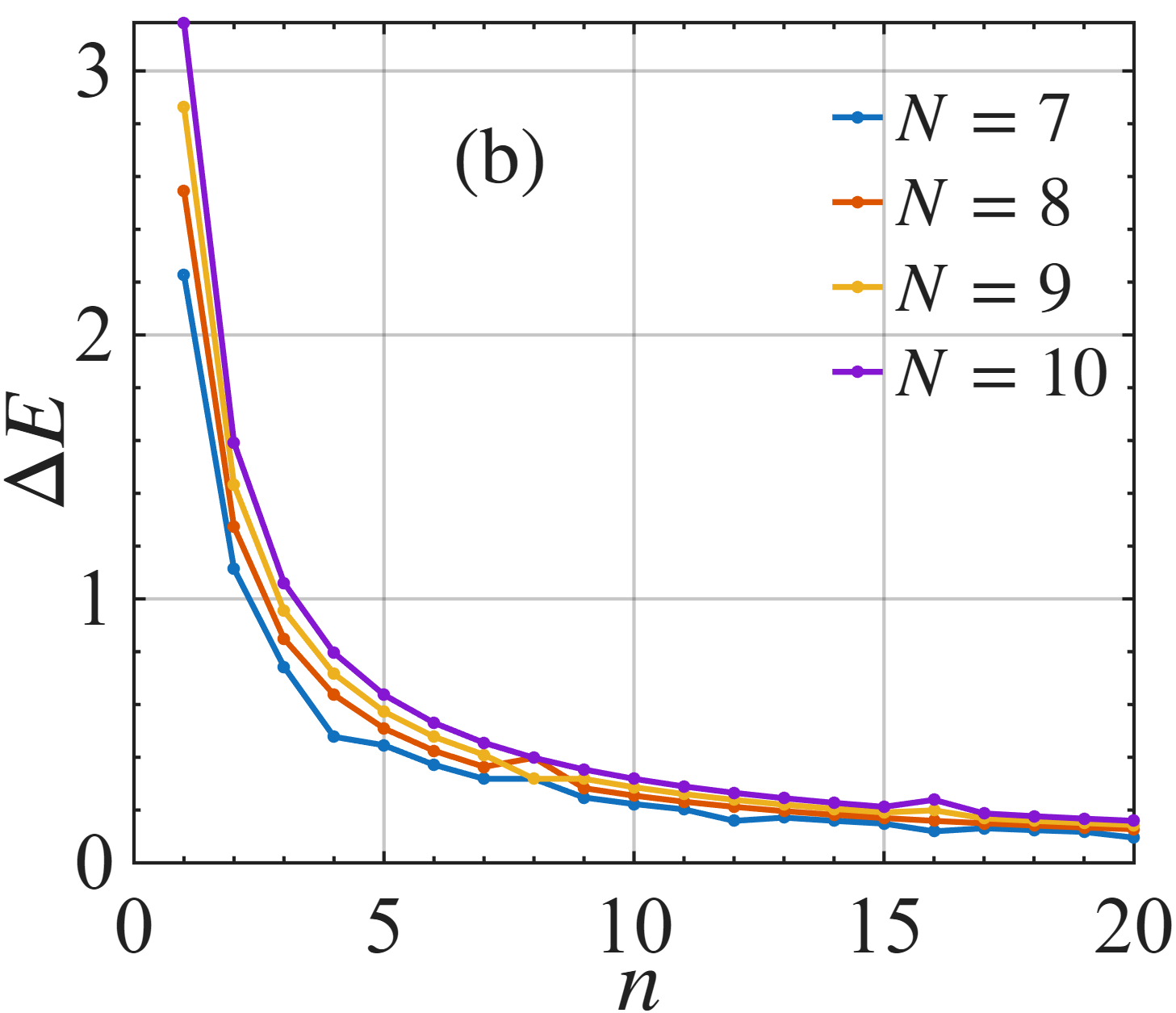}
\includegraphics[width=0.246\linewidth,height=0.18\linewidth]{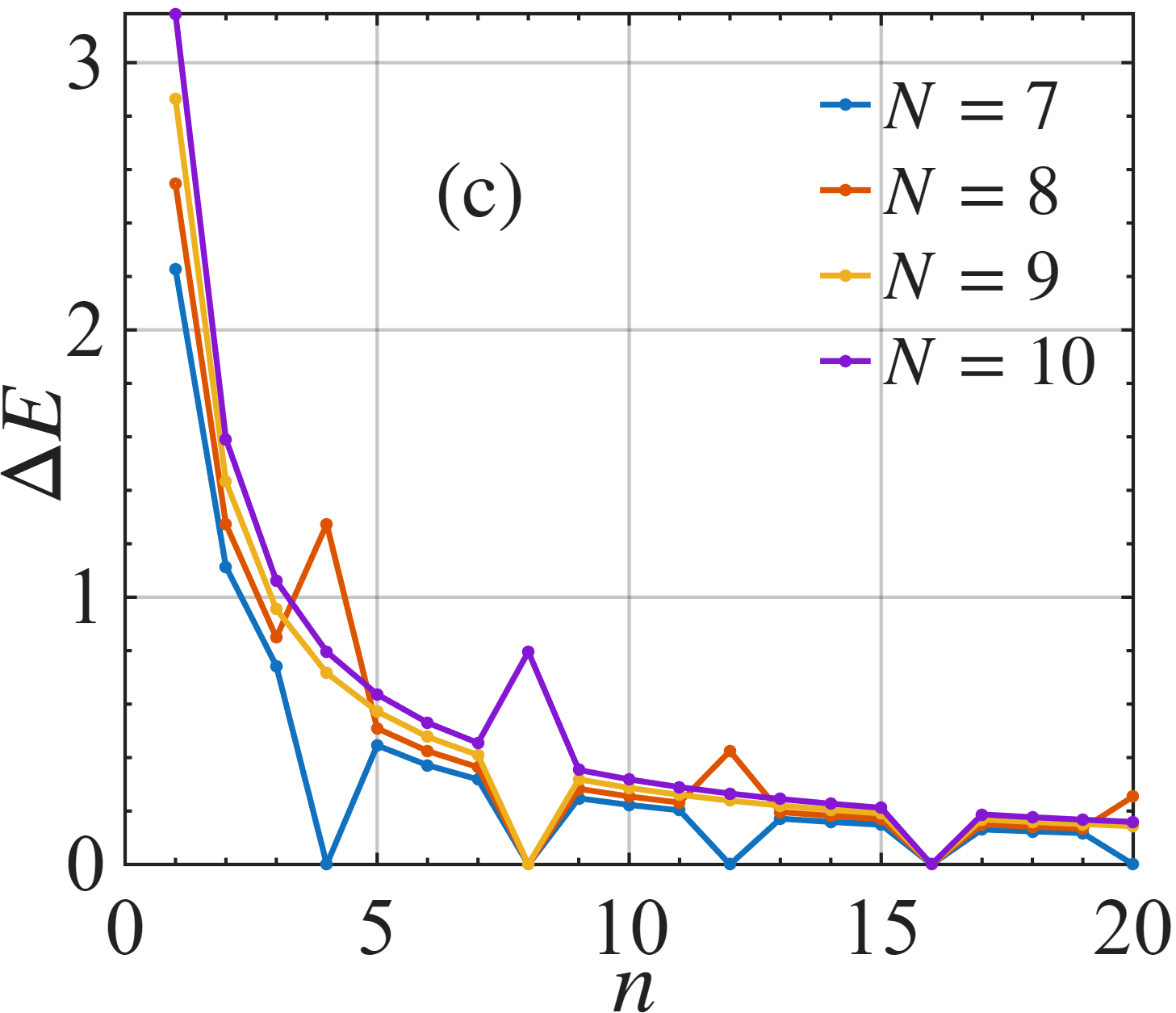}
\includegraphics[width=0.246\linewidth,height=0.18\linewidth]{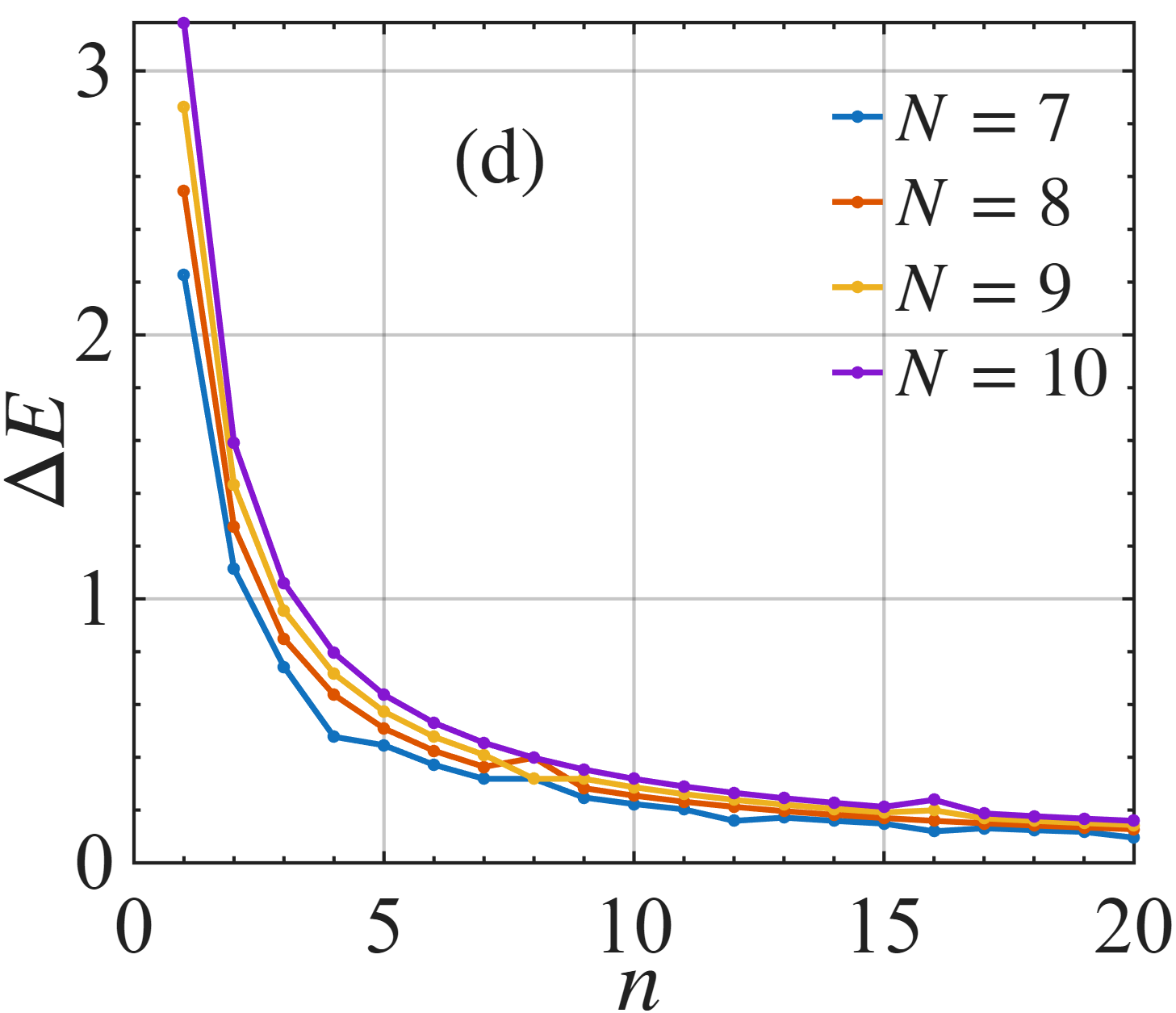}
\caption{Charging power of a noninteracting quantum battery driven by a time-periodic long-range Ising charger for different system size $N$, as indicated in the legend. Panels (a,b) correspond to the integrable charger, while panels (c,d) show the nonintegrable case. PBC are used in panels (a,c), and OBC in panels (b,d). Parameters are $\tau_0=\tau_1=\pi/2$, $J=1$, $h_x=0$ (integrable) or $h_x=1$ (nonintegrable), $h_z=1$, and $\omega=1$.}
\label{E_ATA_N}
\end{figure}
\subsection{Nearest-neighbor interacting charger}

We now examine the charging power generated by a nearest-neighbor interacting charger. Keeping the system size fixed at $N=8$, we vary the symmetric driving period $\tau$ and evaluate the resulting power under both OBC and PBC, for integrable and nonintegrable dynamics. As in the long-range case, the charging power decreases monotonically with increasing $\tau$ across all structural configurations, as illustrated in Figs.~\ref{NN_E_tau}. This consistent trend indicates that the reduction of charging power at longer driving periods is a generic consequence of the Floquet protocol itself, rather than a feature tied to interaction range, boundary geometry, or integrability.
\vspace{0.25cm}
\begin{figure}[H]
    \centering
\includegraphics[width=0.246\linewidth,height=0.18\linewidth]{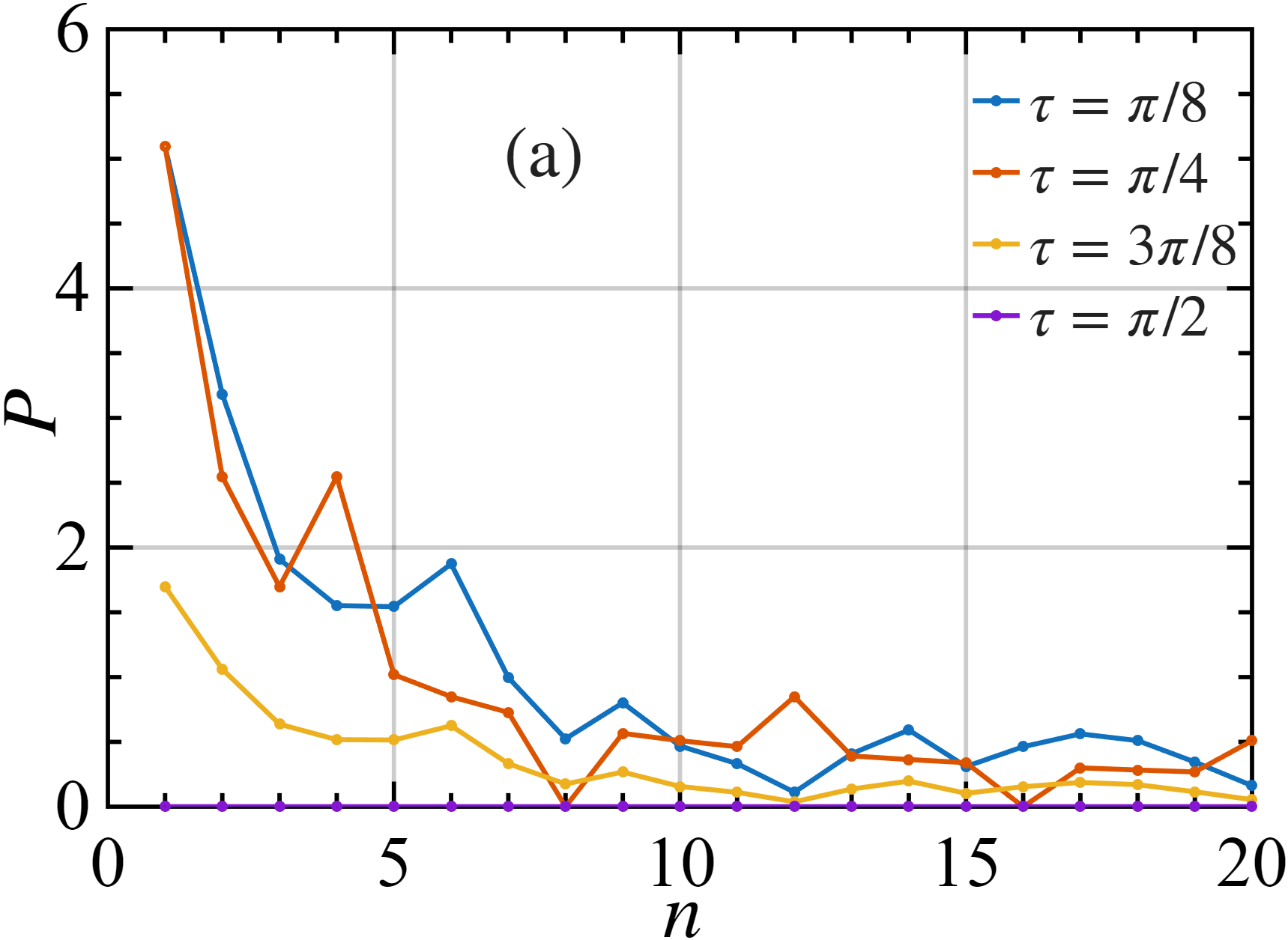}
\includegraphics[width=0.246\linewidth,height=0.18\linewidth]{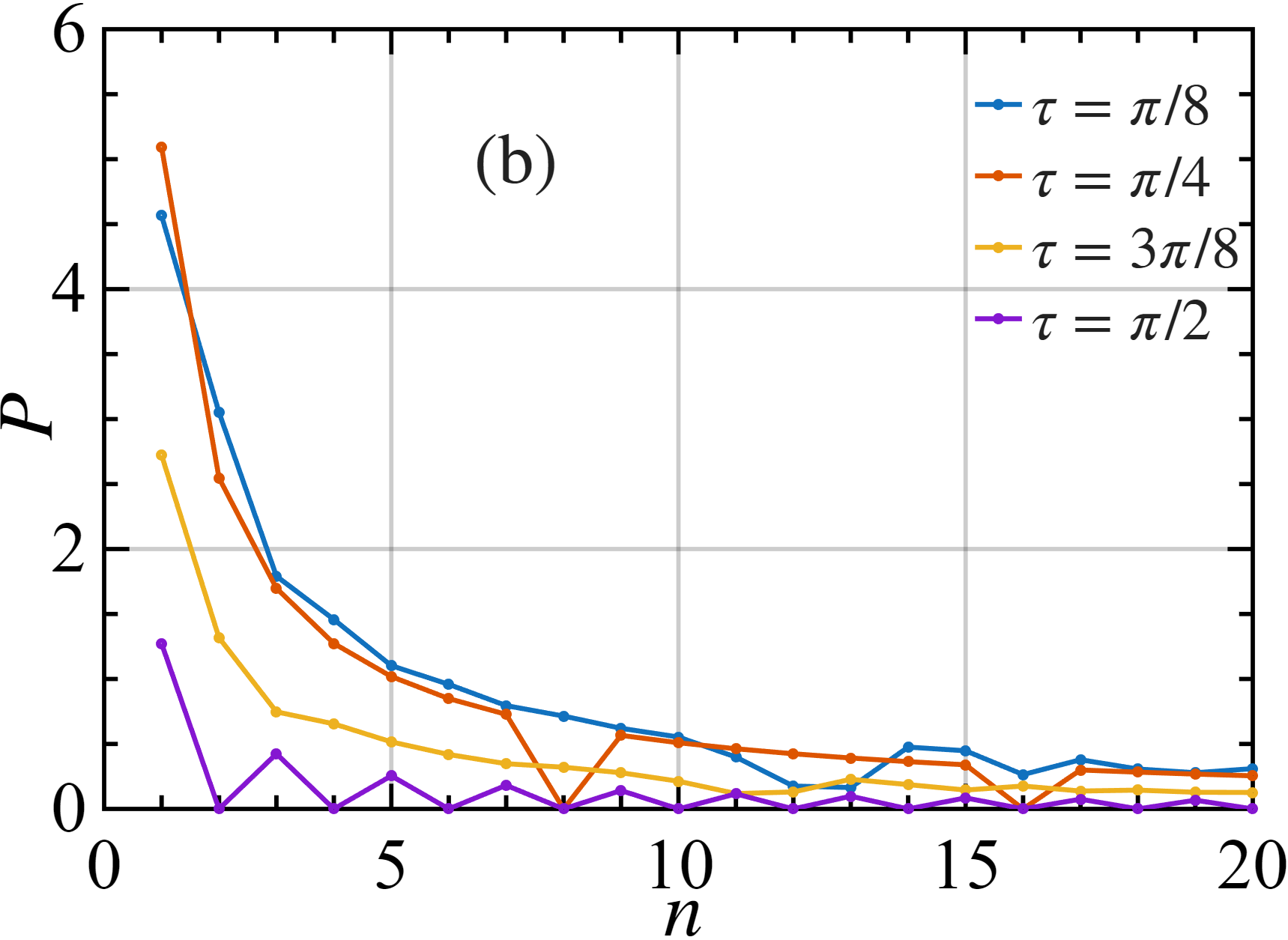}
\includegraphics[width=0.246\linewidth,height=0.18\linewidth]{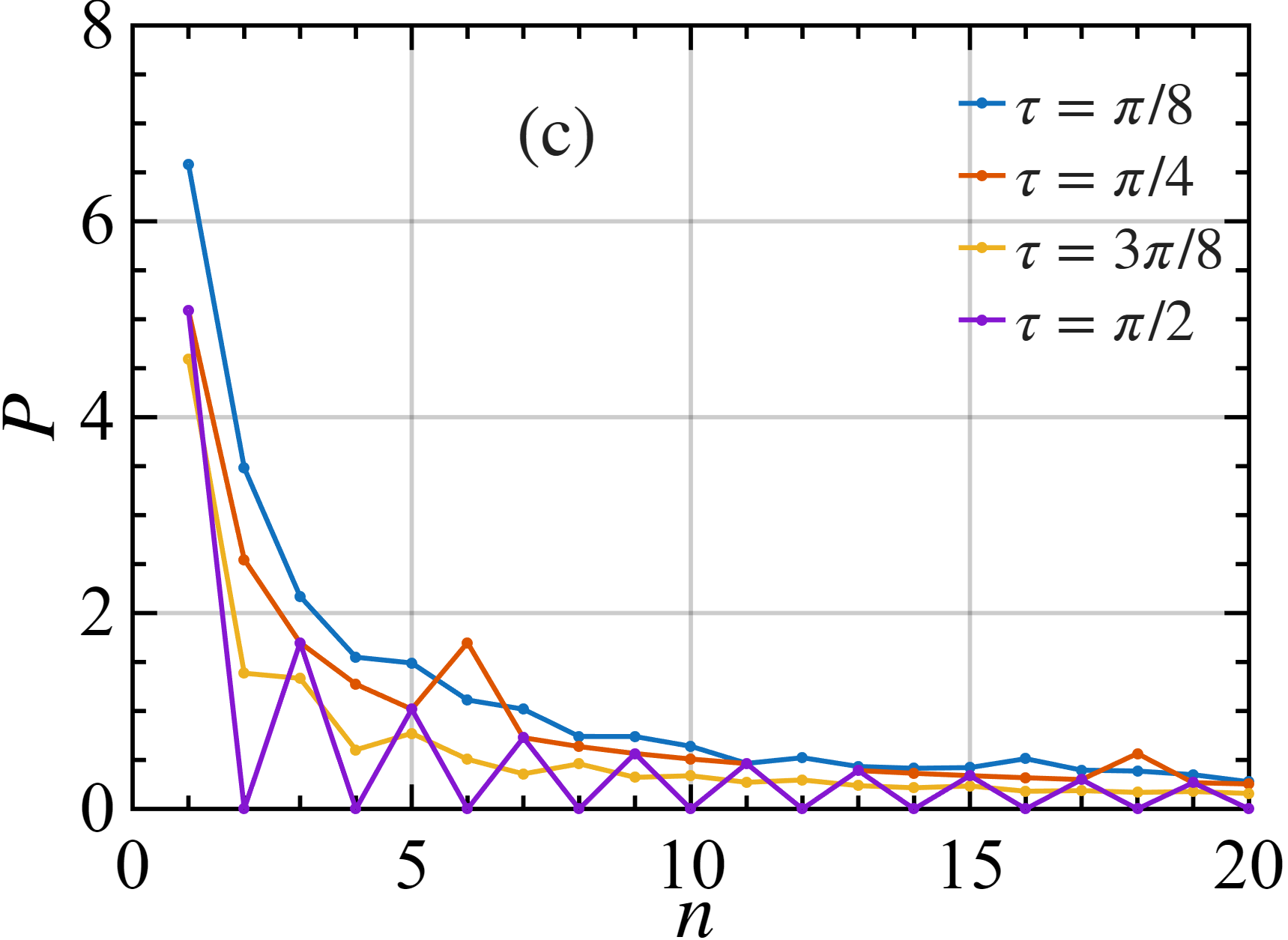}
\includegraphics[width=0.246\linewidth,height=0.18\linewidth]{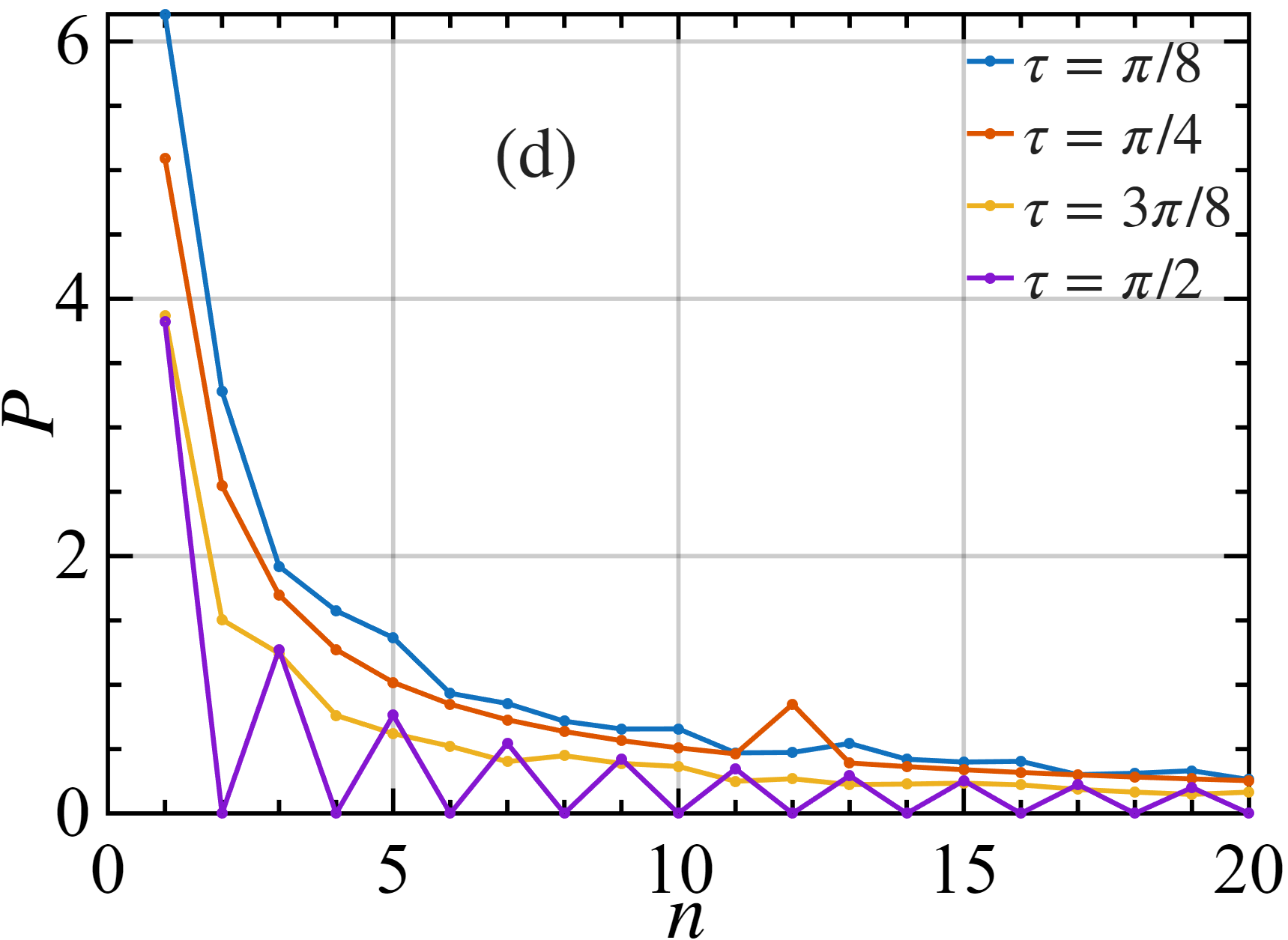}
    \caption{Same as Fig.~\ref{long_E_tau}, but for a nearest-neighbor interacting charger.}
    \label{NN_E_tau}
\end{figure}

Finally, we fix the symmetric driving protocol to $\tau_0=\tau_1=\pi/4$ and investigate how the maximum charging power scales with system size in the nearest-neighbor regime. For all boundary conditions and dynamical settings considered, the peak charging power increases approximately linearly with $N$, as shown in Figs.~\ref{E_NN_N}. Although the absolute magnitude of the charging power in the nearest-neighbour case is generally lower than that achieved with long-range interactions, the observed linear scaling confirms that extensive structural behavior persists even in the short-range limit. This indicates that collective enhancement of charging power does not rely exclusively on long-range couplings but remains robust under nearest-neighbor interactions.
\vspace{0.25cm}
\begin{figure}[H]
    \centering
\includegraphics[width=0.24\linewidth,height=0.20\linewidth]{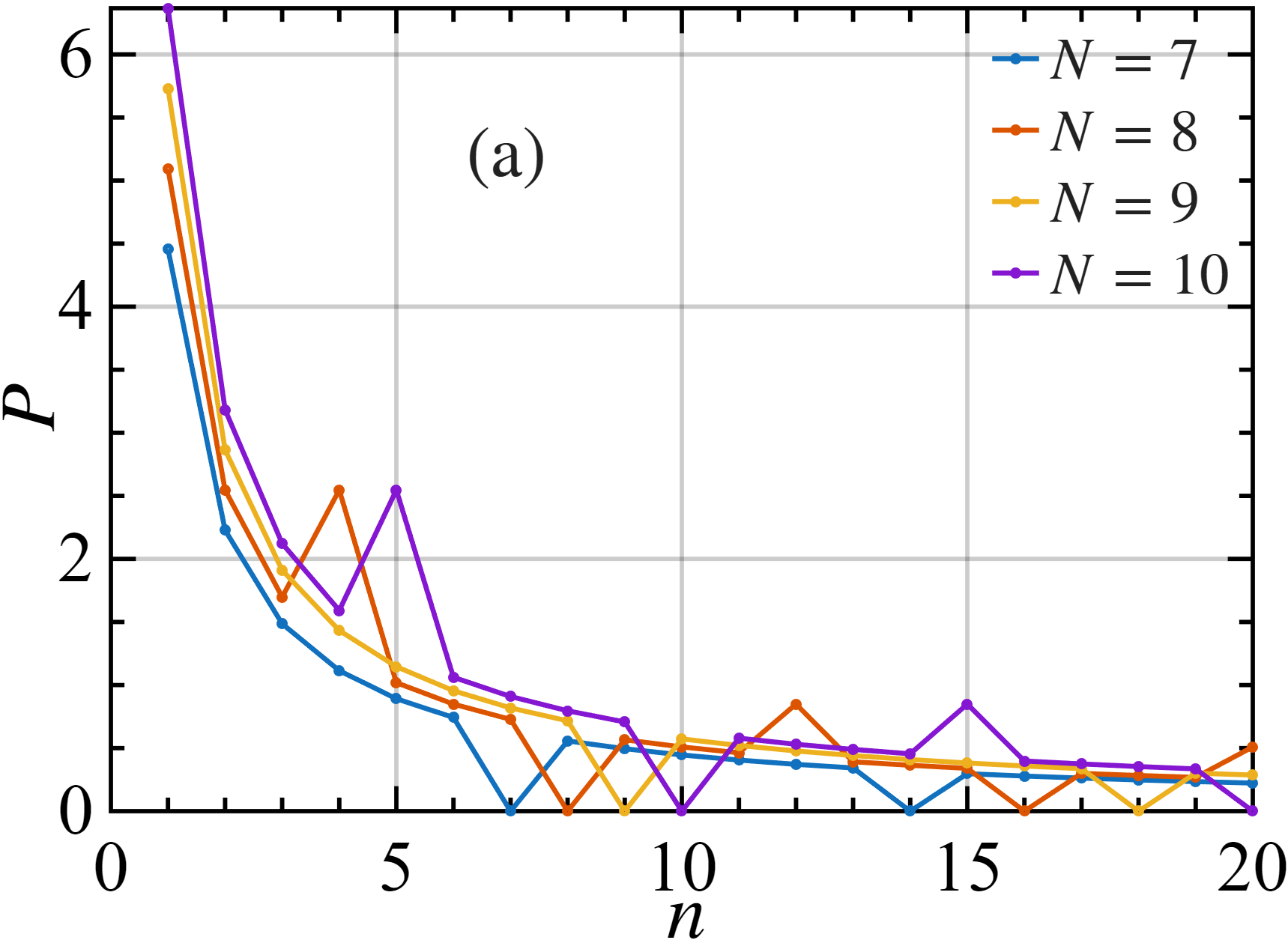}
\includegraphics[width=0.24\linewidth,height=0.20\linewidth]{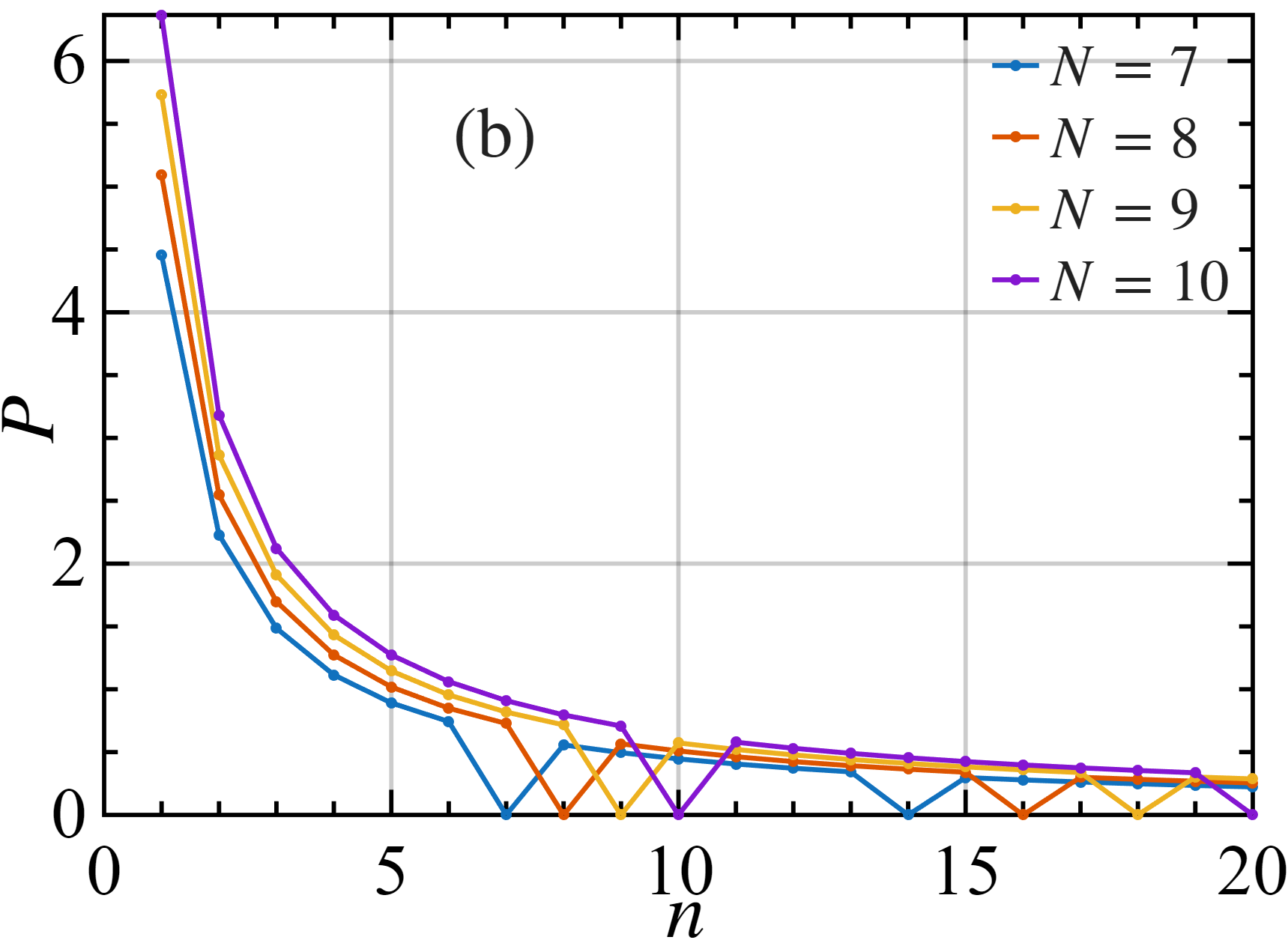}
\includegraphics[width=0.24\linewidth,height=0.20\linewidth]{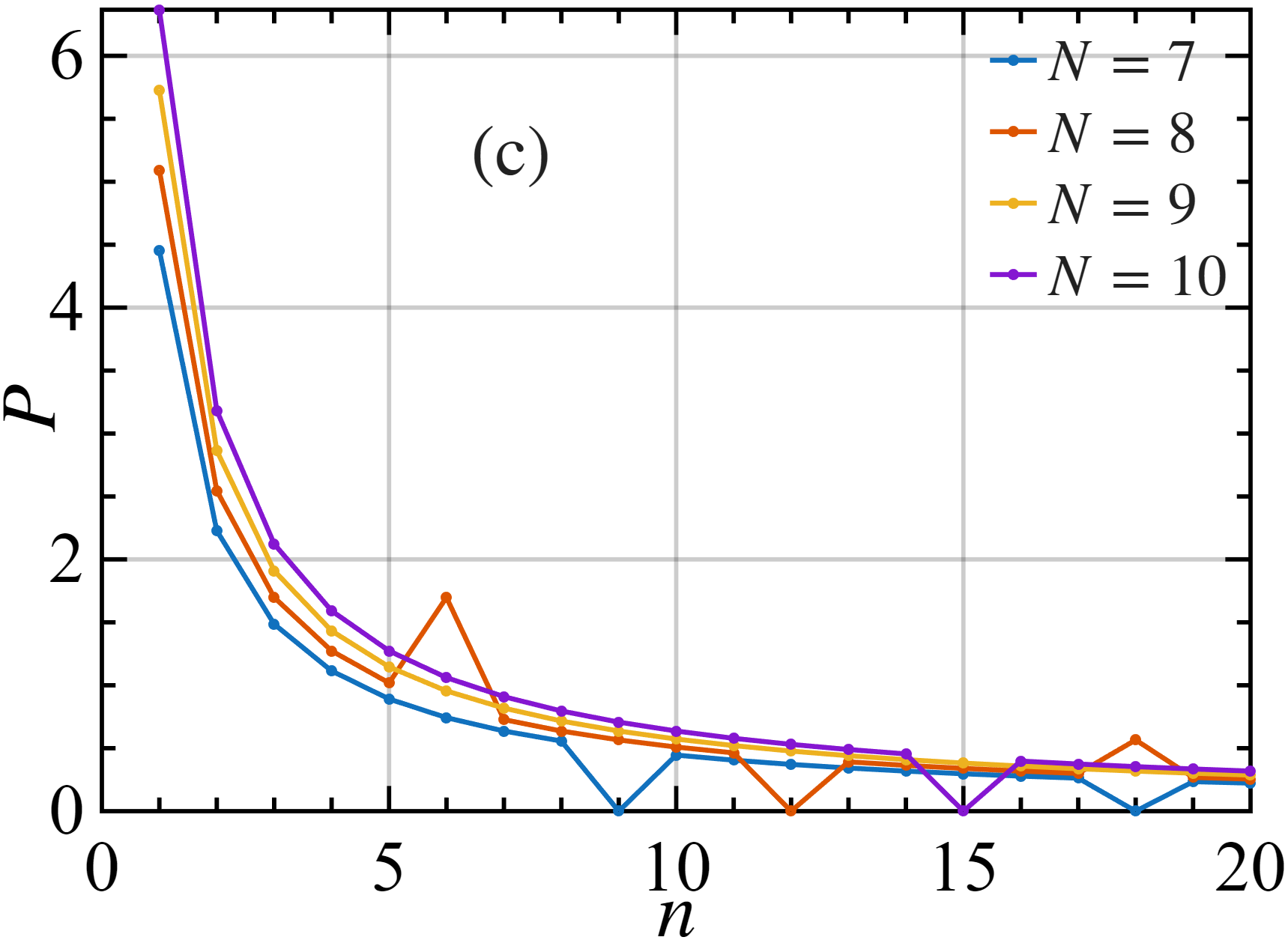}
\includegraphics[width=0.24\linewidth,height=0.20\linewidth]{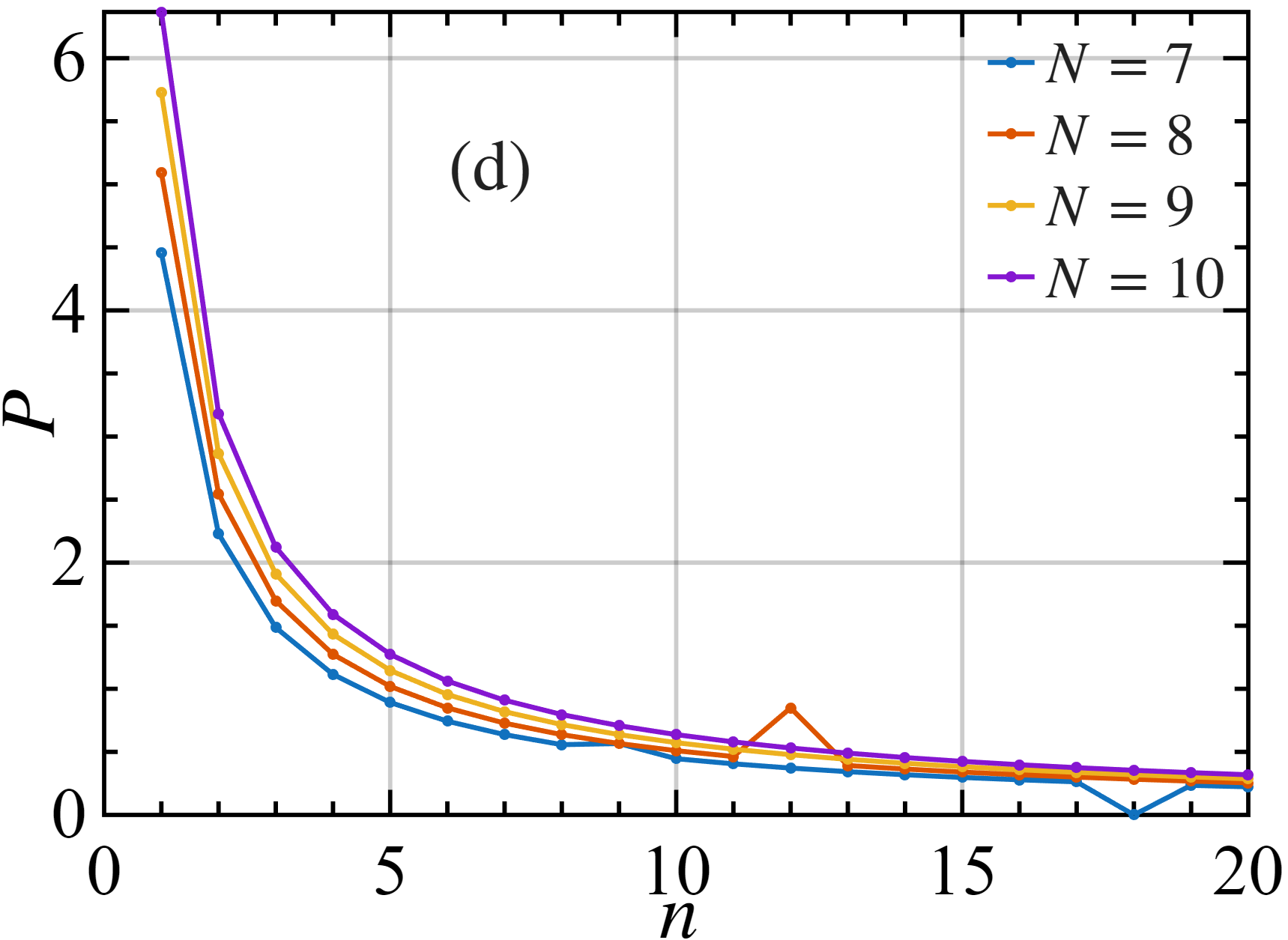}
    \caption{ Same as Fig.~\ref{E_ATA_N}, but for a nearest-neighbor interacting charger.}
    \label{E_NN_N}
\end{figure}

\end{document}